\documentclass[journal, twocolumn]{IEEEtran}

\usepackage{graphicx}
\usepackage{upgreek}
\usepackage{sidecap}
\usepackage{dcolumn}
\usepackage{bm}
\usepackage[utf8]{inputenc}
\usepackage[T1]{fontenc}
\usepackage{mathptmx}
\usepackage{textcomp}
\usepackage{siunitx}
\usepackage{bigints}
\usepackage{setspace}
\usepackage{hyperref}
\usepackage[switch]{lineno}
\usepackage[table,xcdraw,dvipsnames]{xcolor}
\usepackage{booktabs}
\usepackage{float}
\usepackage{xcolor}
\usepackage{ulem}
\usepackage{multirow}
\usepackage{hhline}
\usepackage{tabularx}
\usepackage{wrapfig}
\usepackage{array}
\usepackage{enumerate} 
\usepackage{stfloats}
\usepackage{chngcntr}
\usepackage[table]{xcolor}
\usepackage{soul}
\usepackage[numbers,sort&compress]{natbib}
\usepackage[font=small,labelfont=bf]{caption} 
\usepackage{url}

\begin{document}

\title{Review on spin-wave RF applications}

\author{\IEEEauthorblockN{Khrystyna~O.~Levchenko$^{1}$\IEEEauthorrefmark{1}, Kristýna~Davídková$^{1,2}$,  Jan~Mikkelsen$^{3}$, and Andrii~V.~Chumak$^{1}$\IEEEauthorrefmark{2}}
\\
\IEEEauthorblockA{
$^1$ Faculty of Physics, University of Vienna, 1090 Vienna, Austria.
\\
$^2$ Vienna Doctoral School in Physics, University of Vienna, 1090 Vienna, Austria.
\\
$^3$ Department of Electronic Systems, Aalborg University, 9220 Aalborg, Denmark.
\\
Email: \IEEEauthorrefmark{1}\textit{khrystyna.levchenko@univie.ac.at},  }
       \IEEEauthorrefmark{2}\textit{andrii.chumak@univie.ac.at}}
\maketitle
\begin{abstract}
This review explores the development of spin-wave technology, highlighting magnonics as a promising route for radio frequency (RF) communication systems. The rollout of 5G and the upcoming 6G networks intensifies the demand for devices that can operate at higher frequencies while remaining scalable, compact, and energy-efficient — requirements that spin waves are well suited to meet. The first two sections revisit the fundamentals of magnonics, trace major milestones in spin-wave research, and summarize recent advances in materials and device design. The third section reviews RF applications studied over the past 50 years, with emphasis on key passive components, such as filters, limiters, delay lines, phase shifters, and directional couplers. The final section discusses both the advantages and the open challenges of spin-wave devices, including insertion losses, linearity, and power handling, together with the strategies to address them. By linking fundamental insights with technological needs, this review outlines a path toward practical RF platforms. Spin-wave-based devices, with their scalability, versatility, and potential for low-power operation, hold strong promise for future wireless communication, particularly in the 5G and 6G era.
\end{abstract}

\tableofcontents

\vspace{6pt}
\section{\label{Introduction}Introduction}
{\fontdimen2\font=3pt} 
\subsection{State-of-the-art and challenges of the commercial communication systems}
\vspace{-1pt}
Each generation of mobile communication systems — from GSM to 5G and the forthcoming 6G — has brought increasingly higher demands on bandwidth, latency and functionality. The initial deployment of 5G in the EU relied on frequency range FR1 (sub-6 GHz), particularly the bands n28 (703 – 748 MHz) and n78 (3.3 – 3.8 GHz)  \cite{3GPP, EU5G, FRWiki}. Yet, the maximum channel bandwidth of FR1 is 100 MHz due to the scarcity of continuous band in this crowded range \cite{EU5G}. To extend capacity, current efforts target FR3 (7.125 – 24.25 GHz) \cite{bazzi2025upper,FRWiki}, while the demand for higher throughput and lower latency accelerates the adoption of FR2 (24.25 – 90 GHz) \cite{FR2RS,EU5G}, with 3GPP allocations spanning from n258 (24.25 – 27.5 GHz) to n263 (57 – 71 GHz) \cite{3GPP,FRTool}. Meeting these requirements calls for RF components that are scalable, compact, and energy-efficient — characteristics that spin waves are well suited to provide. Access to higher frequency ranges enables broader channel bandwidths and higher data rates, essential for the Internet of Things (IoT), autonomous systems, and high-definition streaming \cite{EU5G}. \setlength{\parskip}{2pt}

\begin{figure}[ht!]
\centering
 \includegraphics[width=1\columnwidth]{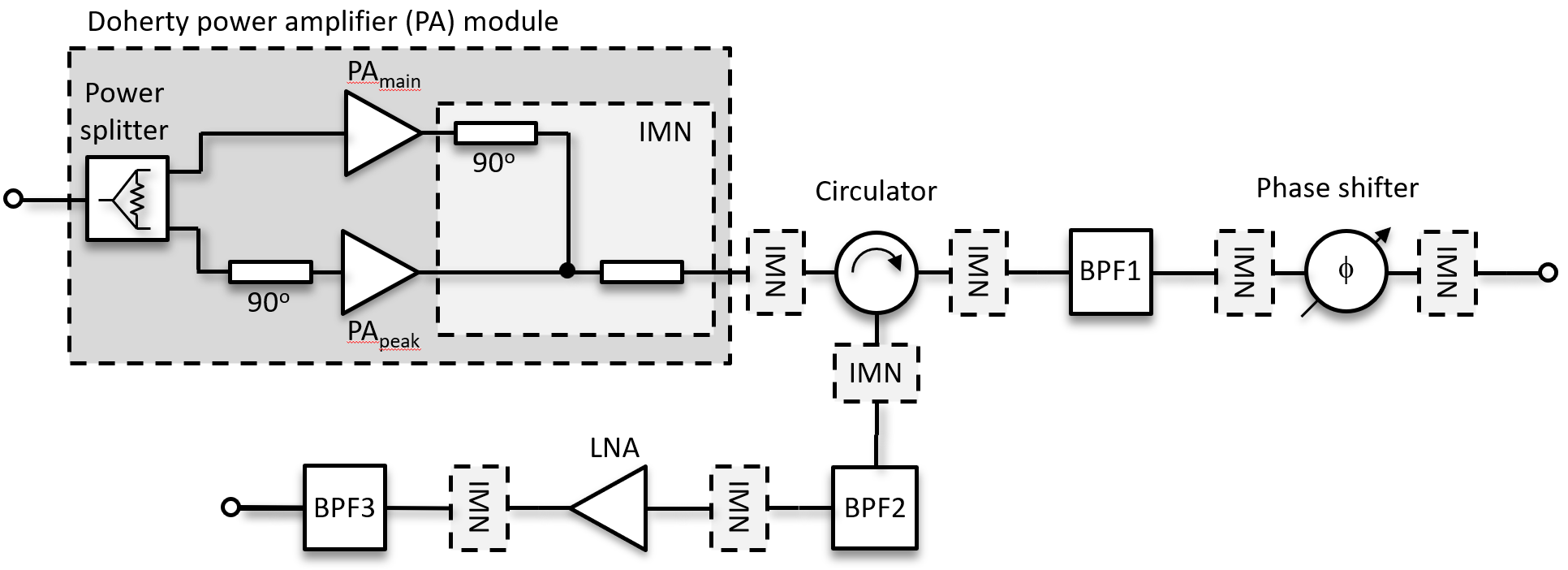}
   \caption{Generic radio channel front-end architecture with typical functional blocks: \textbf{Doherty PA }- power amplifier module, which consists of a power splitter (divides the incoming signal into multiple signals with even or uneven power levels) and two power amplifiers (boost the power of a signal to ensure its transmission) - $\mathrm{PA}$\textsubscript{main} and $\mathrm{PA}$\textsubscript{peak}; \textbf{IMN} - impedance-matching network for efficient power transfer between the signal source and the power amplifier; \textbf{circulator} - isolates high power transmitter from the sensitive receiver; \textbf{BPF}(1/2/3) - passband filter that allows signals within a designated frequency band to pass; \textbf{LNA} - low-noise amplifier, that enhances weak incoming signal from the antenna while adding minimum noise; \textbf{phase shifter} - adjusts the phase of the signal to control the direction of transmission or reception by the antenna.}
   \label{f:1}		
\end{figure}

Currently, a fundamental and widely adopted telecommunications technology is the Multiple-Input Multiple-Output (MIMO) RF system \cite{MIMO, wallace2012guest}. It implements a number of parallel RF channels, each consisting of both passive and active elements connected to multiple antennas (Fig.~\ref{f:1}) in order to simultaneously transmit and receive multiple data streams. Key performance metrics for such systems are power efficiency of the power amplifier module, insertion loss of all passive blocks between the amplifier output and the antenna, manufacturing complexity, size, and costs. The adoption of MIMO provided a significant boost in system capacity, needed for current and future communication generations. However, operation at higher frequencies still presents significant challenges, including increased signal losses, need for advanced materials, and greater power consumption to maintain signal quality \cite{araujo2016massive, rusek2012scaling}.

The performance of active components, such as amplifiers, mixers, modulators (units that require external power to process signals\footnote{Depending on the design, certain RF units can be realized either as active or as passive elements, e.g., filters, mixers, phase shifters.}), in terms of size and signal transmission is largely driven by semiconductor technology. At this stage of development, magnonics can not compete with the semiconductor-based systems, especially regarding insertion losses, but might be a viable long-term candidate due to its potential for reduced power consumption and smaller noise. In MIMO architectures, where numerous RF channels operate in parallel, power efficiency becomes critical at 5G high-band and anticipated 6G frequencies. The combined power dissipation of such MIMO system calls for advanced thermal management and complex cooling solutions, increasing the overall form factor and reducing power efficiency. While isolator-based magnonics can partially mitigate these losses, particularly by stabilizing high-power amplifiers during beam scanning, they address only a small fraction of the overall power dissipation in dense MIMO systems, leaving other active components unaffected.\setlength{\parskip}{2pt}


On the other hand, passive components, such as filters, couplers, power limiters, phase shifters, mixers, transmission lines, etc (units that direct and control RF signals without external power source or amplification) are more 'open' to new technologies and materials due to accumulated issues unresolved with current MIMO approach. These units often rely on monoblock ceramic resonators \cite{tan2023monoblock, carceller2017practical}, made from high-dielectric-constant materials and silver coatings to achieve moderate insertion loss and a high quality factor Q (the ratio of stored to lost energy per oscillation cycle). However, at frequencies above 10 GHz, these benefits diminish due to design complexity, limited scalability, increased insertion losses, and the brittle, inflexible nature of the ceramic materials. Here, magnonics offers potential benefits: with on-chip magnet integration and proper selection of materials and device architecture, it is possible to realize compact, reconfigurable, low-bias SW devices operating efficiently across a wide frequency range beyond 10 GHz (see later sections for details) \cite{louis2016bias, cocconcelli2025self, haldar2021functional}. Other passive components, such as power limiters based on semiconductor PIN diodes, also face challenges at higher frequencies in a form of increased parasitic capacitance and inductance, limited isolation and delayed recovery from charge storage effects. In contrast, magnonic devices offer a charge-free operation and moderate intrinsic losses even in the high-frequency regime \cite{Barman2021, Chumak2022}. Moreover, magnonic power limiters have the potential to outperform PIN-diode-based counterparts in the GHz-to-THz range, as they rely on spin-wave dynamics that evolve on nano- to picosecond timescales and have no recovery delays associated with charge carrier motion. \setlength{\parskip}{2pt}

Consequently, most conventionally used filters, multiplexers and delay lines for wireless data transmission systems are based on the compact Surface Acoustic Waves (SAW) technology \cite{delsing2019, ilderem2020technology}, which has enabled multiband telecommunication devices to meet current demands (e.g., LTE, 5G low band). However, frequency limitations persist, mainly due to increased insertion loss above 3 GHz \cite{delsing2019, ilderem2020technology}, along with the exposure limitations of interdigitated transducers (IDTs)\cite{delsing2019, hara2010super}. Ergo, SAW technology cannot support a reliable operation across the entire 5G frequency range. This bottleneck is partially addressed by the Bulk Acoustic Waves (BAWs) technology, which allows for a reliable operation up to 6-10~GHz, and experimental vertical-BAW pushing the bar up to 25-30~GHz. The main drawbacks of BAW technology include increased damping, inefficient energy transfer between the transducer and the wave within the propagation medium, strict requirements for isolating the wave from the substrate to prevent unwanted energy dissipation, and the resulting complexity of the fabrication process \cite{aigner2008saw}. Unlike SAW, bulk acoustic waves propagate through the volume of the material, which can lead to difficulties in maintaining signal integrity and resonance efficiency at high frequencies. It is worth noting that acoustic technologies targeting the > 30~GHz range are actively evolving, with advanced BAW devices based on periodically poled materials, such as AlScN and LiNbO\textsubscript{3} showing promise due to their high electromechanical coupling and scalability to higher frequencies. However, these solutions still face challenges, including fabrication complexity and limited on-chip tunability, stemming from their fixed resonator geometries and inherent piezoelectric properties.\setlength{\parskip}{2pt}

As performance trade-offs between BAW and SAW RF devices continue to limit overall system efficiency, manufacturers are actively exploring alternatives. Spin-wave technology presents a promising solution by addressing several key limitations of traditional devices: (1) \textbf{a wide frequency range from < 1 GHz to THz} \cite{ChumakHandbook2019}; (2) \textbf{compatibility with the industry-standard fabrication} (e.g., photolithography); (3) \textbf{inherent isolation}, as spin waves are naturally confined within the magnetic material and do not propagate into the substrate or adjacent components, unlike SAWs; (4) \textbf{rich nonlinear functionalities}, such as power limiting and improved signal-to-noise ratio. Furthermore, spin waves offer enhanced \textbf{flexibility}, as their dispersion characteristics can be tuned via material's magnetic parameters, structure modification or external biasing field. This allows the development of reconfigurable RF devices with dynamic control on nanosecond timescales (more on the spin-wave advantages in \textit{Section II B}). While this field is still emerging and presents certain challenges (as outlined in {Section IV}), its rapid development, inherent flexibility, and long-term potential make it a particularly promising direction for future RF technologies. \setlength{\parskip}{2pt}

In this review, we aim to engage both the magnonics and RF engineering communities by examining the challenges of modern RF components (primarily passive elements) through the lens of spin-wave technology. 

\vspace{-1pt}
\subsection{Milestone history of spin-wave research and technology}

\vspace{-3pt}
A \textbf{spin wave (SW)} represents a collective excitation of spin system in a magnetically ordered media \cite{Bloch1930, Holstein1940, Gurevich1996, Stancil}. The quanta of these excitations are called \textbf{magnons}. \textbf{Magnonics}, also known as magnon spintronics, investigates information transport and processing through spin waves as an alternative to, or together with charge currents \cite{Serga2010, kruglyak2010magnonics, demokritov2012magnonics}. Magnons themselves are charge-neutral quasiparticles, unlike electron currents in conventional electronics, which allows them to propagate without energy dissipation, i.e., without Joule heating. Owing to their inherent wave nature, the information in SWs can be encoded simultaneously in amplitude, phase, and frequency. Although the study of SW phenomena is less than a century old, it has become a rapidly developing area of research, giving rise to a variety of advanced device concepts. These developments offer a promising alternative to conventional RF- and CMOS technologies \cite{Wang2020} and pave the way for cutting-edge fields of magnonic neuromorphic computing \cite{Bracher2018, Wang2020Resonator, Papp2021} or quantum computing \cite{Chumak2022, Awschalom2021, Lachance-Quirion2019}.

\vspace{3mm}
\underline{The timeline of milestone discoveries}
\begin{itemize}
 \item[\textcolor{NavyBlue}{\textbullet}] {\bf \textcolor{NavyBlue}{1919:}} The phenomenon of \textbf{ferromagnetic resonance (FMR)} was first experimentally discovered independently by Griffiths and Zavoisky, while its first comprehensive theoretical description was given by Kittel in 1948 \cite{Kittel1948}.  \setlength{\parskip}{2pt}
 
 \item[\textcolor{NavyBlue}{\textbullet}] {\bf \textcolor{NavyBlue}{1921:}} Stern and Gerlach performed the first experimental \textbf{observation of the electron spin} \cite{Gerlach1922}. Silver atoms, while passing through an inhomogeneous magnetic field, were deflected up or down by the same amount, indicating a directional quantization of quantum mechanical angular momentum – spin. \setlength{\parskip}{2pt}
 
 \item[\textcolor{NavyBlue}{\textbullet}] {\bf \textcolor{NavyBlue}{1928:}} \textbf{Theoretical equation for spin 1/2 particles} was developed by Dirac, showing positive and negative energy states \cite{dirac1928quantum}. \setlength{\parskip}{2pt}
 
{\fontdimen2\font=2.8pt \item[\textcolor{NavyBlue}{\textbullet}] {\bf \textcolor{NavyBlue}{1930:}} \textbf{Spin waves} were proposed by Bloch to explain the reduction of spontaneous magnetization in ferromagnets near the Curie temperature, with the famous {\textit{T}\textsuperscript{3/2}} Bloch law being an indirect confirmation of SW existence \cite{Bloch1930}. \setlength{\parskip}{2pt}}
 
\item[\textcolor{NavyBlue}{\textbullet}] {\bf \textcolor{NavyBlue}{1935:}} The development of the theory describing the relaxation dynamics of the magnetization vector by Landau and Lifshitz, which led to the formulation of the fundamental equation of magnetization dynamics – \textbf{Landau-Lifshitz-Gilbert equation (LLG)} \cite{Landau1935,  gilbert2004phenomenological}. \setlength{\parskip}{2pt}

 \item[\textcolor{NavyBlue}{\textbullet}] {\bf \textcolor{NavyBlue}{1940-56:}} Holstein and Primakoff \cite{Holstein1940}, and later Dyson \cite{Dyson1956}, further developed theoretical framework of the field, introducing a concept of spin-wave quanta, now known as \textbf{magnon}. It was predicted that these quanta should behave as weakly interacting quasiparticles obeying the Bose-Einstein statistics. Notable pioneering studies also include SW detection via neutron scattering by Brockhouse \cite{brockhouse1957scattering}, and through inelastic light scattering by Grünberg \cite{grunberg1981magnetostatic}. The subfield of magnetism associated with the quantum magnetic dynamic phenomena was later called \textbf{magnonics}. Nowadays, similarly to electronics, magnonics is not limited by the scale of its medium/quanta, but covers a broad field of SW phenomena \cite{kruglyak2010magnonics}. \setlength{\parskip}{2pt}
  
\item[\textcolor{NavyBlue}{\textbullet}] {\bf \textcolor{NavyBlue}{1946:}} The \textbf{first direct observation of spin waves} by Griffiths \cite{GriffithsJ.H.E.1946} via the FMR spectroscopy of non-propagating SWs with wavevector  \(\textbf{k} = 0\) (uniform precession). Pioneering experiments on the propagating SWs \(\textbf{k} \neq 0\)  were performed two decades later by Fleury et al. \cite{Fleury1966} using Brillouin light scattering (BLS) spectroscopy. \setlength{\parskip}{2pt}
 
\item[\textcolor{NavyBlue}{\textbullet}] {\bf \textcolor{NavyBlue}{1952:}} \textbf{Microwave (MW) magnetic devices} debuted on the applied scene with Hogan’s work \cite{hogan1952ferromagnetic} on the gyrator using Faraday rotation. This low-loss (7 dB) broadband device demonstrated potential for one-way transmission systems, microwave circulators, microwave switches, electrically controlled variable attenuators and modulators. \setlength{\parskip}{2pt}
 
\item[\textcolor{NavyBlue}{\textbullet}] {\bf \textcolor{NavyBlue}{1952-57:}} The \textbf{first non-linear measurements} were reported by Bloembergen and Damon \cite{Bloembergen1952}, and Bloembergen and Wang \cite{Bloembergen1954} at the FMR of the material, while ‘pumping’ high-power RF field transverse to the applied magnetic field. As explained by Suhl \cite{Suhl1956,Suhl1957}, certain SW modes are parametrically excited when the RF field induces uniform magnetization precession and, simultaneously, the input RF power exceeds a threshold. Interest to nonlinear physics peaked in the 80s, but it is already making a swift comeback to cutting-edge research since it provides valuable guidelines for the efficient generation and amplification of SWs on the nanoscale – a key to developing advanced magnonic networks. \setlength{\parskip}{2pt}

{\fontdimen2\font=3pt 
\item[\textcolor{NavyBlue}{\textbullet}] {\bf \textcolor{NavyBlue}{1956:}} Synthetic dielectric ferrimagnet \textbf{yttrium iron garnet (Y\textsubscript{3}Fe\textsubscript{5}O\textsubscript{12} – YIG)} was first fabricated by Bertaut and Forrat \cite{bertaut1956structure}. A comprehensive investigations of the crystallographic and magnetic properties of single-crystal YIG spheres was first conducted a year later by Geller and Gilleo \cite{Geller1957}. This is an indispensable material for magnonics, akin to silicon in semiconductor physics, due to its wide range of accessible frequencies (MHz - GHz), smallest SW losses and the narrowest FMR linewidth.  Bulk YIG supports centimeter-scale SW propagation with lifetimes up to 1~\(\upmu\)s at room temperature. The Gilbert damping parameter of YIG (both for bulk and  for \(\sim \)100~nm-thick films) reaches values of about 10\textsuperscript{-5}, allowing the development of RF devices with a high-quality factor {\textit{Q}} in the order of 10\textsuperscript{5} under ideal conditions. Practical realization, however, should also consider geometry, fabrication tolerances, and coupling losses. These properties justify YIG's growing importance in RF applications \cite{Harris2012} (e.g., in filters, Y-circulators, microwave generators) and in high-fidelity experimental physics, particularly for exploring novel phenomena \cite{Yttrium, Chumak2021A}. The operating principles of many modern YIG-based devices rely on either non-propagating SWs (FMR) or on the modification of the electromagnetic wave (EMW) properties, which propagates in a medium with a YIG sphere. In this review, we focus primarily on RF devices operating with propagating SWs. \setlength{\parskip}{2pt}

\item[\textcolor{NavyBlue}{\textbullet}] {\bf \textcolor{NavyBlue}{1959:}} The \textbf{first publication on the characteristics of ferrite RF power limiters} by Uebele \cite{Uebele1959}. The study introduced several techniques for tuning the insertion loss of ferrite-loaded waveguide structures at high RF powers, and described the operating characteristics of a device. \setlength{\parskip}{2pt}

\item[\textcolor{NavyBlue}{\textbullet}] {\bf \textcolor{NavyBlue}{1960:}} The \textbf{discovery of magnetostatic surface spin waves (MSSWs)}, also known as Damon-Eshbach SWs, named after the pioneering scientists \cite{Eshbach1960}, marked a milestone in SW physics. Authors demonstrated how in an in-plane magnetized slab, SW modes transition from a volume to a surface-localized state as the frequency increases. These surface modes, absent in bulk but present in bounded media, exhibit exponential decay into the film and extend the SW spectrum beyond that of harmonic plane waves. Their work unified the understanding of volume and surface spin dynamics in magnetically ordered films within the magnetostatic limit. \setlength{\parskip}{2pt}

\item[\textcolor{NavyBlue}{\textbullet}] {\bf \textcolor{NavyBlue}{1970-80s:}} A peak in high-profile investigations dedicated to the \textbf{spin-wave-based passive and active RF devices} \cite{Owens1985, Stitzer1983}. One of the most influential and comprehensive overviews of \textbf{magnetostatic spin wave (MSW)} RF devices was published in \textit{IEEE} \textbf{76}(2) \cite{Adam1988, Glass1988, Ishak1988, Rodrigue1988, Schloemann1988}, where each article addressed a different aspect of the field. Among the discussed topics were SW-based filters, transducers, delay lines, directional couplers, phase shifters, parametric amplifiers, signal-to-noise enhancers and circulators, to name the few. A detailed discussion is provided in later sections of the review. Following the peak, several research groups continued to optimize the performance of SW RF devices, e.g., \textbf{by reducing the insertion loss of RF filters to as low 2.5 dB} \cite{bobkov2002microwave}. \setlength{\parskip}{2pt}

\item[\textcolor{NavyBlue}{\textbullet}] {\bf \textcolor{NavyBlue}{2001-now:}} A promising direction of universal magnonic components emerged with the introduction of \textbf{magnonic crystals} by Nikitov et al. \cite{nikitov2001spin} in 2001, building on earlier work by Sykes, Adam, and Collins in 1976 \cite{sykes1976magnetostatic}. Magnonic crystals are artificial materials with a spatially periodic variation of magnetic properties for versatile applications, including RF, logic and data-processing (e.g., filters \cite{Chumak2017, merbouche2021frequency}, sensors \cite{inoue2011investigating, metaxas2015sensing}, transistor \cite{Chumak2014}, resonators \cite{reed1985current}, logic gates \cite{nikitin2015spin}, etc). Their key properties can be tailored by (often simultaneously): (1) the use of different materials with suitable magnetic properties (e.g., saturation magnetization \cite{gubbiotti2010brillouin, obry2013micro, mruczkiewicz2017spin} ); (2) the choice of the periodic pattern \cite{wang2013design} and (3) external factors, such as the bias magnetic field \cite{chumak2009current, levchenko20251d} or temperature \cite{vogel2015optically}. Since then, a significant progress has been made in developing various types of magnonic crystals, improving their functionality, and understanding the underlying physics \cite{krawczyk2014review, Chumak2022}.\setlength{\parskip}{2pt}

\item[\textcolor{NavyBlue}{\textbullet}] {\bf \textcolor{NavyBlue}{2004-now:}} The idea of encoding binary data into the SW amplitude was first stated by Hertel et al. \cite{Hertel2004}, marking the beginning of \textbf{spin-wave digital data-processing} era \cite{Mahmoud2020}. The experimental development of \textbf{spin-wave logic} began with the work of Kostylev et al. \cite{Kostylev2005}, and was followed by Schneider et al.  \cite{Schneider2008} demonstrating a proof-of-principle XNOR logic gate and a universal NAND logic gate. The prototypes of NOT, NOR and NAND logic gates were numerically simulated and experimentally realized by Lee et al. \cite{Lee2008}. In all aforementioned devices, the data were encoded in SW amplitude, where a finite amplitude defines logic “1” and zero amplitude corresponds to logic “0”). Alternatively, Khitut et al. \cite{Khitun2010}, proposed to use the SW phase, rather than the amplitude, to digitize information. This approach allows a straightforward implementation of a NOT gate in magnonic circuits and supports the realization of \textbf{majority gate} in the form of a multi-input SW combiner \cite{Khitun2010}. A key advantage of this architecture is the ability to operate simultaneously with spin waves of different wavelengths, paving the way for single-chip parallel computing \cite{Khitun2012}. An experimental prototype of a majority gate based on a macroscopic YIG structure was shown later by Fischer et al. \cite{Fischer.2017}, while a \textbf{chip-ready inline majority gate} was developed by Imec in 2020 \cite{talmelli2020reconfigurable}. A major breakthrough came with the demonstration of \textbf{magnon transistor} by Chumak et al. \cite{Chumak2014} enabling all-magnon digital data processing. Following this path, Wang et al. \cite{Wang2020} designed and experimentally verified a \textbf{magnonic directional coupler for integrated magnonic half-adders} using single-mode nanoscale YIG waveguides. The proposed concept, developed with 30-nm technology, offers a footprint comparable to a 7-nm CMOS half-adder, with \(\approx 10\)x smaller energy consumption. \setlength{\parskip}{2pt}

\item[\textcolor{NavyBlue}{\textbullet}] {\bf \textcolor{NavyBlue}{2008-now: }}~\textbf{The miniaturization of magnonic structures from millimeter to micrometer lateral dimensions has advanced significantly for both metallic systems \cite{demidov2008linear, vlaminck2008current, chumak2009spin, vlaminck2010spin} and YIG films }\cite{Chumak2014, sebastian2015micro}. Further progress in downscaling YIG structures to sub-\(\upmu\)m dimensions was achieved by the revolutionary development of high-quality, nm-thick YIG films via liquid-phase epitaxy (LPE) \cite{hahn2013comparative, althammer2013quantitative, Pirro2014, Dubs2017, Dubs2020}, pulsed laser deposition (PLD) \cite{sun2012growth, d2013inverse, yu2014magnetic, Onbasli2014, schmidt2020ultra}, or sputtering \cite{liu2014ferromagnetic, chang2017sputtering, ding2020sputtering, schmidt2020ultra}, accomplishing sub-10 nm thicknesses. Importantly, in nm-thick YIG waveguides, below a certain critical waveguide's width, the exchange interaction starts dominating over a dipolar, resulting in a strong quantization of the energy levels and in the appearance of distinct SW bands in the magnon spectrum. Operating in such a \textbf{single-mode regime} minimizes parasitic scattering into higher width modes, significantly advancing applied nanomagnonics \cite{Wang2020, Wang2020Resonator}. \setlength{\parskip}{2pt}

\item[\textcolor{NavyBlue}{\textbullet}] {\bf \textcolor{NavyBlue}{2016-now:}} \textbf{The optimization and development of novel spin-wave transducers}. Since the efficiency of SW transducers impacts energy consumption in data-processing magnonic devices and contributes to insertion loss in RF devices, an extensive investigation on their improvement has been carried out \cite{bailleul2003propagating, vanderveken2022lumped, connelly2021efficient, bruckner2025micromagnetic, erdelyi2025design}.  \setlength{\parskip}{2pt}

\item[\textcolor{NavyBlue}{\textbullet}] {\bf \textcolor{NavyBlue}{2020-now:}} The research \textbf{revival of spin-wave RF applications driven by 5G technology requirements}. Recent advancements include the demonstration of on-chip MSW resonators \cite{dai2020octave}, stopband filters \cite{feng2023micromachined} and high-power RF pulse measurements via microstrip YIG power limiters \cite{dai2020octave}. A cutting-edge nanoscale three-in-one RF device, integrating a frequency-selective limiter, filter, and delay line, was recently developed by Davídková et al. \cite{davidkova2025nanoscale} further advancing  applied magnonics. \setlength{\parskip}{2pt}

 {\fontdimen2\font=2.7pt} 
\item[\textcolor{NavyBlue}{\textbullet}] {\bf \textcolor{NavyBlue}{2020-now:}} The emergence of a novel \textbf{quantum magnonics} field has unified prior research efforts into a clear-sighted direction. A kick-off moment was likely to the work of Lachance-Quirion et al. \cite{Lachance-Quirion2020}, as it showed the detection of a single magnon with a quantum efficiency up to 0.71 using an entangled hybrid structure of a superconducting qubit and a ferrimagnetic crystal. Currently, a major focus in the field is given to exploration and utilization of \textbf{propagating magnons}, as opposed to non-propagating Kittel mode, to enable dynamic quantum systems with enhanced functionality \cite{karenowska2015excitation, Knauer2023}. One of the most prominent subfields is \textbf{hybrid magnonics}, which explores the coupling of magnons with other physical systems, such as photons, phonons, or qubits, thereby offering a versatile platform for the development of next-generation quantum technology for information processing, wave-based computing, and sensing applications \cite{lachance2019hybrid, haas2020sensitivity, li2020hybrid, Awschalom2021, haas2022development, yuan2022quantum, li2022hybrid, xu2023quantum}.} \setlength{\parskip}{2pt}

\item[\textcolor{NavyBlue}{\textbullet}] {\bf \textcolor{NavyBlue}{2021-now:}} The integration of \textbf{inverse design and machine learning in magnonics} has
introduced modern artificial intelligence (AI)-based tools for solving versatile tasks in the field. Notable developments include non-linear switches, (de-)multiplexers and circulators shown by Wang et al. \cite{wang2021inverse}, as well as neuromorphic computing unit for the vowel sound recognition developed by Papp et al.\cite{Papp2021}. A significant progress was recently achieved by Zenbaa et al. \cite {zenbaa2025universal, zenbaa2025realization} with a first experimental demonstration of reconfigurable, lithography-free, and simulation-independent inverse-design devices capable of implementing diverse RF components and logic units. With the growing power of AI, the field of inverse-design magnonics is advancing rapidly and holds great promise for future magnonic technologies. 
\end{itemize}

\subsection{Selected fundamental principles of magnonics}
\vspace{-2pt}
\subsubsection{\textit{Key definitions}} this \textit{Section} is based on the foundational works of Serga et al.~\cite{Serga2010} and Chumak~\cite{ChumakHandbook2019}. In an unstructured, continuous magnetic system, the magnetization precession \textbf{\textit{M}} is described through the Landau-Lifshitz-Gilbert (LLG) equation of motion:

\begin{equation} \label{LLG}
    \frac{d\textbf{\textit{M}}}{dt}=-\gamma \mu_0 (\textbf{\textit{M}} \times \textbf{\textit{H}}_{\textup{eff}}) + {\frac{\alpha}{M_{\textup{s}}} \Big(\textbf{\textit{M}} \times {\frac{d\textbf{\textit{M}}}{dt}}}\Big) \,\,,
\end{equation}
where \( \gamma\) is the gyromagnetic ratio - constant of proportionality between magnetic moment and angular momentum, \(\mu_0\) - the permeability of free space, \(\textbf{\textit{H}}_{\textup{eff}}\) - the effective magnetic field, \(\alpha\) - the dimensionless Gilbert damping constant, $M_\mathrm{s}$ - saturation magnetization.

\begin{figure}[hbt]
\centering
    \includegraphics[width=0.78\columnwidth]{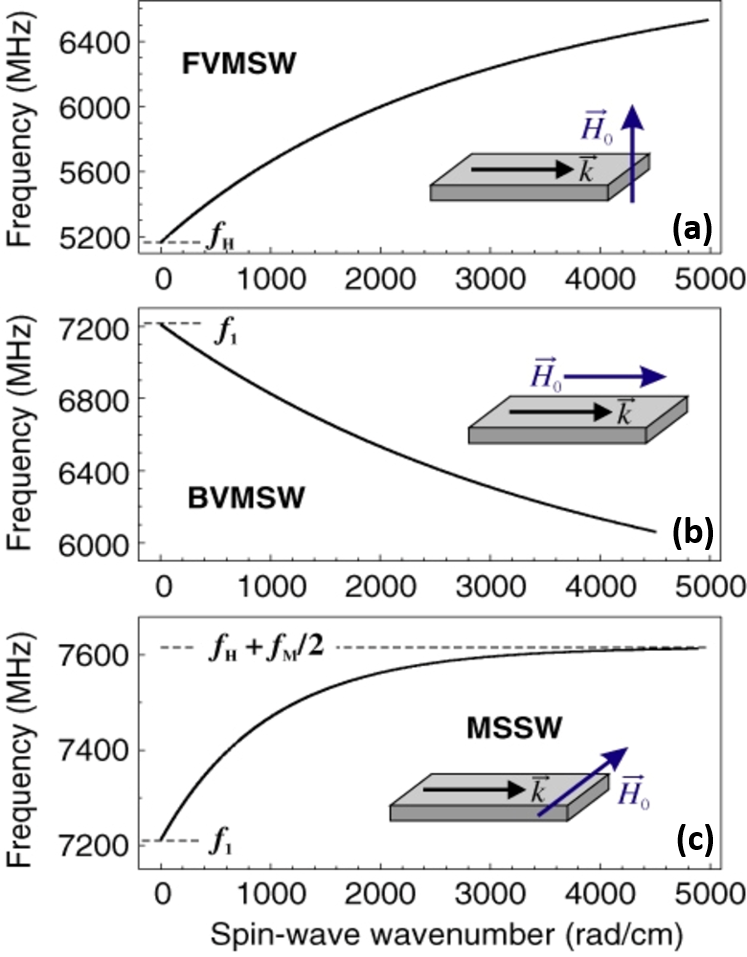}
    \caption{Calculated dispersion characteristics (lowest order thickness modes) using Eq.(2-4) for (a) Forward volume magnetostatic spin wave (FVMSW), (b) Backward volume magnetostatic spin wave (BVMSW), (c) Magnetostatic surface spin wave (MSSW). Bias magnetic field \textit{\textbf{B}} $=$ 184.5~mT, saturation magnetization \(M_{\textup{s}}\) $=$ 175~mT, film thickness $d = 5~\upmu \mathrm{m}$. Adapted from \cite{ChumakHandbook2019}.}
    \label{f:2}		
\end{figure}

Solutions to Eq.~\ref{LLG} in the macrospin approximation include uniform and non-uniform modes of magnetization precession. The uniform mode corresponds to the in-phase precession of individual spins and leads to the Kittel equation for ferromagnetic resonance. The non-uniform mode describes spin waves – spins oscillating at the same frequency but with different phases. 

Three fundamental SW configurations can be excited in a magnetic medium depending on the orientation of applied magnetic field with respect to the direction of SW propagation (Fig.~\ref{f:2}). Each one is characterized by a distinct dispersion relation \(f(k)\), describing the dependence of the SW frequency \(f\) on its wavevector $k$ (wavenumber). Spin waves are called \textbf{magnetostatic} when their behavior is dominated by long-range dipole interactions. In this regime, their dynamics can be described using magnetostatic approximations, and their group velocities are orders of magnitude lower than those of electromagnetic waves \cite{Stancil}.
\setlength{\parskip}{8pt}

\textcolor{NavyBlue}{\scalebox{0.75}{$\bullet$}} \textit{Forward Volume Magnetostatic Spin Wave (FVMSW)} - configuration, corresponding to a magnetic film magnetized normal to its plane; see Fig.~\ref{f:2}(a). The dispersion relation for FVMSW is given by:
\begin{equation}\label{FVMSW}
  f =  \frac{\gamma}{2 \pi} \sqrt{(B- \mu_0 M_\mathrm{s}) \Bigg[ B - \mu_0 M_\mathrm{s} \Bigg(1-\frac{1-e^{-kd}}{kd} \Bigg) \Bigg]},
\end{equation}
where $d$ is the film thickness, which primarily determines the slope of the dispersion. An important distinguishing feature of FVMSWs is the out-of-plane dispersion isotropy, as the spin waves always propagate normally to the external bias field $B$. 
\vspace{2pt}

\textcolor{NavyBlue}{\scalebox{0.75}{$\bullet$}}  \textit{Backward Volume Magnetostatic Spin Wave (BVMSW)} - configuration, where a magnetic film is magnetized in-plane and the bias field is applied parallel to the SW propagation; see Fig.~ \ref{f:2}(b). It is a volume mode, meaning that the amplitude of magnetization precession has a cosinusoidal distribution across the film thickness. The defining feature of BVMSW is the negative slope of the dispersion curve, which results in a negative group velocity. This unusual physics implies that the phase and group velocities are counter-propagating, and increasing the wavevector \textbf{k} leads to a decrease in the SW frequency. BVMSWs follow a dispersion relation: \begin{equation}\label{BVMSW}
  f =  \frac{\gamma}{2 \pi} \sqrt{B \Bigg[ B + \mu_0 M_\mathrm{s} \Bigg(\frac{1-e^{-k d}}{k d} \Bigg) \Bigg]}.
\end{equation}
\vspace{2pt}

\textcolor{NavyBlue}{\scalebox{0.75}{$\bullet$}}  \textit{Magnetostatic Surface Spin Wave (MSSW)} - in this configuration a magnetic film is also magnetized in film's plane, but the applied field is perpendicular to the direction of SW propagation; see Fig.~\ref{f:2}(c). Unlike volume spin waves, MSSWs (also known as Damon–Eshbach modes) are confined to a single surface of the magnetic film along which they propagate. Their precessional amplitude decays exponentially across the film thickness, reaching a maximum at one surface. The dispersion relation for MSSWs is given by:
\begin{equation}\label{MSSW}
  f =  \frac{\gamma}{2 \pi} \sqrt{B\Bigg(B + \frac{\mu_0 M_\mathrm{s}}{2}\Bigg)^2 - \Bigg(\frac{\mu_0 M_\mathrm{s}}{2}\Bigg)^2 e^{-2kd}}.
\end{equation}

\begin{figure}[h]
\centering
    \includegraphics[width=1\columnwidth]{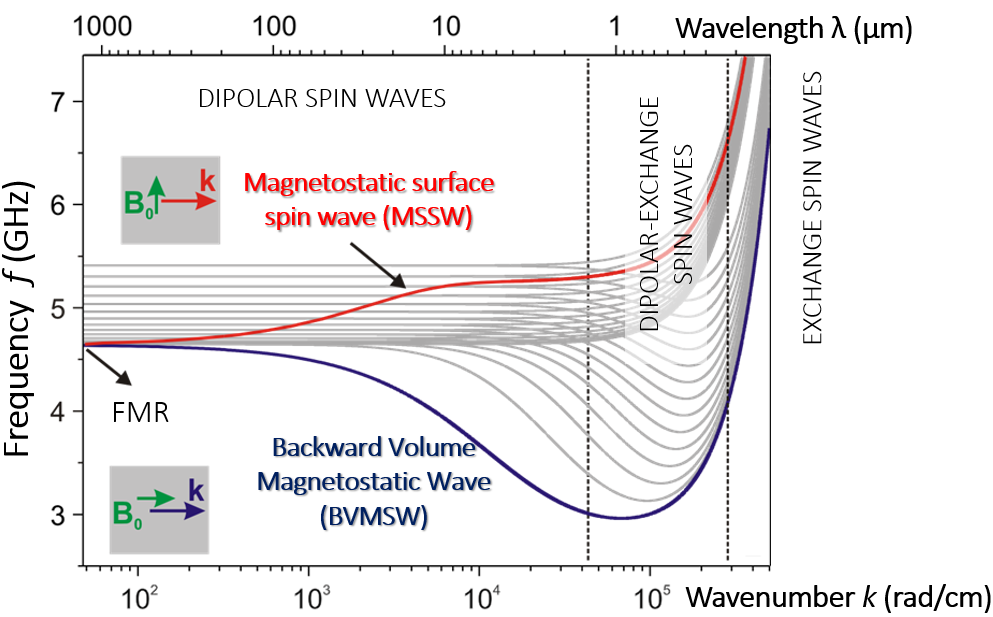}
    \caption{Dispersions of in-plane MSWs and a transition between dominant interactions (dipolar, dipolar-exchange, and exchange) as a function of increasing wavenumber $k$. Adapted from~\cite{ChumakHandbook2019}.}
    \label{f:3}		
\end{figure}

\vspace{-3pt}
In the simplest case, there are two main contributions to the SW energy: \textbf{weak, long-range dipole-dipole} and \textbf{strong, short-range exchange interactions} \cite{Gurevich1996, Stancil}, see Fig. \ref{f:3}. Since in the magnetostatic regime dipolar interactions dominate the SW dynamics (their wavelength is much longer than the exchange length), the dispersion relations for in-plane magnetized films are strongly anisotropic (directionally dependent). These dispersions are notably complex and deviate from the linear behavior typical of light in a uniform media. There are selected device architectures, where magnetic anisotropy offers an additional degree of freedom and can be beneficial; in others, it plays a minor role. However, for most data-processing and RF applications, both the strong anisotropy and the relatively low group velocity of MSW may impose limitations. 

\subsubsection{\textit{Spin-wave parameters important for applied science}} Important SW parameters, such as group velocity \(v_{\textup{gr}}\), lifetime \(\tau\), mean free path \(l_{\textup{free}}\) and others, can be obtained from the analysis of the dispersion relations: \setlength{\parskip}{2pt}

\textcolor{NavyBlue}{\scalebox{0.75}{$\bullet$}} \textbf{Group velocity} of a spin wave is defined as \(v_\mathrm{gr} = 2 \pi \cdot \partial f/\partial k \). As mentioned above, the group velocity is negative for the BVMSWs at small wavenumbers, then it crosses zero in the MSSW region, and increases monotonically in the exchange region, reaching values around 20~km/s (in YIG waveguide of width 1 \(\upmu\)m and 100~nm thickness, magnetized along by a 100~mT bias field \cite{ChumakHandbook2019}). From an application standpoint, \(v_{gr}\) is important metrics, as it determines the speed of data transfer and defines delays between different magnonic elements. \setlength{\parskip}{3pt}

\textcolor{NavyBlue}{\scalebox{0.75}{$\bullet$}} The main parameters influencing \textbf{spin-wave lifetime \(\tau\),} are the SW frequency \textit{f}, and Gilbert damping constant \(\alpha\), of a magnetic material (\(1/\alpha\) defines approximately the number of precession periods before the wave vanishes). The lifetime of the uniform precession mode in an infinite medium or a sphere is simply expressed as \(\tau_0 = 1 / \left(\alpha 2 \pi f \right) = -2 / \left(\gamma \mu_0 \Delta H \right)\) \cite{Stancil}. An explicit expression for the lifetime, appropriate for the particular mode and geometry is \(\tau_k = -1 / \mathrm{Im} \{ 2 \pi f \} = \tau_0 \cdot \frac{ \partial \left(2 \pi f_0 \right)}{\partial\left( 2 \pi f \right)}\), where \({\partial \left(2 \pi f_0 \right)} = -\gamma \mu_0 H_{\textup{eff}} \) \cite{Stancil}. \setlength{\parskip}{3pt}

\textcolor{NavyBlue}{\scalebox{0.75}{$\bullet$}} \textbf{Spin-wave free path} \(l_{\textup{free}}=v_{\textup{gr}} \cdot \tau\) is the distance a SW propagates before its amplitude decays to 1/e of its original value. The free path of long-wavelength BVMSW is usually large, often reaching several hundreds micrometers. For MSSW \(l_{\textup{free}}\) is approximately proportional to the film thickness, hence, is much larger in \(\upmu\)m-thick YIG samples. Importantly, parasitic losses in magnetic devices are inversely proportional to \(l_{\textup{free}}\). \setlength{\parskip}{3pt}

\textcolor{NavyBlue}{\scalebox{0.75}{$\bullet$}} The general profile of both \(v_{\textup{gr}}\) and \(l_{\textup{free}}\) is quite complex, featuring regions of increase and decline. Yet, shorter wavelengths do not necessarily require long free paths. \textbf{The ratio} \(l_{\textup{free}} / \lambda\), which shows how many wavelengths (i.e., how many unit elements) a wave propagates before it relaxes, is also important. For exchange SW with nm-scale wavelengths, this ratio can exceed 3000 \cite{ChumakHandbook2019}, offering compact device architectures. \setlength{\parskip}{3pt}

\textcolor{NavyBlue}{\scalebox{0.75}{$\bullet$}} Usually, spin waves are studied in spatially localized structures, such as thin films or waveguides, magnetized in-plane by an external field. The SW dispersion is determined by the waveguide's geometry (\textbf{thickness} \textit{d} and \textbf{width} \textit{w}), material parameters (\textbf{saturation magnetization} \({\textit{M}}_{\textup{s}}\) and \textbf{exchange constant} \({\textit{A}}_{\textup{ex}}\)) and the \textbf{applied magnetic field} \(\mu_0\textit{H} \).

\vspace{4pt}
\subsubsection{\textit{Choosing magnonic materials}} 
The choice of a material is critical in both fundamental and applied magnonics. Key requirements for a promising candidate include: (i) low Gilbert damping for long SW lifetimes; (ii) high saturation magnetization to enable operation in the high-frequency region and support fast SW propagation; (iii) high Curie temperature for thermal stability; (iv) simplicity in the design and fabrication of magnetic films and, if applicable, their excitation transducers (antennas).  

 A detailed overview of commonly used materials in fundamental magnonics, as well as those with high potential for applications, is provided by Chumak \cite{ChumakHandbook2019, Chumak2017}. Among others, Permalloy (Py), CoFeB, and the Heusler compound CMFS are discussed. Nevertheless, the most promising magnonic material to date remains monocrystalline YIG, typically grown by a high-temperature LPE on a gadolinium gallium garnet (GGG) substrate \cite{glass1976attainment, Dubs2017, Dubs2020, dubs2025magnetically}. This ferrimagnet exhibits the lowest known magnetic losses, enabling SW lifetimes of several hundred nanoseconds, and thus is widely used in research \cite{SagaYIG, YIGmagnonics}. 
YIG's low loss stems from two main factors: (i) as a magnetic dielectric, YIG lacks free electrons, minimizing phonon scattering and associated thermal dissipation \cite{SagaYIG}; (ii) its magnetism arises primarily from Fe\textsuperscript{3+} ions with no net orbital angular momentum, resulting in a weak spin-lattice coupling (i.e., minimal spin-orbit interaction). As a result, YIG's magnetic properties depend purely on spin and gyromagnetic ratio \cite{Stancil}. Furthermore, high-quality LPE-grown YIG single crystals contain very few inhomogeneities, suppressing two-magnon scattering and further reducing losses \cite{Dubs2017, Dubs2020}. For much of its history, high-quality YIG was limited to micrometer-thick films, preventing nanoscale patterning. Yet, in the last decade, progress in the fabrication methodologies — including PLD \cite{sun2012growth, d2013inverse, yu2014magnetic, Onbasli2014, Hahn2014, balinskiy2017spin, schmidt2020ultra}, sputtering \cite{schmidt2020ultra, liu2014ferromagnetic, jakubisova2015optical, ding2020sputtering} and LPE  \cite{Dubs2017, Dubs2020, hahn2013comparative, dubs2025magnetically} — have enabled the growth of high-quality nanometer-thick YIG films. While their quality and performance remains slightly inferior to traditional micrometer-thick films, it is already sufficient for many magnonic applications and continues to outperform alternative materials \cite{MagnonSpintronics}.\setlength{\parskip}{2pt}

Recently, a unified theory for the saturation magnetization \(M_\mathrm{s}(T)\) in multi-sublattice ferrimagnets (e.g., YIG) was developed by Serha et al. ~\cite{serha2025theory}, providing a single-framework description across the full temperature range by joining the low-temperature SW (Bloch) regime to the near-\(T_\mathrm{c}\) mean-field. This approach provides a predictive backbone for device modeling in magnonics, spintronics, and quantum magnonics.

 \vspace{0.5mm}
\noindent\rule{\columnwidth}{0.4pt}
\vspace{0.4mm}
  
\section{Modern magnonics: general overview}
\subsection{Current state of the magnonic field} 
\vspace{-3pt}
The modern field of magnonics is broad and expands rapidly, attracting researchers from the neighboring fields of spintronics, quantum optics, superconductivity, materials science, and nanotechnology. Probably the most convenient way to illustrate its scope is the roadmap figure (Fig.~\ref{f:4}) by Barman and Gubbiotti \cite{Barman2021}. Magnonics covers a wide range of topics; although most directions remain fundamentally oriented, these studies are aimed at technologies with strong industrial potential \cite{Chumak2022, Barman2021}. 

\vspace{-2pt}
\subsection{Vectors of the cutting-edge developments}
\vspace{-3pt}
Spin waves are already attracting considerable attention as potential data carriers in energy-efficient computing devices and as a storage medium \cite{Wang2020, Chumak2021A, dutta2015non, khitun2013magnonic, gertz2015magnonic}. Since their wavelengths are orders of magnitude smaller compared to EMWs of the same frequency, they allow for the design of micro- and nanoscaled devices \cite{Barman2021, mahmoud2020fan, Dieny2020}. On the other hand, SWs in the GHz range are of particular interest for telecommunication and radar applications \cite{ChumakHandbook2019}. This potential, combined with a wide range of linear and nonlinear properties, also makes SWs ideal for exploring fundamental physics \cite{Serga2010}. One- and two-dimensional soliton formations \cite{serga2004self, wu2004generation}, non-diffractive SW caustic beams \cite{schneider2010nondiffractive}, wave-front reversals \cite{serga2005parametric}, room-temperature Bose-Einstein condensation of magnons \cite{schneider2021control, schneider2021stabilization, Schneider2020}, quantum magnonics \cite{Lachance-Quirion2019, Knauer2023} and parametric generation of SWs in magnonic nanoconduits \cite{Heinz2022} is just a small selection.

Within modern magnonics, 3D nanomagnetism represents an emerging frontier. A 2025 roadmap \cite{gubbiotti20252025} highlights how advances in fabrication, imaging, and modeling of 3D architectures are opening pathways toward next-generation technologies, including ultrahigh-density storage, logic, spintronics, and neuromorphic computing.

\renewcommand {\thefigure}{4}
\begin{figure}[ht!]
\centering
    \includegraphics[width=1\columnwidth]{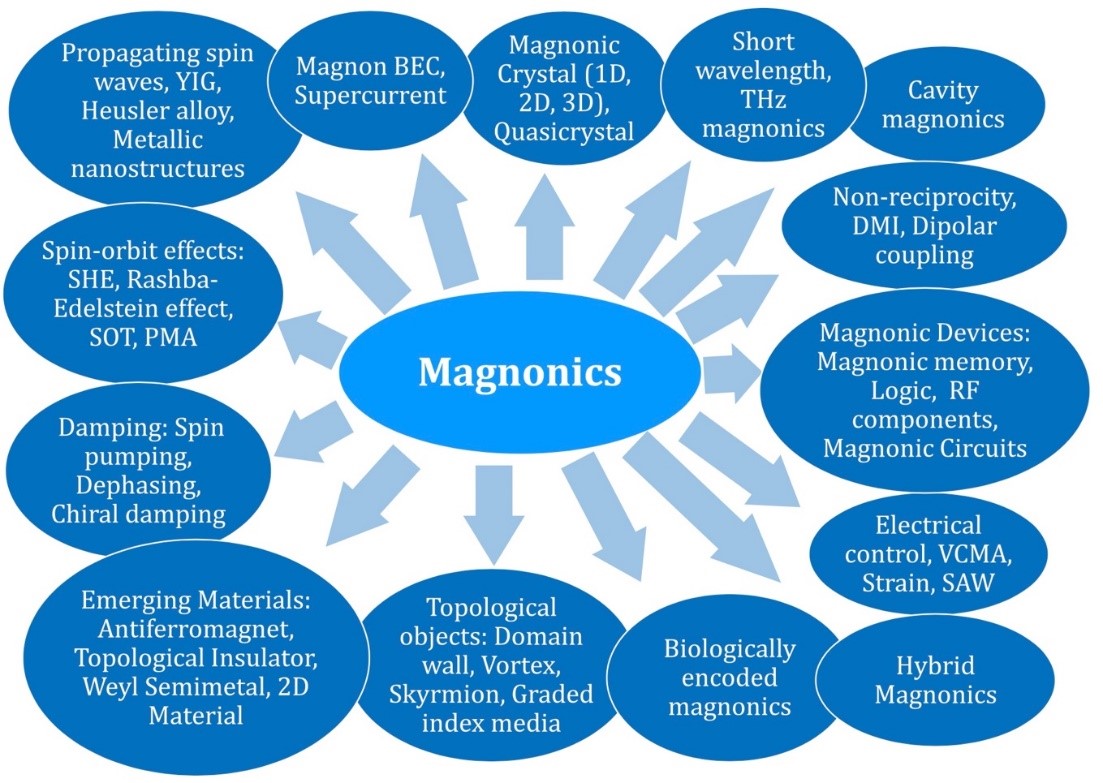}
    \caption{The many branches of magnonics. Adapted after \cite{Barman2021}.}
    \label{f:4}		
\end{figure}

This topic has been partially discussed in \cite{Chumak2022, Barman2021, gubbiotti20252025}. Below, we summarize key current and emerging developments in magnonics, relevant to SW-based RF devices: \setlength{\parskip}{6pt}

\subsubsection{\textit{Notable progress in microwave transducers design}} The fundamental method to excite SWs is by means of external dynamic magnetic field generated by alternating current (AC) in a transducer \cite{MagnonSpintronics}\footnote{In the description of spin-wave excitation "antenna" is used interchangeably with "transducer".}. The AC creates an alternating Oersted field via Ampère’s law, which in turn exerts torque on the magnetization in an adjacent magnetic medium. At frequencies above the FMR, this Oersted field can then excite spin waves in the material \cite{Mahmoud2020}. The direct coupling between the RF field and the magnetic moments allows for the conversion of electromagnetic energy into SW energy \cite{Rezende2020}.The energy carried away in SWs propagating perpendicular to the antenna is related to electromagnetic energy propagating along the antenna by an equivalent radiation resistance \cite{ganguly1975radiation}. The shape and dimensions of inductive antennas have a strong impact on performance parameters, such as bandwidth, insertion loss, and excited modes. In typical YIG-based devices with unoptimized excitation and detection antennas, signal losses primarily stem from the impedance mismatch between the circuit and antennas' radiation resistance, as well as from antennas' ohmic resistivity and geometric losses \cite{erdelyi2025design}. Additional attenuation occurs as a result of magnetic damping in the YIG film, and only a fraction of the original power is ultimately detected. These losses are especially critical in miniaturized systems, where the high resistivity of micron-scale antennas can significantly limit transduction efficiency. Overall, narrower antennas excite a broader wavevector spectrum, allowing to reach short-wavelength (high-\textit{k}) SWs and higher SWs modes. In this case, generated magnetic field strength is smaller, yet more localized beneath the antenna. In contrast, wider antennas produce a stronger and more uniform magnetic field over a larger area, which enhances efficient excitation of long-wavelength (low-\textit{k}) SWs, but limits the excitation to a narrower wavevector range. Wider antennas are generally easier to fabricate with conventional lithographic techniques. Antennas spacing also requires optimization: they must be close enough to support efficient SW transmission, yet sufficiently separated to avoid cross-talk and interference dominating the spectrum. \vspace{-3pt}

\renewcommand {\thefigure}{5}
\begin{figure}[!h]
\centering
    \includegraphics[width=0.66\columnwidth]{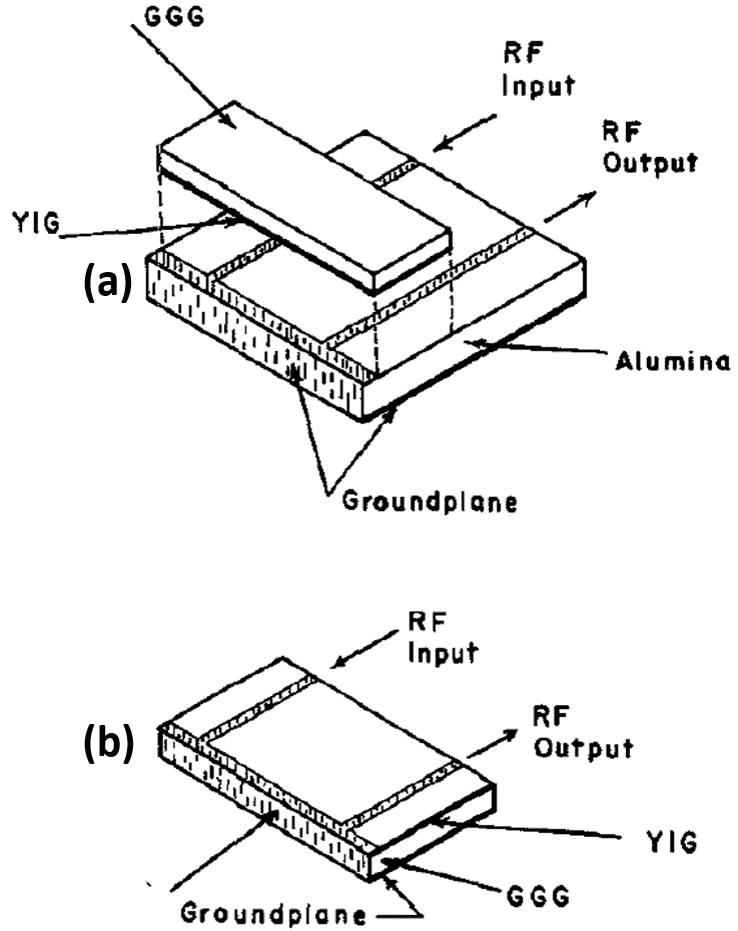}
    \caption{Microstrip excitation of magnetostatic SWs a) with a standard alumina substrate; b) with a GGG substrate. Adapted after \cite{ganguly1975microstrip}}
    \label{f:5}		
\end{figure}

Various types and designs of transducers have been developed recently, each offering distinct advantages depending on the magnonic application. Convenient excitation of spin waves could be achieved using a simple and compact microstrip antenna - a metallic strip on a dielectric substrate with a ground plane underneath (Fig.~\ref{f:5}). Due to strong coupling between electromagnetic and spin waves, a single strip is sufficient for macroscale operations. This is in contrast to SAW excitation, where a meander interdigital transducer is necessary \cite{ganguly1975radiation, ganguly1975microstrip}. The single-strip SW excitation (0.18~mm-wide strip on a 0.25~mm-thick substrate) in a measurement series of 1.7-6.25~$\upmu$m-thick YIG samples revealed the following trends \cite{ganguly1975microstrip}: (i) for a given bias field and antenna geometry, the excitation bandwidth increases with YIG thickness and saturation magnetization; (ii) maximum radiation resistance increases slightly in thicker YIG films; (iii) increasing the bias field narrows the excitation bandwidth but raises the maximum radiation resistance. 
Consequently, for optimal broadband excitation, the microstrip width should be comparable to the film thickness, enabling impedance matching and reducing ohmic losses. However, as device scales approach macro and nanosize, such antennas become highly resistive, leading to degraded RF performance \cite{erdelyi2025design}. Furthermore, microstrip offers limited directionality and mode control, is more susceptible to reflections and reduced coupling efficiency due to its single-conductor design. This presents a challenge for integration, indicating the importance of design optimization. \vspace{-4pt}

An excellent overview of key magnonic transducers (Fig.~\ref{f:extra3}) together with the rules for their low-insertion-loss design are given by Connelly et al.  \cite{connelly2021efficient}. Beyond microstrips, multi-conductor transmission-line antennas (e.g., CPW, Fig.~\ref{f:extra3}(b)) in a form of metallic stripes patterned above the magnetic film, are commonly used in magnonics. They generally outperform single-conductor designs, offering higher radiation resistance, better impedance matching, and narrower excitation profiles in \textit{k}-space, leading to more efficient coupling to target wavenumbers. However, CPWs suffer from reduced radiation efficiency at higher wavenumbers and are not optimal for wideband operation. Narrowband multi-conductor antennas Fig.~\ref{f:extra3}(c, d) have high radiation and match efficiencies over narrow bandwidths, even for high-\textit{k} SWs. In particular, meander antennas excel in terms of bandwidth, matching, efficiency and power dissipation due to their higher radiation resistance. Though, they should be considered carefully in power limiters and delay lines as nonlinear film-behavior is reached at a lower RF input powers. Also, their complex geometries may be challenging to fabricate and they can introduce spatial constraints on chip integration. Wideband multi-conductor transducers Fig.~\ref{f:extra3}(e, f) provide efficient and broadband performance by changing the dimensions of the "finger lines" (chirping or fanning). Fanned lines introduce spatially distributed launch points, while chirped lines enable a uniform wavefront across the target frequencies and have higher radiation resistance than fan gratings of equal length. These antennas are optimal for broadband applications, such as filters, delay lines and wave-based computing, but they require advanced nanofabrication and careful alignment. Based on extensive analysis, the authors \cite{connelly2021efficient} also proposed an optimized meander design for short-wavelength SWs excitation in 100~nm thick YIG film ($M_{\mathrm{s}}$ = 140 kA/m) with an efficiency of -4.45 dB and a 3 dB-bandwidth of 134 MHz. \vspace{-4pt}

Another fundamental approach was proposed by Vlamick et al.  \cite{vlaminck2010spin}, who simulated and experimentally verified SW transmission in 1-15 GHz frequency range on Py strips ($\alpha \approx $ 0.012-0.014, $ \upmu_{\mathrm{0}} M_{\mathrm{eff}}$ = 880 mT) using various antennas. In addition to CPW, seven meander designs with varying size (corresponding to the SW wavelength of 0.8 and 1.6 $\upmu$m), number (3 and 5), distance between the two antennae (5-15 $\upmu$m), as well as thickness (10 and 20 nm) and width (2, 3.5 and 8 $\upmu$m) of the Py strip were tested. By relating the antenna microwave response in the reciprocal space to the gyromagnetic response of the magnetic film, authors accurately described propagating SW spectra in various geometries. Several other prominent designs, including CPW, ladder and meander antennas, have been realized in the following years \cite{bailleul2003propagating, vanderveken2022lumped, mahmoud2020fan, zhang2018antenna}.\vspace{-5pt}

\renewcommand {\thefigure}{6}
\begin{figure*}[!ht]
\centering
    \includegraphics[width=0.67\paperwidth]{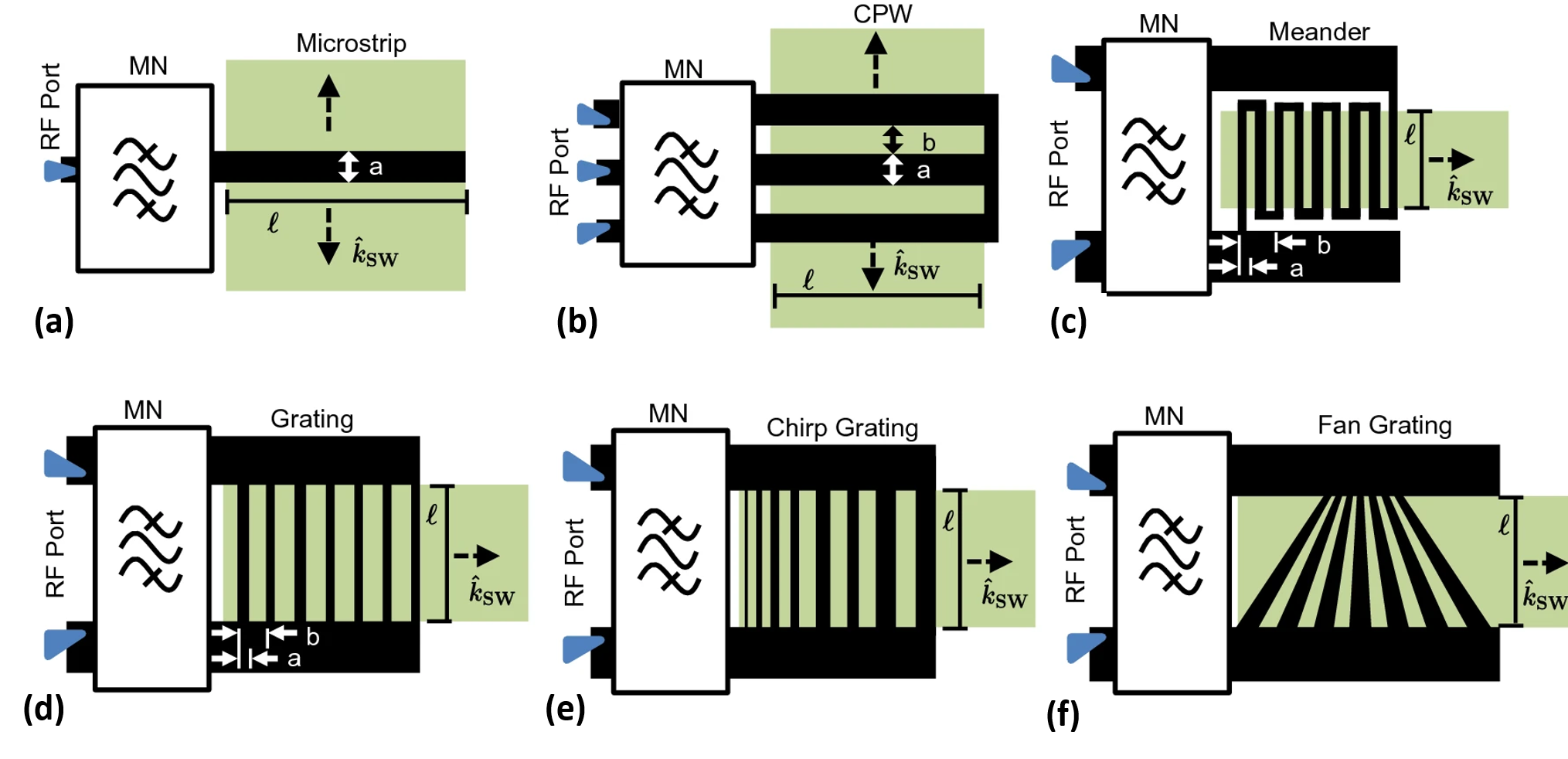}
    \caption{Key types of magnonic transducers: a) narrow-band single-conductor (microstrip); b) narrow-band multi-conductor (CPW); c), d) multi-conductor transmission-line (meander and grating, respectively); e), f) wide-band multi-conductor for chirp and fan gratings, respectively. Here, \textit{a} is a width of a signal line; \textit{b} - separation between the lines in multi-conductor designs; l - length of transducers, measured as indicated on each figure individually; $k_{\mathrm{SW}}$ - wavevector of excited SW; MN - matching network. Adapted after \cite{connelly2021efficient}}
    \label{f:extra3}		
\end{figure*}

Beyond conventional transducers, new designs are explored. For example, magnetoelectric antennas, based on piezoelectric-magnetostrictive bilayers, have been proposed as compact and power-efficient alternatives to conventional antennas \cite{yao2015bulk, domann2017strain}. When driven by RF signal, these antennas generate an oscillating magnetic dipolar field that acts as a source of electromagnetic radiation, with performance further enhanced by acoustic and magnetic resonances. Alternatively, spin-torque nanooscillators \cite{demidov2010direct, madami2011direct} or spin Hall oscillators \cite{chen2016spin} can be used, but they lack the required bandwidth and spectral purity for computing, while inductive techniques \cite {vlaminck2010spin, wu2006} offer broad bandwidth, but are energy-inefficient. Emerging methods, like inverse magnetostriction devices \cite{chen2017voltage} show potential, but require improvements in energy efficiency. Spin-orbit torques, which can rapidly switch magnetization, present an intriguing alternative for SW excitation. As reported by Talmelli et al. \cite {talmelli2018spin}, the specific design of thin in-plane magnetized Ta/(Co,Fe)B waveguides minimized parasitic coupling between emitter and detector, enabling the excitation of SW with wavevectors up to 6 rad/$\upmu \mathrm{m}$. Finally, an impactful lumped-element model was shown by Vanderveken et al. \cite{vanderveken2022lumped} to describe antennas in terms of their Ohmic resistance, self-inductance, and additional inductance from SW or FMR excitation in the magnetic layer.\vspace{-4pt}

In 2025, a few breakthrough frameworks for the design of magnonic transducers were introduced. Erdélyi et al. \cite{erdelyi2025design} integrated a circuit-level model with micromagnetic simulations to analyze complex antennas geometries. Their model predicted a transducer pair with an insertion loss of 5~dB over a 100~MHz bandwidth in a 800 nm-thick YIG/GGG-based device ($\alpha \approx 3.86 \times 10^{-4}$, $ M_{\mathrm{s}}$ = 153.1 kA/m). The fabricated GSG (Ground-Signal-Ground) CPW antennas consisted of a double layer of 150~nm Au/10~nm Ti (titanium is an adhesion layer) and were 100~$\upmu$m-long, 2~$\upmu$m-wide and 10~$\upmu$m apart from each other. The low losses were achieved by applying antenna radiation resistance scaling rules to optimize the energy transfer between  the electric and magnetic domains. In parallel, Brückner et al. \cite{bruckner2025micromagnetic} developed a milestone micromagnetic simulation framework to compute the SW impedance and optimize antennas of arbitrary geometries, potentially reaching up to 75\% efficiency in YIG systems. Further improvements are expected during the device-specific research and development.\vspace{-4pt}

Solutions to the inefficient MW - SWs conversion caused by their vastly different GHz-range wavelengths were proposed by Ivzhenko et al. \cite{ivzhenko2025influence}. In this theoretical work, a planar inverse split-ring resonator coupled to a thin Py film was designed to exploit strongly localized near fields at the resonator’s anti-gap and photon–magnon coupling, enabling up to 4.2× higher energy transfer for the fundamental and 4× for a width-quantized SW mode. The design is compact, tolerant to impedance mismatch and suitable for on-chip integration, with potential for further coupling enhancement and quantum-oriented applications.

\subsubsection{\textit{Advancement in the understanding of nonlinear spin-wave physics}} Spin-wave nonlinearity refers to phenomena in which SWs interact with each other, considering energy and momentum exchange. In magnetically ordered media, the inherent nonlinearity of spin systems allows thermally excited SWs to couple with the magnetization mode driven by an external microwave field. When the amplitude of this driving (pumping) field exceeds a critical threshold, the energy absorbed from it starts to compensate intrinsic SW losses, leading to an exponential growth in SW amplitude - a process known as instability \cite{Gurevich1996, Stancil, kalinikos2013nonlinear, zheng2023tutorial, Rezende2020, Bloembergen1952, Bloembergen1954, serha2025ultra}. On a fundamental scale, such a parametric excitation of SWs is driven by multimagnon scatterings under the momentum and energy conservation laws. In the case of \textbf{perpendicular pumping}, microwave field transverse to the static magnetization excites a uniform FMR precession $\textbf{k}_p = 0$ at the pumping frequency $\textit{f}_p$, which couples to magnons with non-zero wavevectors. Above a threshold amplitude, SW instabilities emerge, as energy from the pumped FMR mode exceeds damping losses. In first-order Suhl instability (three-magnon scattering) \cite{kurebayashi2011controlled, ge2024nanoscaled, Suhl1956, Suhl1957, ge2024nanoscaled, serha2025ultra}, the FMR mode decays into two counter-propagating magnons with $\textit{f}_1 = \textit{f}_2 = \textit{f}_p/2$, $\textbf{k}_1 = - \textbf{k}_2$. In the second-order instability (four-magnon scattering) \cite{schultheiss2012direct, olson2007ferromagnetic}, when the main mode approaches saturation, two counter-propagating SWs with frequencies equal to pumping frequency are driven by two FMR mode magnons, i.e.,  $\textit{f}_1 = \textit{f}_2 = \textit{f}_p$, $\textbf{k}_1 = - \textbf{k}_2$. Alternatively, in \textbf{parallel pumping} process, when the microwave field is applied parallel to the static field, it couples directly to the spin system due to the ellipticity of magnetization precession \cite{schlomann1960recent, Gurevich1996}. Then, similarly to perpendicular pumping, when the amplitude of the microwave field reaches a threshold, its energy compensates SW losses, causing exponential growth of the SW amplitude and triggering a three-magnon scattering. This process produces pairs of counter-propagating magnons at half the pump frequency. In general, parametric pumping is the only common technique for amplifying both dipolar and exchange SWs in YIG films, covering frequencies from a few to several tens of GHz, with maximum gains exceeding 30 dB at room temperature \cite{Serga2010}. \vspace{-4pt}

From an application point of view, the rich nonlinear SW properties can offer functionalities like improved signal processing and amplification, logic operations, power limitations or signal-to-noise ratio enhancement. Previously, nonlinearity was often viewed as a limiting factor due to a lack of understanding, as it restricts a maximum operational power of RF devices (especially critical for the power amplifier). However, the magnonics community has since gained sufficient insight to understand and control the phenomenon \cite{Serga2010}. For instance, efficient nonlinear switching of a directional coupler was recently demonstrated by Wang et al. \cite{Wang2020}, where the output SW intensity depends on the input microwave power. Innovative approach, developed by Chumak et al. \cite{Chumak2014} used nonlinear four-magnon scattering as an operational principle in the magnonic transistor. In this prototype, signal magnons were scattered by localized gate magnons, resulting in a controlled attenuation before reaching the drain. The implementation of a magnonic crystal enhanced the localization of gate magnons, thereby increasing scattering probabilities and amplifying the nonlinear response. Stimulated three-magnon scattering was utilized in a nanoscale magnon transistor a decade later by Ge et al. \cite{ge2024nanoscaled}. The proposed device is based on 20~$\upmu$m long, 100~nm wide and 50~nm thick YIG nanowaveguides under an in-plane bias, where a gate magnon at 14.6~GHz is converted into an idler at 10.4~GHz and a source at 4.2~GHz. A directional coupler mixes the gate and source signals, while a dual-band magnonic crystal suppresses gate and idler at the output, yielding a gain of 9 while preserving a phase. Such a design presents a step forward in the amplification of propagating SWs in nanowaveguide-based systems and allows integration into a magnon circuits. In another key development, Jungfleisch et al. \cite{jungfleisch2015thickness}, showed that reducing the thickness of YIG films suppresses undesirable nonlinear multimagnon scattering, allowing for higher operational power.
\vspace{-4pt}

Spin-wave nonlinearity has also enabled new computing operations. Wang et al. \cite{wang2023deeply} demonstrated how nonlinear excitation in 200 nm-wide YIG waveguides produces a self-normalizing FVSW mode with above 2 GHz frequency shift and wavelengths down to 200 nm, enabling robust on-chip magnonic signal generation (see \textit{Section III-E} for details). Following year, Breitbach et al. \cite{breitbach2024nonlinear} developed an all-magnonic erasing process in Ga:YIG film \cite{dubs2025magnetically}, highlighting potential for logic and neuromorphic systems. Serga et al. \cite{serga2007parametrically} showed that RF signals carried by dipolar SWs in a tangentially magnetized film can be stored as standing dipole-exchange modes and later recovered using double-frequency parametric pumping. Furthermore, using a perpendicular pumping, breakthrough lifetimes of >18~$\upmu$s were achieved for short-wavelength magnons in ultra-pure YIG spheres by Serha et al. \cite{serha2025ultra}. The lifetimes were extracted from the three-magnon-splitting threshold, establishing long-coherence magnons as viable carriers for quantum magnonics.\vspace{3pt}

\subsubsection{\textit{Nanoscaling}} "Smaller, faster, more efficient" has long been a motto for technological progress, leading to the widespread computerization of daily life. For many years, applied magnonics could address only the last two goals, as downscaling fabrication and structuration of its key material (YIG) posed significant challenges. Recent advances in LPE \cite{Dubs2017, Dubs2020} allowed to grow YIG films with thickness down to 9~nm and kick-started the miniaturization of YIG-based structures down to lateral dimensions of 50~nm. The effects of waveguide downscaling on SW spectra were explored by Wang et al. \cite{wang2019spin} and Heinz et al. \cite{Heinz2020, Heinz2021}. When the width of an YIG waveguide is sufficiently small, the exchange interaction dominates over the dipolar one, leading to the unpinning of SW modes. This alters the quantization condition and shifts higher-order width modes to higher frequencies, effectively providing a single-mode regime \cite{wang2019spin}. The progress is driven by the exceptional structural quality and magnetic performance of the epitaxial nanofilms \cite{Pirro2014, Dubs2017}. These developments pave the way for advanced nanoscale SW networks for neuromorphic, stochastic, and reservoir computing, as well as quantum information processing. Wang et al. \cite{wang2024nanoscale} provided an in-depth review of such emerging architectures, highlighting their potential for the next-generation computing. In parallel, 2D van der Waals magnetic materials are gaining attention, as their atomic-layer thickness, tunable properties and potential for integration into existing technologies position them as promising candidates for "beyond-YIG" scalable magnonic structures \cite{zhang20242d, ahn2024progress}. \vspace{4pt}

\subsubsection{\textit{Micromagnetic simulation tools of spin-wave dynamics}} The growing complexity of fundamental investigations constantly increases demands for powerful simulation tools. A key approach in magnonics is the micromagnetic semiclassical theory, where magnetization is represented by a continuous unit-vector field, and interactions are described by nonlinear partial differential equations \cite{Chumak2022}. This framework allows for the resolution of domain walls and spin waves without considering individual atomic moments, though analytical solutions are often not feasible. To bridge the gap between real systems and simulations, numerical methods such as the finite-difference and finite-element methods are used to solve micromagnetic equations \cite{suess2002time, abert2019micromagnetics, bruckner2023magnum}. To reduce the computational complexity and cost, modern tools increasingly rely on parallel computing architectures, like GPUs \cite{kakay2010speedup, chang2011fastmag, dvornik2014thermodynamically}, or novel algorithms, such as machine learning \cite{wang2024nanoscale}. \setlength{\parskip}{2pt}

Micromagnetics in the frequency domain is often preferred in magnonics due to its higher efficiency compared to time-domain simulations. This approach is based on linearizing the LLG equation to extract dynamic modes and power-spectral densities, and has recently been extended to include nonlinear effects, making it a powerful tool for designing SW prototypes \cite{d2009novel, Bruckner2019, Perna2022}. Additionally, magnonic devices often integrate spintronics effects, such as spin pumping and the inverse spin-Hall effect, requiring modifications to the simulations. For example, a self-consistent coupling of micromagnetics with spin-diffusion models has been discussed in \cite{Chumak2022}. While macroscopic systems can be reasonably predicted, simulation of the entire device remains challenging. For a design with few variables, simple approach like binary search algorithm with a fast forward solver has proven effective, as demonstrated by Wang et al. \cite{wang2021inverse}. More complex, high-dimensional problems, however, call for advanced optimization techniques. Many of the excellent micromagnetic simulation tools in magnonic community are open-source and user-friendly, e.g., Magnum \cite{magnumfd}, MuMax$^3$ \cite{mumax1}, TetraX \cite{TetraX, korberFiniteelementDynamicmatrixApproach2021a}, Boris Computational Spintronics \cite{borisspintronics}, NeuralMag \cite{abert2024} and OOMMF \cite{oommf}. \setlength{\parskip}{2pt}

Simulation platforms such as COMSOL Multiphysics, Ansys HFSS, and CST Microwave Studio are increasingly used to study SW dynamics within RF systems and are already well-established within the RF engineering community. Although COMSOL was not originally developed for micromagnetic modeling, recent introduction of frequency-domain module (finite element method) that solves LLG equation have enabled the simulation of SW modes and dispersions \cite{zhang2023frequency, yu2021micromagnetic, hua2024micromagnetic}. This approach provides reasonably accurate results for calculating spin wave dispersions and mode profiles, but lacks support for simulating magnetization switching, domain wall dynamics, nonlinear effects and is not publicly available. Similarly, HFSS (frequency-domain solver) and CST (time-domain solver) \cite{AnsysHFSS, CST} can model electromagnetic interactions with magnetic materials and support limited spin dynamics, they are not optimized for micromagnetics.  \vspace{6pt}

\subsubsection{\textit{Machine learning and inverse design}}
Inverse design is a revolutionary approach in which the desired functionality is specified first, and a feedback-based algorithm then determines the device structure. Although only recently introduced to magnonics, it has already proven to be a powerful tool for optimization of the devices design, while also reducing fabrication costs and development time.
The universality of this approach was validated by Wang et al. \cite{wang2021inverse} with a proof-of-concept device based on a patterned 100-nm-thick YIG film. The authors demonstrated linear, nonlinear, and nonreciprocal magnonic functionalities, and subsequently inverse-designed  a (de-)multiplexer, a nonlinear switch, and a circulator. Kiechle et al. \cite{kiechle2022experimental} further experimentally demonstrated a SW lens designed via machine learning, which focuses spin waves by shaping an effective magnetization profile in thin YIG film. Validated by MuMax$^3$ simulations, this proof-of-concept demonstrates a pathway for developing advanced magnonic devices, such as SW processors or neuromorphic systems.

Papp et al. \cite{Papp2021} advanced the photonic design methods into the magnonic domain, enabling the inverse design of SW-based convolvers, spectrum analyzers and RF components. SWs naturally reach a nonlinear regime at higher amplitudes (than EM waves), allowing the resulting interference-based devices to realize weighted sums and interconnections within a single magnetic film. Neuromorphic computing, inspired by brain-like architectures, relies on such distributed parallel information processing, which is particularly suited for fast and energy-efficient tasks in IoT and RF domains. Important realization of all-magnonic neurons in Ga:YIG films by Breitbach et al. \cite{breitbach2025all} further demonstrated how nonlinear SW excitation enables threshold activation, cascadability and multi-input integration. Proposed prototype designs were validated through neural-network simulations with accuracies exceeding 97\%, highlighting them as robust low-power building blocks for scalable wave-based neuromorphic hardware. Owing to SW group velocity and natural interference, magnonic neuromorphics is also capable of pattern recognition and adaptive learning \cite{Csaba2017, Papp2021, breitbach2025all}.

Alongside neuromorphic computing, Ising machines have emerged as a complementary approach for solving optimization problems \cite{litvinenko202550, gonzalez2024global} by mapping them onto interacting spin networks, where the lowest spin configuration represents the optimal solution. In SW Ising machines, delay lines play a key role by separating RF SW pulses in time, so that each pulse represents an artificial Ising spin. By routing these pulses through lines of different lengths, controlled interference is created. Considering intrinsically slower SW group velocity, these delay lines can be realized on the millimeter, rather than the kilometer scale, required in optical systems. Such a compact magnonic Ising machine was experimentally realized by Litvinenko et al.~\cite{litvinenko2023spinwave} from a $5~\upmu\text{m}$-thick YIG film-based delay lines and  off-the-shelf microwave components. The device can support 8-spin optimization problem and solve it within $\approx 4~\upmu\text{s}$ at an energy cost of only $\approx 7~\upmu\text{J}$. The demonstrated prototype has proven to be a compact, energy-efficient, and versatile platform for high-performance combinatorial optimization solvers. \setlength{\parskip}{1pt}

A milestone in inverse-design magnonics was achieved by Zenbaa et al. \cite {zenbaa2025universal}  who developed a versatile reconfigurable magnonic inverse-design device with \(10^{87}\) degrees of freedom, capable of supporting a wide range of functionalities. Two independent optimization algorithms were used to realize key RF components, such as a tunable notch filter with 5 MHz bandwidth and up to 48 dB suppression, and an RF demultiplexer on a single YIG-based chip, targeting 5G/6G technologies. In a follow-up study, Zenbaa et al. \cite{zenbaa2025realization} further explored the nonlinear functionality in the inverse-designed logic gates, proving their relevance for reservoir and binary logic. The device provided SW signal modulation over four orders of magnitude, highlighting its versatility and high performance. To address the computational memory constrains, Voronov et al. \cite{Voronov2024Inverse} proposed a topology optimization framework based on level-set and adjoint-state methods, enabling efficient design of complex magnonic structures. 
We believe that, given the recent breakthroughs in inverse-design magnonics, this approach holds strong potential to solve or mitigate long-standing challenges in SW-based RF technology discussed below, and stands out as a key tool for advancing applied magnonics. \vspace{5pt}

\subsubsection{\textit{Nonreciprocity}} 
Spin-wave nonreciprocity refers to the asymmetrical propagation of waves with wavevectors \textit{+k} and \textit{-k} within a magnetic material (with different frequencies or amplitudes). The asymmetry stems from and can be tailored by different mechanisms: (1) interfacial Dzyaloshinskii–Moriya interaction due to broken inversion symmetry and strong spin-orbit coupling \cite{bracher2017creation}; (2) geometric curvature combined with dipole interactions, as in nanoscale magnetic grating of Co nanowires on top of ultrathin YIG film \cite{chen2019excitation}; (3) dipolar coupling in multilayers with structural asymmetry \cite{grassi2020slow, gallardo2019reconfigurable, wojewoda2024unidirectional}; (4) magnetization configuration, i.e., MSSWs have intrinsic amplitude nonreciprocity due to the asymmetry of dynamic dipolar fields across the film \cite{gladii2016frequency}. In-depth on-going research is dedicated to applied nonreciprocal magnonics. The ability to inhibit signal flow in one direction while allowing in the opposite is essential for protecting RF devices from reflections, isolating transmitters from receivers in radar systems, or shielding qubits from environmental disturbances in quantum architectures. An example of such a magnonic isolator was given by Zenbaa et al.~\cite{zenbaa2025yig} using a YIG(100~nm)/SiO$_2$(5~nm)/CoFeB(40~nm) stack with pronounced nonreciprocal SW propagation. Here, the isolator functionality arises from a dipolar coupling and complementary magnetic properties of the layers: YIG ensures low damping and efficient SW transport, while CoFeB provides strong magnetic anisotropy. At $f \approx 7.23~\mathrm{GHz}$ under a $\pm 200~\mathrm{mT}$ bias field, the magnon decay length is $\sim 10.92~\upmu\mathrm{m}$ (positive bias) versus $\sim 3.78~\upmu\mathrm{m}$ (negative bias), confirming isolator-like behavior and suitability for on-chip magnonics. Due to inherent nonreciprocity, MSSW modes may also have lower insertion losses than BVMSW in transmission-based devices, as in the latter half the energy is lost to backward propagation. \vspace{5pt}

\subsubsection{\textit{Quantum magnonics}} 
Quantum magnonics is a leading research area focused on the quantum properties of magnons and their implementation in quantum computing, including entanglement, hybrid-system coupling and Bose-Einstein condensation \cite{wang2024nanoscale}. The field’s emergence less than a decade ago was heavily influenced by Lachance-Quirion et al. study \cite{Lachance-Quirion2020} of the entanglement between the superconducting qubit and YIG sphere in a 3D microwave cavity. The observed negative Wigner function in the study by Xu et al.  \cite{xu2023quantum} provided credible evidence of quantum behavior. Despite paving the way for efficient single-magnon detection, critical for sensing and hybrid quantum systems, this work was limited to non-propagating FMR mode. To overcome the limitations of 3D microwave cavities and enable chip-scale architectures, recent works studies planar (2D) superconducting resonators that couple to magnons in Permalloy films \cite{Li2019} as well as to the fundamental modes of YIG spheres \cite{Li2022,Song2025}. The next important milestone, as predicted by Wang et al. \cite{wang2024nanoscale}, is the excitation and detection of propagating single magnons, which are essential for realizing magnon-based quantum information transport and processing. 
\setlength{\parskip}{1pt}
 
To advance quantum magnonics, it is crucial to accurately predict behavior of target magnetic material in a cryogenic environment. Prior studies of the widely used YIG/GGG systems at temperatures down to 50 K \cite{karenowska2015excitation, danilov1989low, Karenovska2018, jermain2017increased} and 4.2 K \cite{danilov1989low}, reported changes in magnetic damping, as well as shifts in frequency of propagating SWs and FMR mode compared to room temperature. These drawbacks at low temperature were often associated with the enhanced contributions from the paramagnetic GGG substrate. More recently, Serha et al. \cite{serha2024magnetic, serha2025damping} studied FMR frequency shift and linewidth broadening down to mK, which is particularly interesting temperature range for quantum magnonics, and connected it with the influence of inhomogeneous stray magnetic field in partially magnetized GGG, as well as the change of crystallographic anisotropy in YIG with decreasing temperature. The latest study by Schmoll et al. \cite{schmoll2025elimination} addresses the issue of substrate-induced FMR linewidth broadening in mK temperatures by microstructuring the YIG film, so that it occupies only regions of homogeneous stray field from the GGG substrate. The dependence of SW damping on wavenumber in ultracold temperatures was further investigated by Schmoll et al. \cite{Schmoll2025}, while Knauer et al. \cite{Knauer2023} reported on a first experimental analysis of a propagating SW in a wide temperature range (room - 45 mK) in a nm-thick YIG film. Careful experimental planning mitigated the strong impact of GGG, confirming that large-scale integrated YIG nanocircuits remain viable at cryogenic temperatures. Recently, Serha et al. \cite{Serha2025YSGAG} and Gugushev et al. \cite{guguschev2025novel} have published key articles for YIG-based quantum magnonics, establishing a new diamagnetic YSGAG substrate and eliminating the parasitic contribution from GGG. Further, in 2025, Serha et al. \cite{serha2025ultra} made a breakthrough reporting >18 \(\upmu\)s lifetimes for short-wavelength magnons in ultra-pure YIG spheres, highlighting magnons’ viability as long-coherence carriers for quantum magnonics.\setlength{\parskip}{2pt} 
A current limitation of the field is the need for low temperatures to suppress thermal magnons (\(10^{18}~\mathrm{cm}^{-3}\) in equilibrium with the phonon bath). Nevertheless, rapid progress indicates strong potential for quantum sensing, information processing, and hybrid quantum technologies. \setlength{\parskip}{4pt}

\subsubsection{\textit{Alternative materials}} The choice of a material is crucial for the functionality of magnonic devices. To support fast SW propagation at the industry-competitive nanoscale, materials with a large SW exchange stiffness \(\lambda \textsubscript{ex}\) are essential. In addition, the fast exchange-dominated SWs, as opposed to dipolar SWs, not only allow operation with microscale antennas, but also offer greater flexibility in prototype engineering due to their highly isotropic dispersion. In this context, \textbf{ferrimagnetic insulators near magnetic compensation} are particularly promising, as their low saturation magnetization \(M_{\textup{s}}\) tends to increase exchange constant \(\lambda \textsubscript{ex}\) and SW group velocity. A notable example is LPE-grown partially-substituted nm-thick Ga:YIG \cite{carmiggelt2021electrical, Boettcher2022, dubs2025magnetically}, that has shown reduced \(M_{\textup{s}}\) and a strong stress-induced out-of-plane uniaxial anisotropy. As a result, Ga:YIG films support SWs with group velocities up to 3.4 faster than those in pure YIG (at \(k > \) 20 rad/\( \upmu \)m), while also demonstrating isotropic propagation, making this material a strong candidate for fast and scalable magnonic networks. \setlength{\parskip}{2pt}

\textbf{Hexaferrite} is another promising candidate for high-frequency applications, particularly for 5G high-band (24.25-27.5~GHz) and 6G communication technologies. This material is known for exceptional tunability of magnetic parameters with respect to required application and strong internal magnetic anisotropy, which enables broad frequency operation with minimal bias fields and relatively low losses \cite{Harris2012}. Therefore, hexaferrites are suitable for non-reciprocal RF components (circulators \cite{zhang2021novel}, isolators \cite{zuo2003self}) and magnetostatic devices (filters \cite{song2010self}, resonators \cite{popov2020nonlinear}). A flagship representative of this class, M-type barium hexaferrite (BaFe\textsubscript{12}O\textsubscript{19}), supports high-frequency operation (50-65 GHz undoped) under zero or low magnetic fields \cite{song2010self, zhang2017epitaxially, chen2006screen}, and reaches even wider range (20-100 GHz) with doping and moderate biasing <~0.7~T \cite{sharma2018rare, popov2011sub}. Despite sub-optimal growth conditions, BaM films with Gilbert damping as low as \(\alpha\)$\sim$ 7\(\cdot\)~10\textsuperscript{-4} \cite{li2016spin} were already demonstrated, which is significantly lower than in most magnetic metals. Given the ongoing efforts to optimize growth methodology in order to develop high-quality, nanometer-thick BaM films, further improvement of parameters are well within reach.

 \vspace{0.5mm}
\noindent\rule{\columnwidth}{0.4pt}
 \vspace{0.2mm}
 
\section{\label{Methods}Spin-Wave technology for RF applications}
\vspace{-2pt}

As discussed in \textit{Section I}, spin waves offer sub-100 nm scalability, a broad MHz-THz operational frequency range, nonreciprocity and nonlinearity. These advantages are best realized in low-damping magnetic insulators, such as high-quality magnetic oxides, where key SW parameters can be tuned via strength or direction of a bias field, or through material engineering \cite{Harris2012, Ishak1988}. In this context, YIG remains the key material due to its exceptionally low FMR linewidth, which directly translates to low microwave losses and efficient SW propagation \cite{Ishak1988}. Most RF devices are based on YIG films, which provide better uniformity than bulk crystals. In the magnetostatic regime, YIG demonstrates low propagation losses, outperforming SAW on lithium niobate, e.g., 20 dB/\(\upmu\)s vs 100 dB/\(\upmu\)s respectively at 9~GHz \cite{Ishak1988}. In addition to their relatively low losses, SWs operational frequencies can pick immediately after SAWs begin to lose efficiency, i.e., above 3~GHz \cite{ganguly1975radiation}, with bandwidths roughly scaling with saturation magnetization and tuned via bias field. For YIG films, the bandwidth theoretically can span from tens of~MHz up to a few~GHz \cite{Adam1988}, though actual performance is influenced by film's dimension, impedance matching and antennas' efficiency. Since SWs are two to five orders of magnitude slower than EMWs, compact devices from sub-100 nm thick films with propagation lengths of up to 580 \(\upmu\)m \cite{yu2014magnetic, qin2018propagating} are feasible, allowing for simpler fabrication. Spin-wave advantages for RF applications are outlined in \textit{Section IV}; here, we address their implementation and selected prototypes.


\vspace{2mm}
It is important to distinguish between two conceptually different approaches to SW RF applications: 

{\bf \textcolor{NavyBlue}{\textbf{1)}}} In the first approach, the RF signal (energy) remains in the electromagnetic domain, and only at specific frequencies is absorbed or re-emitted via magnetization precession in magnetic spheres or thin films \cite{Harris2012, dai2020octave, feng2023micromachined,  gillette2011active, elliott1974broadband, yang2022x}. Devices based on this principle operate at \textit{\textbf{k}} = 0 (FMR or standing SW modes), making them suitable for stopband filters, tailored passband filters, resonators, and microwave sources. Delay lines and frequency-selective limiters based on absorption are not feasible with this approach, as information is still carried at the speed of light. However, these devices offer very low insertion losses and are comparable to conventional nonmagnetic RF components with over 30 dB isolation and < 0.1 dB insertion losses \cite{Harris2012}. \setlength{\parskip}{6pt}

{\bf \textcolor{NavyBlue}{\textbf{2)}}} Another approach is based on the excitation of propagating spin waves by an RF field applied to an input transducer, followed by their detection using an output transducer. This approach is similar to SAW-based technology and is particularly suitable for delay lines, as the delay is determined by the SW group velocity, tunable via material parameters and bias field \cite{Adam1988, Ishak1988, bobkov2002microwave, adam2004msw}. The main drawback lies in the increased insertion loss, accumulated over three sources: conversion efficiency at the input transducer, SW propagation losses, and back-conversion efficiency at the output transducer. Nevertheless, analysis of the 2.5~dB insertion loss reported in \cite{bobkov2002microwave} suggests a total transduction efficiency above 80\%: -2.5 dB corresponds to a power ratio of  0.562, and assuming equal contributions from all three loss sources, this yields an estimated efficiency of \(\sqrt[3]{0.562}\) = 82.5\%. Further improvements would make SW RF devices even more compelling. \setlength{\parskip}{2pt}


The first peak of interest in magnonics from both scientific and industrial communities was in 70s-80s, as seen in historical outline. A significant part of these studies was focused on data transmission systems, where MSWs were proposed as complementary to SAWs \cite{ganguly1975radiation, ganguly1975microstrip}. Spin waves were explored for a broad range of applications \cite{Adam1988}, with particular emphasis on broadband microwave receivers (e.g., frequency channelizers, filters, delay lines, frequency-selective limiters) and on beam steering in phased array antennas by variable time delay \cite{Adam1988}. The most comprehensive industry-oriented overview of SW-based microwave devices was given in the special issue of \textit{Proceedings of the IEEE}, \textbf{76}(2) (1988). The summary below draws on various contributions from that issue (all rights to text and figures belong to the original authors). At the time, experimental progress in device development often outpaced theoretical understanding, making it difficult to predict or explain performance reliably \cite{Rodrigue1988}. With the benefit of today’s knowledge and technological advances, many of these earlier designs can now be refined and adapted to meet scientific and industrial requirements. In the following, a brief overview of selected SW devices is discussed - their role, state of technology, and modern magnonic counterparts. 
\vspace{-2pt}
\subsection{Filters}
\vspace{-2pt}
RF filters are essential components in communication systems designed to control the flow of signals by allowing certain frequencies to pass while blocking others. By selectively filtering frequencies, they prevent interference, improve signal clarity, and ensure spectral selectivity across a wide range of applications, such as telecommunications, broadcasting, and wireless communication. For years, high-performance passive RF filters have mostly relied on silver-coated ceramic monoblocks, offering a balance between attenuation, insertion loss, and compactness. Advances in material science have improved these ceramics by reducing dielectric losses tan \(\delta\), and increasing the dielectric constant \(\epsilon_r\), enabling significant miniaturization. However, their bulkiness (typically in the millimeter range) remains a major limitation for integration into compact mobile RF systems. Efforts to downscale them by combining ceramic dielectrics with magnetic materials and optimizing the structure are promising, but introduce manufacturing challenges and risk performance degradation. Moreover, the primary bottleneck has shifted from dielectric loss to conductor loss in the silver coating, signaling the need for alternative solutions. \setlength{\parskip}{2pt}

Surface acoustic waves \cite{delsing2019, ilderem2020technology} technology helped address some of the limitations of conventional RF filters, leading to modern multi-band, multi-standard mobile communication systems. The SAW filters market is extensively growing, including such tech giants as Abracon, API Technologies, Kyocera, Microchip Technologies, Murata Manufacturing, Qorvo, Qualcomm Technologies, etc. A key advantage of SAWs is their significantly lower group velocity compared to EMWs, allowing bulky RF devices (due to their cm to m wavelengths) to be miniaturized by a factor of 100,000. This enables compact, chip-based RF signal processing within the tens of MHz to several GHz range. Yet, as discussed in \textit{Section I}, the high-frequency limitations persist due to increased damping and insertion losses above 40~dB at 15~GHz \cite{yamanouchi19975}, as well as the exposure limitations of interdigitate transducers (IDTs) \cite{delsing2019, hara2010super}. 

\renewcommand {\thefigure}{7}
\begin{figure}[ht!]
\centering
    \includegraphics[width=0.62\columnwidth]{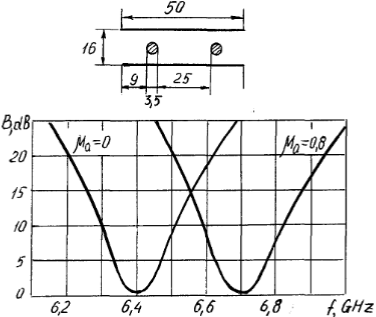}
    \caption{Two-order ferrite filter response. Adapted after \cite{yu1990tunable}.}
    \label{f:7}		
\end{figure}

Compared to SAW, bulk acoustic wave filters have smaller losses at frequencies above 3 GHz due to their higher resonator Q-factor \cite{delsing2019}. Emerging technologies like temperature-compensated SAW and piezoelectric-layer-based filters may shift the balance back in favor of SAW technology, yet it is still uncertain whether the resulting filters will successfully extend their reach beyond the 3~GHz. In contrast, BAW RF filters are already available for mid-range frequencies (up to 6~GHz), though their high-frequency range (30~GHz) is remains unexplored. Key limitations include increased damping, inefficient energy transfer between the transducer and the acoustic wave at higher frequencies, strict requirements for BAW isolation from the substrate to prevent unwanted energy dissipation, and the associated fabrication complexity \cite{hara2010super}.\setlength{\parskip}{2pt}

Magnonics offers a superior operational frequency region and high technological flexibility. One of the first proposed magnonic concepts was the SW delay line (see \textit{Subsection D} for details), which can be repurposed as a bandpass filter by adjusting transducer dimensions and YIG-ground plane separation. A functional prototype of a narrow-band SW filter with a bandwidth down to 30~MHz in the microwave regime was demonstrated by Ishak et al.  \cite{Ishak1988}. Using two transducers on a 20~\(\upmu\)m-thick YIG film, and introducing a 250 \(\upmu\)m-thick dielectric spacer between the YIG film and the gratings narrowed the 30~MHz passband and kept the ripple amplitude below 0.1~dB while tuning across 3-7 GHz. To realize a wideband RF filter, Adam et al. \cite{Ishak1988, adam1985msw} proposed to increase the transducer width, decrease YIG thickness, and to taper the waveguide edges in order to suppress width-mode coupling. Resulting FVMSW RF filter showed more than 45 dB out-of-band rejection and 16–34 dB insertion loss in the 0.3–12 GHz range. \setlength{\parskip}{2pt}

\renewcommand {\thefigure}{8}
\begin{figure}[hbt]
\centering
\includegraphics[width=1.0\columnwidth]{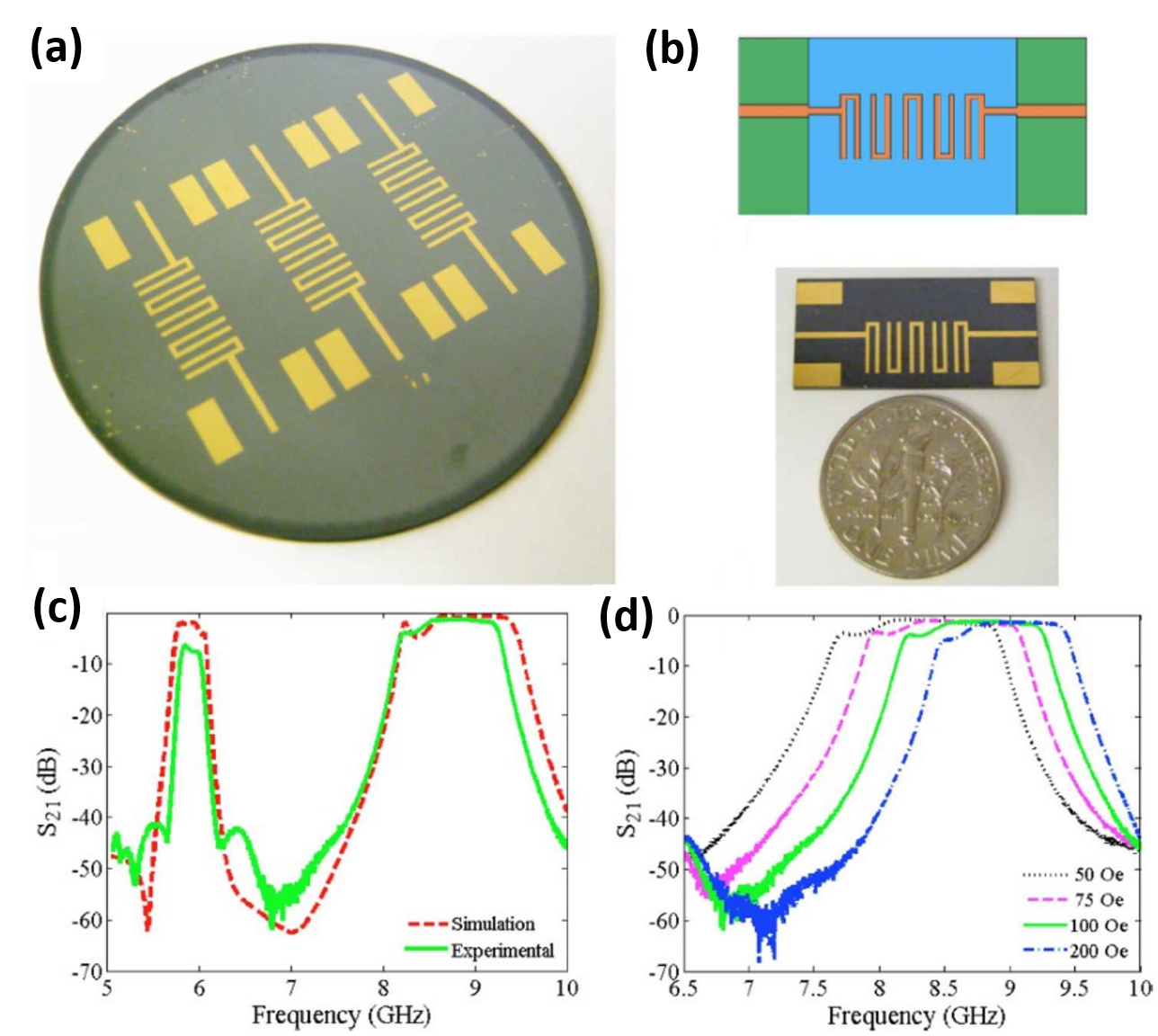}
\caption{(a) Photograph of patterned 50 mm diameter YIG wafer. (b) Design and photograph of fabricated ferrite microstrip hairpin line coupled resonator bandpass filter. (c) Simulated and experimentally obtained passband characteristics of the RF filter. Measured spectrum is collected under applied magnetic field of 10 mT. (d) Passband characteristics of the filter under applied magnetic field from 5 to 20~mT. Adapted after \cite{Harris2012, gillette2011active}.}
    \label{f:8}		
\end{figure} 

The tunable microwave filters for communications systems were proposed in 1990 by Kapilevich and Safonov \cite{yu1990tunable}, with an exemplary two-order filter shown in Fig. \ref{f:7}. Here, polycrystalline YIG cylinders of 3.5 mm diameter each were symmetrically placed 25 mm apart and 9 mm from the waveguide input. Frequency tuning of up to 300 MHz was achieved by varying the bias magnetic field up to 20 kA/m, with measured passband insertion loss of only 0.5 dB and a Voltage Standing Wave Ratio (VSWR) of 1.5. The implementation of polycrystalline ferrite boosts the microwave power passing through the filter by three orders of magnitude, while nonlinear effects are not detected up to 2~W. The passband center is tuned mainly by the cylinder diameter (coarse tuning), while the bias field provides fine adjustment, allowing the same structure to serve both for transmitter and receiver channels of mobile relay equipment. The resonator is large enough that EMWs vary across it, so the usual simplified models are no longer accurate. This regime of forced and free oscillations in large magnetic bodies was later explained analytically \cite{Gurevich1996}. \setlength{\parskip}{2pt}

Since the 2010s, interest in RF filter has been renewed. Among the first new-wave designs was a magnetically tunable bandpass filter proposed by Gillette et al. \cite{Harris2012, gillette2011active}. The device, shown in Fig. \ref{f:8}, consists of a five-pole Chebyshev bandpass filter implemented in microstrip hairpin-line coupled resonator geometry, and fabricated on a 1 mm-thick polycrystalline YIG substrate ($M_{\mathrm{s}}$ = 139.3 kA/m, FMR linewidth $\Delta H =$ 1.5 mT). The filter is actively tuned via a small in-plane field (5–20~mT), which modifies the magnetic permeability of the substrate above the FMR. The compact planar geometry enables passband center frequency shift from 8.3 to 9.0 GHz and low insertion losses between 1 and 1.4 dB, yet remains compatible with modern photolithographic fabrication. Its low tuning field, good power handling, and intrinsic radiation tolerance make it promising for space and mobile communication \cite{Harris2012}.

\renewcommand {\thefigure}{9}
\begin{figure}[h!]
\includegraphics[width=1\columnwidth]{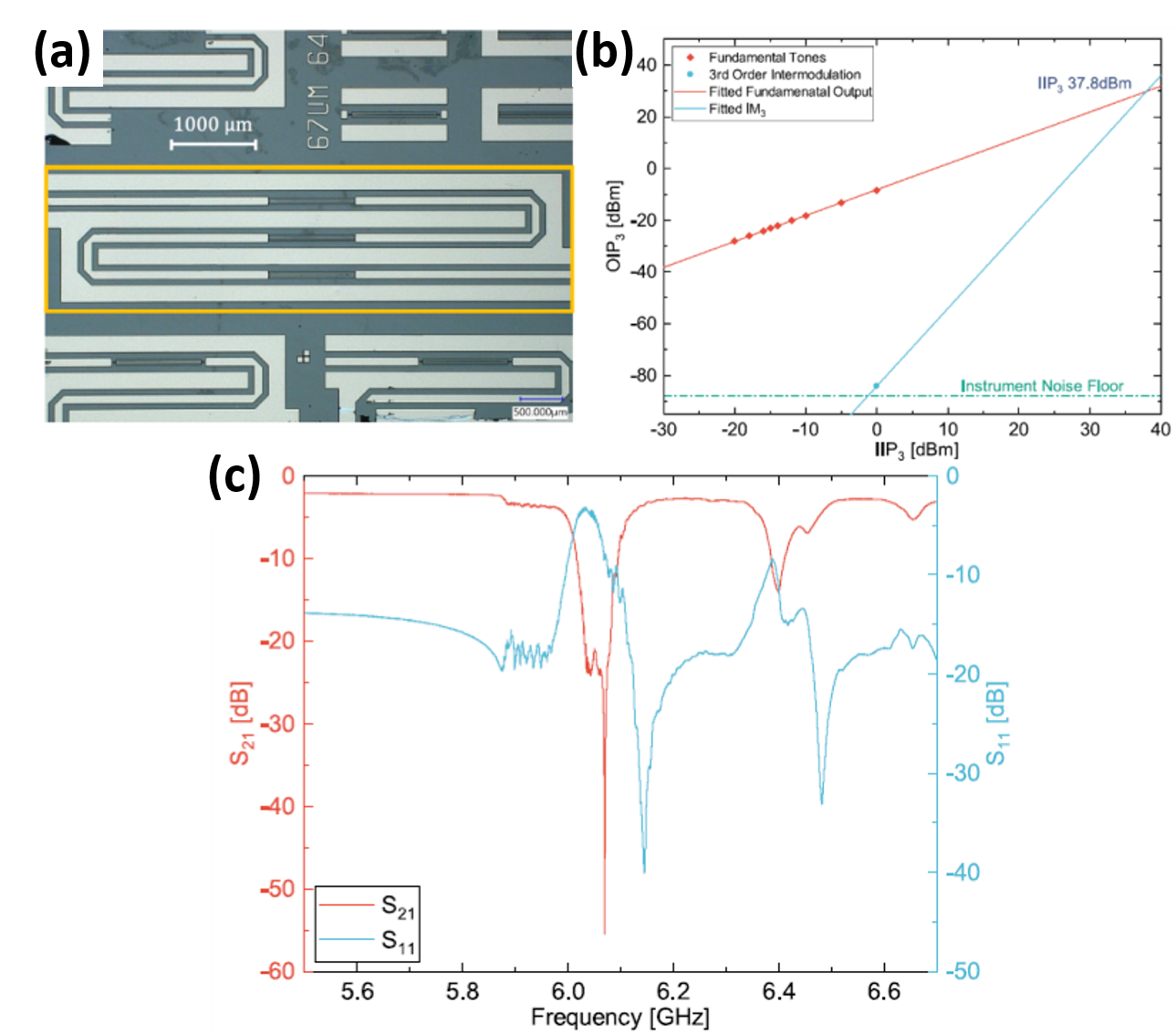}
\caption{(a) Optical image of the fabricated third-order filter (yellow box) with three YIG resonators of slightly different widths  W\textsubscript{1} = 73~\(\upmu\)m, W\textsubscript{2} = 70 \(\upmu\)m, and W\textsubscript{3}  = 67~\(\upmu\)m. (b)  Measurement of the passband IIP3; the IIP3 point is estimated around 37.8~dBm. (c) \textit{S}-parameters of the YIG bandstop filter with a center frequency of 6.07~GHz under a 390~mT bias field. Adapted from~\cite{feng2023micromachined}.}
\label{f:9}	
\end{figure}

The latest achievement in the field is a planar monolithic Chebyshev stopband filter by Feng et al. \cite{feng2023micromachined} fabricated via micromachining technology (Fig.~\ref{f:9}). The device consists of 3~\(\upmu\)m-thick YIG ($M_{\mathrm{s}}= 140$ kA/m) mesas ion-milled on a GGG wafer and aluminum transducers deposited directly on the mesas for strong FVMSW coupling. Under 390~mT out-of-plane magnetic field, the filter showed 55~dB maximum stopband rejection at a center frequency of 6~GHz, with 1.5–3.1~dB passband insertion loss and +37.8~dBm passband IIP3 (Third order Input Intercept Point). By applying different bias fields, the stopband frequency was tuned from 4 to 8~GHz, while maintaining above 30~dB rejection. If incorporated with compact electromagnet, this reconfigurable, low-loss, and high-rejection filter can attenuate spurs across the 5G and X-bands \cite{feng2023micromachined}. \setlength{\parskip}{2pt}

Using deep anisotropic etching of YIG/GGG, Tiwari et al.~\cite{tiwari2025high} realized MSW filters based on 3-$\upmu$m-thick YIG films grown on 500-$\upmu$m GGG substrates. The design consisted of a transducer with hairpin-type YIG resonators, microfabricated by thinning and etching the GGG substrate to depths up to 100~$\upmu$m. This innovative architecture enabled zero-power-bias filters tunable across 3-20~GHz and demonstrated improved resonator coupling ($>8\%$ above 6~GHz), surpassing the $<3\%$ coupling previously reported for top-electrode designs.

\renewcommand {\thefigure}{10}
\begin{figure}[hbt]
\includegraphics[width=1\columnwidth]{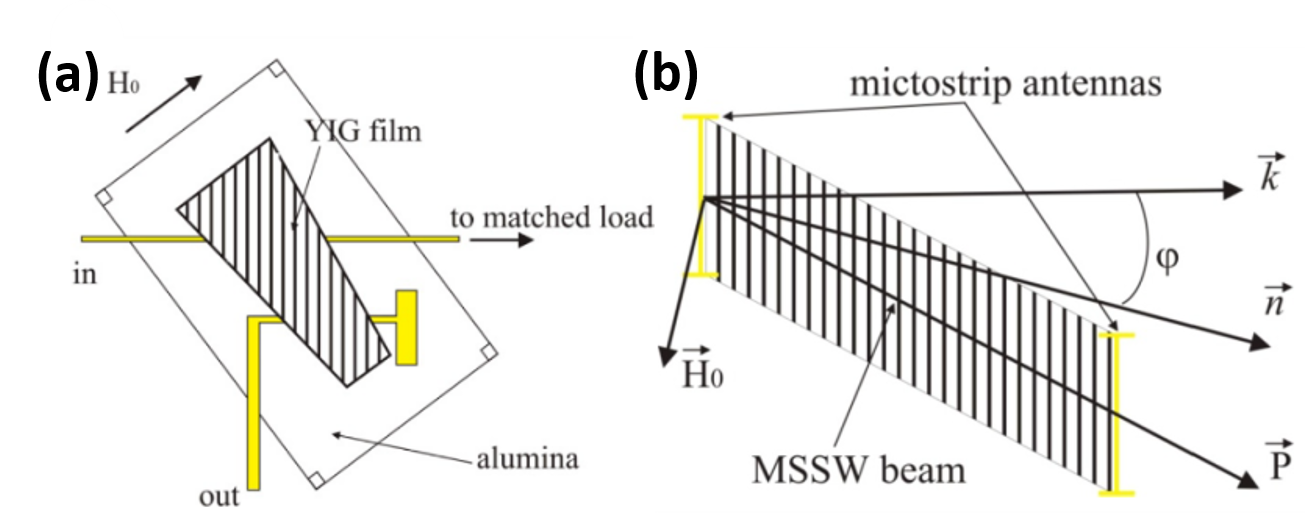}
\centering
\caption{(a) Magnetostatic surface spin wave RF filter. (b) Scheme of the MSSW filter working principles. Adapted from \cite{Yttrium, bobkov2002microwave}}
\label{f:10}
\end{figure}

\renewcommand {\thefigure}{11}
\begin{figure*}[ht!]
\centering
\includegraphics[width=0.65\paperwidth]{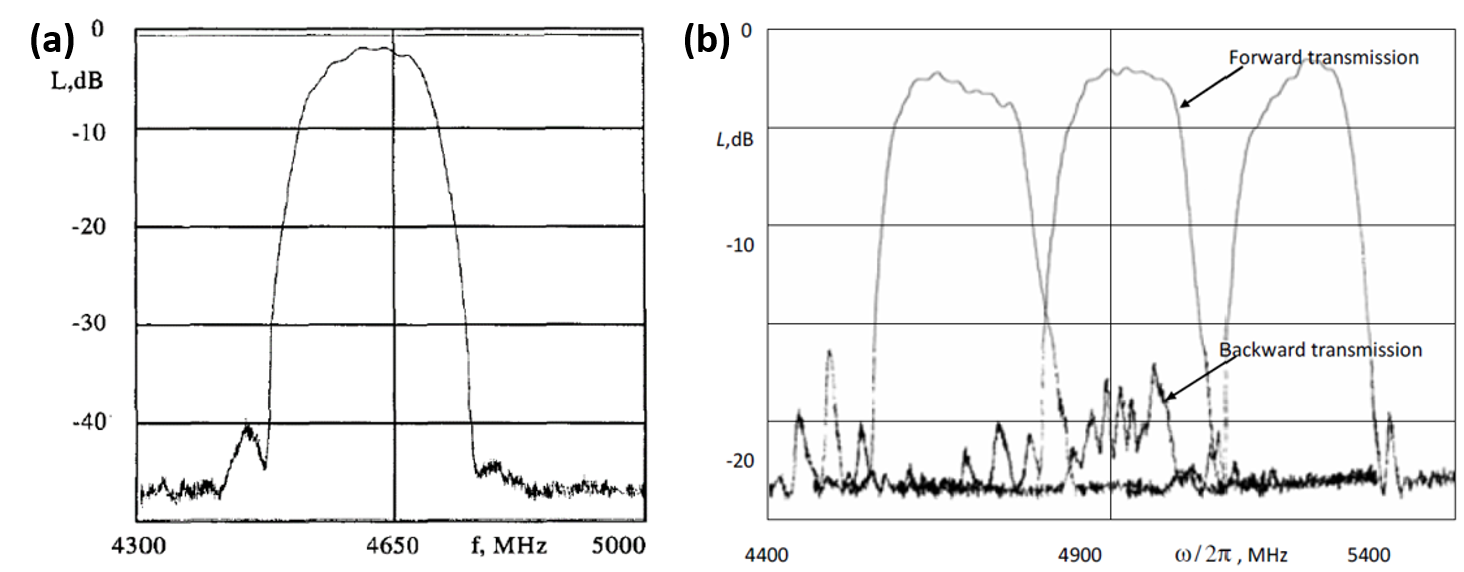}
\caption{(a) Insertion losses of the YIG-based \(\upmu\)m-thick MSSW RF filter. (b) Amplitude–frequency response of a 16-channel MSSW filter. Adapted from \cite{Yttrium, bobkov2002microwave}.}
    \label{f:11}		
\end{figure*}

\renewcommand {\thefigure}{12}
\begin{figure*}[!hb]
\centering
\includegraphics[width=0.63\paperwidth]{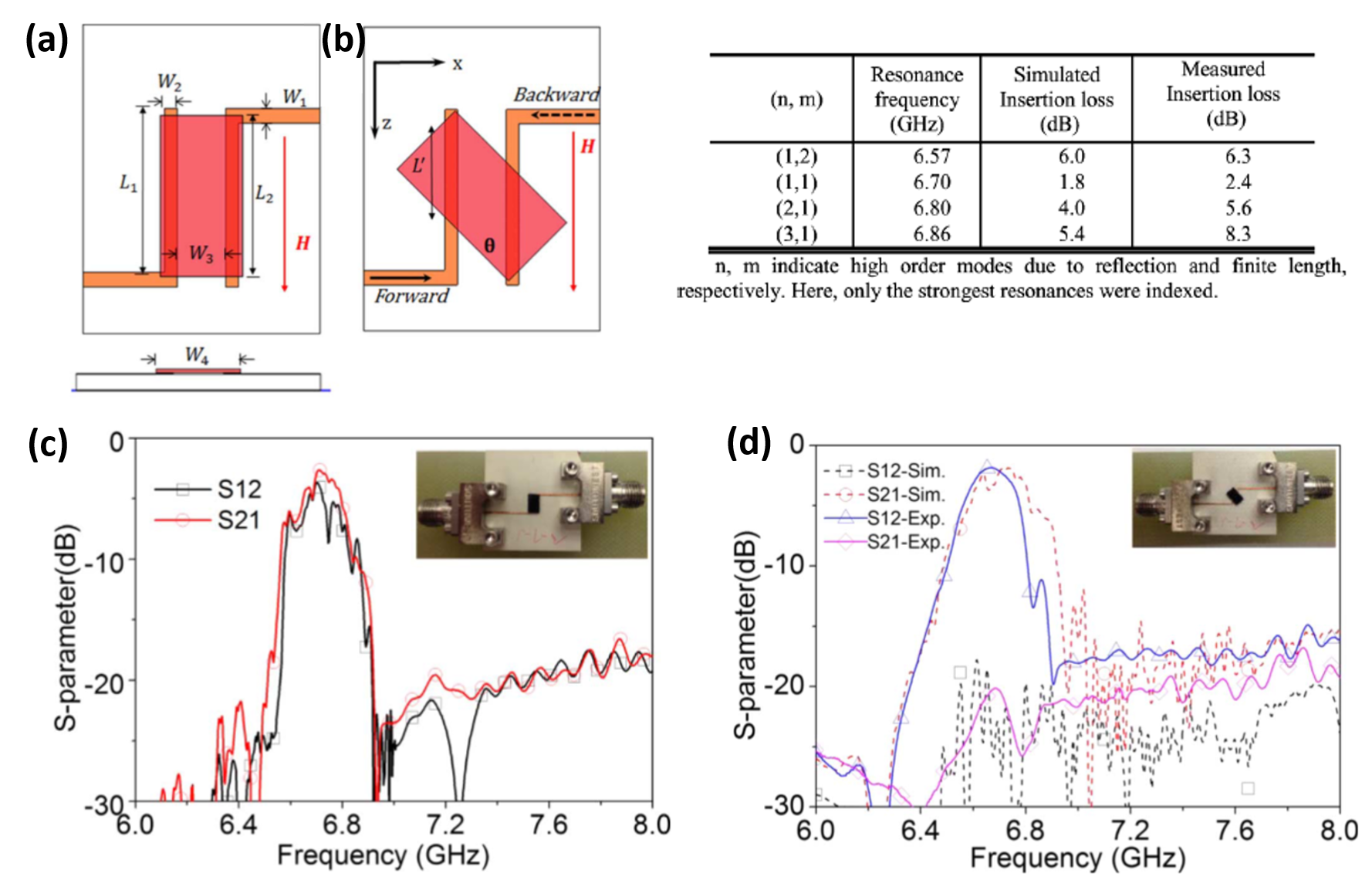}
\caption{(a) Geometry of the proposed passband filters consisting of inverted-L-shaped microstrip antennas and parallel-aligned \(\upmu\)m-thick YIG film with key geometrical parameters:  W\textsubscript{1} = 0.37 mm, W\textsubscript{2} = 0.32 mm, W\textsubscript{3} = 1.2 mm, W\textsubscript{4} = 2 mm, L\textsubscript{1} = 4 mm, L\textsubscript{2} = 3.6 mm. (b) Antennas with rotated YIG film, L\textsuperscript{'} = 2.6 mm and \(\theta\) = 45\textdegree. \textit{Table}: resonance modes indexing. (c) Measured passband response for parallel YIG alignment (scheme (a)) under a 160 mT perpendicular bias. (d) Simulated and measured results of the bandpass filter with the YIG resonator rotated 45° relative to the transducers, as shown in scheme (b). Adapted from \cite{wu2012nonreciprocal}.}
\label{f:12}	
\end{figure*}

One of the best SW-based filters in terms of insertion loss was reported in 2002 by Bobkov et al. \cite{bobkov2002microwave}, Fig. \ref{f:10}. The authors realized a device based on a metal-dielectric-ferrite stack in which the MSSW propagates at an acute angle \(\upvarphi\) to the in-plane bias field. Transducers for SW generation and detection were designed in a shape of a microstrip and fabricated from a dielectric substrate with high permittivity \(\epsilon_r\) \(\sim\) 10. A YIG ferrite layer was placed on top of the antennas, as shown in Fig. \ref{f:10}(a). The parameters of the single-crystal epitaxial film were not explicitly stated, but expected to be within 20-55 \(\upmu\)m in thickness, around $M_{\mathrm{s}}$ = 139.3~kA/m in saturation magnetization and $\Delta H =$ 0.04..0.06~mT in FMR linewidth. \textbf{Demonstrated single-channel filter achieved insertion losses <~2.5~dB at the center frequency and out-of-band rejection >~40~dB} (Fig. \ref{f:11}(a)), while a 16-channel filter (filterbank, channelizer) maintained losses <~11~dB with 125~$\pm$~5~MHz channels over 4 to 6~GHz range \cite{Yttrium}. The insertion loss for the device could be further reduced to 8~dB with an addition of protective bands between the channels. Proposed approach allows to tune bandwidth independently of center frequency, which is a key advantage for multichannel filters. Detailed explanation is provided in the respective chapter of „\textit{Yttrium: Compounds, Production, and Applications}” \cite{Yttrium}. Authors initially identified MSSW as the most suitable SW mode for efficient RF signal processing due to its intrinsic nonreciprocity and high excitation efficiency using simple transducers. These advantages were capitalized on in the development of the RF passband filters and filterbanks \cite{Yttrium}.

The core operating principle of the proposed devices is based upon MSSW wavevector \textbf{k} lying in the plane of the structure at an arbitrary angle \(\upvarphi\) to vector \textbf{n}; \textbf{k} is dictated by the transducer’s structure (Fig.~\ref{f:10}(b)). Vector \textbf{n} is orthogonal to \textbf{H}\textsubscript{0}, while the beam propagates along \textbf{P}. As the angle \(\upvarphi\) within the \(\uppi/2 < \upvarphi \leq \uppi\) area narrows, the MSSW frequency range decreases as well, enabling efficient control of a passband. The microstrips for MSSW excitation and detection were placed on the dielectric layer surface adjacent to the ferrite layer, maximizing the SW excitation efficiency \cite{Yttrium}. The MSSW spectrum can be shifted by adjusting the static field \textbf{H}\textsubscript{0}, which is controlled by a magnetic shunt made of soft magnetic material. By varying the distance between the small magnet and the shunt from 0 to 3 mm, the field can be tuned  in the 80-140~mT range, leading to about 4-6~GHz variations in the filter's central frequency. An important characteristic of the considered prototype is its nonreciprocity, i.e., a strong attenuation when the signal is transmitted at reverse direction (from output to input). As such, the device performs both filtering and isolating functions simultaneously. Fig. \ref{f:11}(b) shows the corresponding amplitude-frequency response under reverse connection for the central band, with attenuation reaching at least 34 dB, making it practical for RF isolation applications \cite{Yttrium}. \setlength{\parskip}{2pt}

 Among the lowest losses yet were reported by Wu et al. \cite{wu2012nonreciprocal} in nonreciprocal tunable bandpass filters with ultra-wideband isolation. MSSW were excited in a single-crystal YIG film ($M_{\mathrm{s}}$ = 139.3~kA/m) using an inverted L-shaped microstrip pair, as shown in Fig.~\ref{f:12}(a,b). Simulations of S\textsubscript{21} and S\textsubscript{12} responses for the parallel YIG configuration predicted reciprocal transmission with losses $\approx1.8$~dB at the central frequency of 6.7~GHz and overall 3~dB losses over the 170 MHz bandwidth. 
 However, measurements showed slightly higher values of $\approx2.4$~dB (Fig.~\ref{f:12}(c)) due to edge roughness from the fabrication (Fig.~\ref{f:12} – Table). Rotating the YIG film by 45\textdegree~suppressed standing-wave resonances and introduced nonreciprocity, making the passband smoother and shifting the propagation losses to depend on the slab width rather than its resonance condition. In this configuration, the forward transmission showed reduced insertion loss of 1.65~dB at 6.7~GHz with a broader bandwidth of 220~MHz, while in reverse the signal was isolated and dropped down to 22~dB (Fig. \ref{f:12}(d)). The device operated in 5.2 – 7.5~GHz range under 110 – 190~mT in-plane field. Additionally, power handling of >30~dBm was achieved, making this MSSW filter architecture promising for C-band RF front-end and other components. 

\renewcommand {\thefigure}{13}
\begin{figure*}[!t]
\centering
\includegraphics[width=0.7\paperwidth]{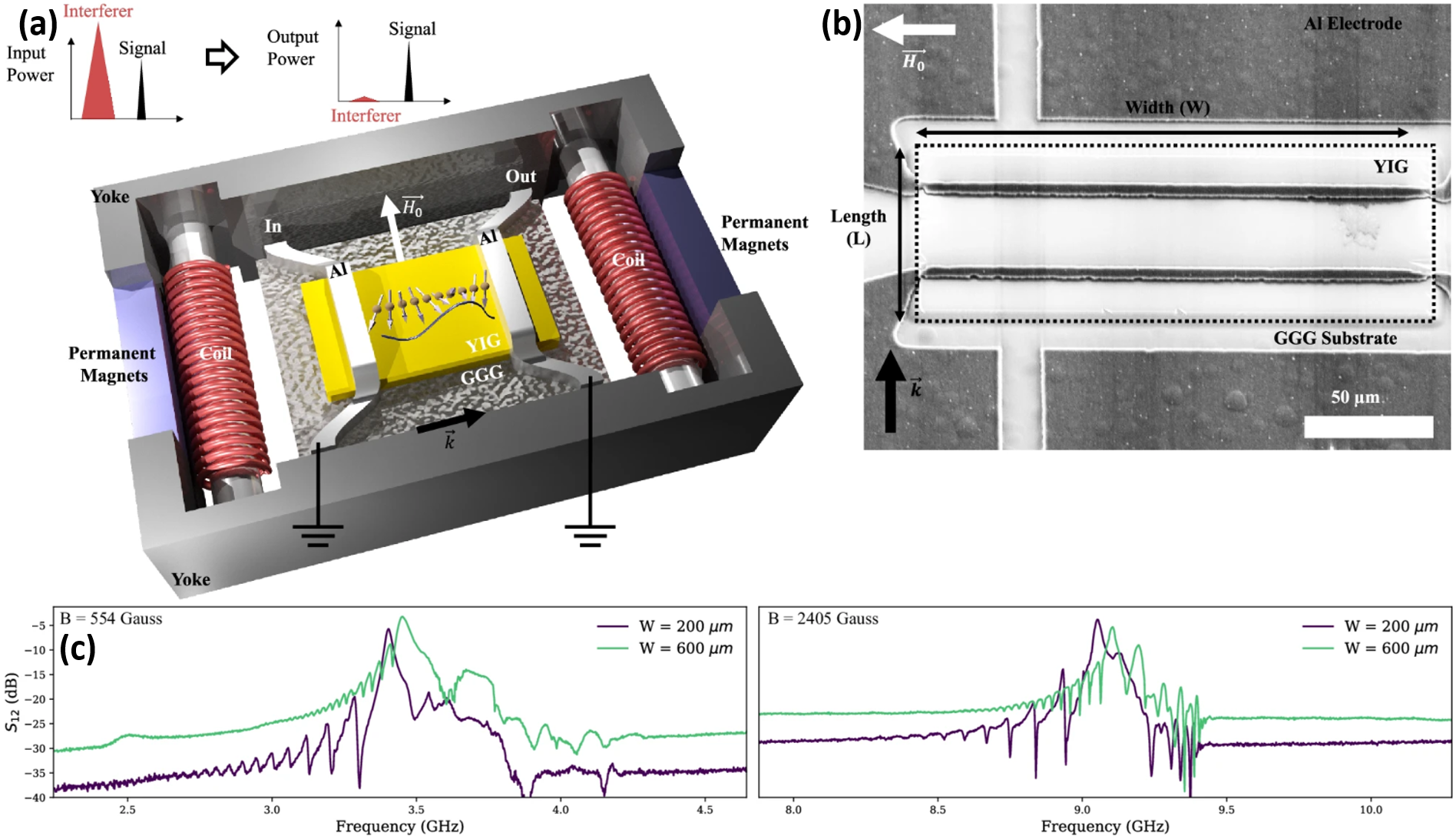}
\caption{(a) Reconfigurable MSSW bandpass filter. (b) SEM image of Al transducers on top of the YIG cavity; device's width = 200~$\upmu$m, length = 70~$\upmu$m. (c) The impact of YIG cavity width on the frequency response with a constant length of 70~$\upmu$m.  Adapted from \cite{du2024}.}
\label{f:13}	
\end{figure*}

Miniature ($<2~\mathrm{cm}^3$) frequency-tunable MSSW bandpass filters with zero static-power biasing were recently developed and tested by Du et al.~\cite{du2024}. The devices were realized from micromachined resonant cavities in 3.3-$\upmu$m-thick YIG films with 2-$\upmu$m-thick, 7-$\upmu$m-wide aluminum microstrips, as shown in Fig.~\ref{f:13}(a,b). The magnetic bias system consisted of two permanent magnets, two shunt magnets wound with coils, and two high-permeability yokes to concentrate the flux into the filter. This versatile design enables operation in the 3.4 – 11.1~GHz range using sub-ms current pulses applied to a nonvolatile bias assembly, achieving 3.2 – 5.1~dB insertion loss Fig.~\ref{f:13}(c). The MSSW filters also demonstrated strong suppression of out-of-band interference while maintaining high linearity (out-of-band third-order intercept point IIP3 above 41~dBm), hence reducing the signal degradation associated with interferences.

A cutting-edge reconfigurable YIG-based SW ladder filter was reported by Devitt et al. in 2025 \cite{devitt2025spinwave}. The key idea is to lithographically engineer a resonance offset between series and shunt resonators, combining wave-vector selection with geometry-dependent demagnetizing fields, so that the entire ladder runs under a single uniform out-of-plane bias. The measured third-order filters have shown low insertion loss of 2.54~dB, up to 663~MHz bandwidth and multi-octave center-frequency tuning from 7.08 to 21.6 GHz, while keeping a high linearity with an input-referred IIP3 >11~dBm in the passband.

\renewcommand {\thefigure}{14}
\begin{figure*}[!ht]
\centering
\includegraphics[width=\textwidth]{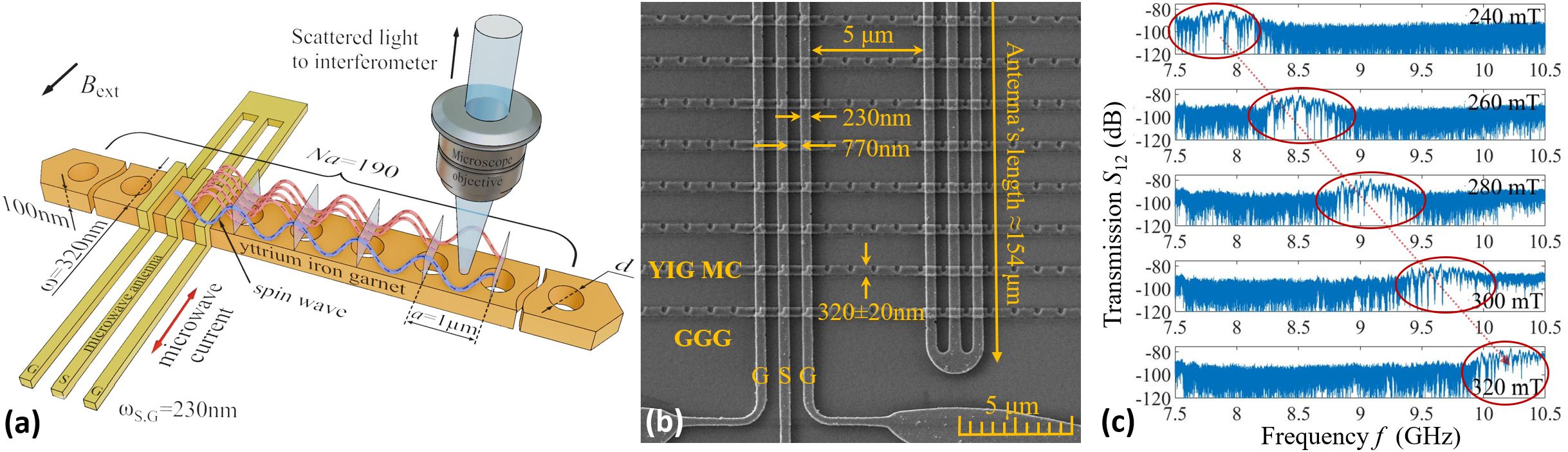}
\caption{(a) Sketch of 1D waveguide-based MC periodically modulated with round holes. Spin waves excited by the CPW are shown in blue, while Bragg-reflected waves - in red. (b) SEM image of the typical 1D MC fabricated from a 100 nm-thick LPE-grown YIG/GGG film, considering $\ n_\mathrm{wg}$ = 100 waveguides per structure and the distance between the antennas – 5 $\upmu\mathrm{m}$; (c) SW transmission signal $\ S_{12}$ of magnonic crystal shown at (b) for varying bias magnetic fields in a frequency range 7.5 GHz–10.5 GHz. Adapted from \cite{levchenko20251d}.}
    \label{f:1dmc}		
\end{figure*}

An alternative strategy involves magnonic crystals, which can function as microwave filters due to their tunable SW transmission characteristics (bandwidth, stopband, passband) through the structure with a periodic change of magnetic properties \cite{merbouche2021frequency, reed1985current, krawczyk2014review, levchenko20251d}. For example, in 2025 Levchenko et al. \cite{levchenko20251d} experimentally realized a nanoscale 1D YIG magnonic crystal (Fig.~\ref{f:1dmc}) by patterning periodic nanoholes (150~nm in diameter, 1~\(\upmu\)m period) into 320 nm-wide, 100 nm-thick YIG waveguides. MSSWs propagation over 5~\(\upmu\)m was shown and the resulting spectra exhibited pronounced Bragg band gaps (up to 26~dB rejection efficiency). Demonstrated functionality can be utilized in a compact notch filter with field-tunable center frequency and gap width. It should be noted that the nanoscaling enables a 100~MHz single-mode transmission window (which can be enhanced by further narrowing the waveguides), but at the cost of higher insertion loss due to reduced magnetic volume and greater sensitivity to fabrication imperfections. With improved fabrication and optimized antennas, magnonic crystals would be promising candidates for low-energy, high-frequency RF applications. As highlighted in \textit{Section B}, magnonic crystals are also strong candidates for other microwave components, including resonators \cite{hartemann1984magnetostatic}, \cite{djafari2011one}, phase shifters \cite{zhu2014magnonic}, delay lines \cite{bankowski2015magnonic} and multi-magnon processing devices \cite{frey2020reflection}, making them versatile building blocks for applied spintronics and magnonics. 

\renewcommand {\thefigure}{15}
\begin{figure*}[!b]
\centering
\includegraphics[width=0.83\paperwidth]{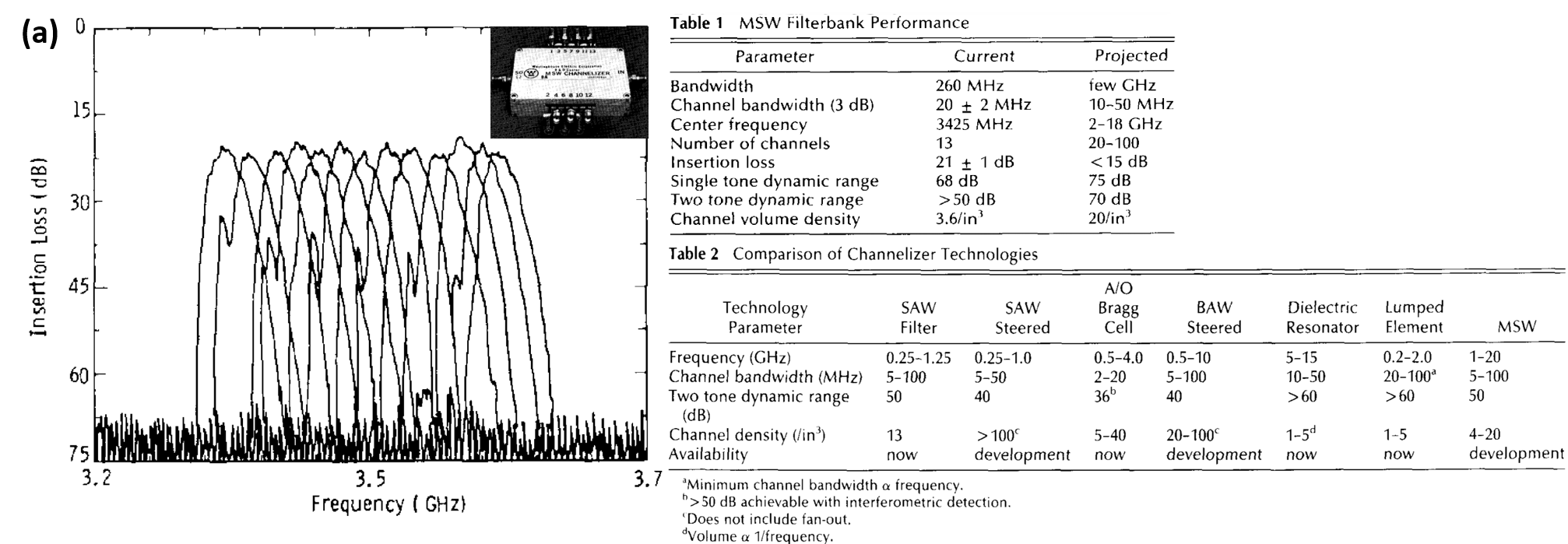}
\caption{(a): Insertion loss as a function of frequency measured for the 13-channel FVMSW filterbank (channelizer). \textit{Inset}: Photo of a device. \textit{Table~1}: Current parameters of the designed device and a projections of its future performance. \textit{Table~2}: Comparison of the channelizer technologies. Adapted from \cite{Adam1988}.
}
\label{f:15}		
\end{figure*}

\vspace{-3pt}
\subsection{Channelizer (filterbank)}
\vspace{-3pt}
As mentioned in previous subsection, channelizers offer a promising route for RF signal processing. They are sophisticated devices, that divide an input signal into distinct frequency bands, allowing for simultaneous filtering and analysis of multiple channels, which is essential for telecommunications and radar systems. SW-based channelizers, in particular, offer compact size, wide operation frequency range and superior dynamic range compared to Bragg cells or compressive receivers \cite{Adam1988}. If to follow a naive design of a filterbank as an array of discrete filters, a rather complex, large and expensive structure would be required to cover the expected frequency range. Instead, they typically use a single filter with a 1-2 GHz bandwidth, covering a broader RF spectrum by folding multiple bands into this range. The frequency ambiguity is resolved by broadband auxiliary detectors within each band. Alternatively, the filterbank can be time-shared across bands to reduce hardware complexity. A prominent example is the YIG-based MSW filterbank \cite{Adam1988}, which consists of 13 narrow-band delay lines arranged along an input. Each line consists of a 1-mm wide, 12-\(\upmu\)m thick strip of LPE-grown YIG, biased perpendicularly to the film surface to excite FVMSW. Narrow passbands (Fig. \ref{f:15}(a)) were obtained using 0.5 mm-wide transducers spaced 160 \(\upmu\)m from the YIG film, and reflective terminations to generate constructive interference at the channel center frequency and destructive interference at off-resonant frequencies. The resulting filterbank has uniform channel spacing, an insertion loss of 21~$\pm$~1 dB at center frequencies, and out-of-band rejection over 50 dB. \cite{Adam1988}. 
\vspace{-1pt}

\subsection{Phase shifters}
\vspace{-3pt}
Phase shifters are important components in RF systems, designed to adjust the phase of a signal without altering its amplitude. They are critical for phased-array antennas, radar systems and wireless communication, where precise phase control is required for beam steering, signal synchronization, and interference mitigation. The performance of RF phase shifters is characterized by the phase-shift range, insertion loss and phase resolution, among others. Typically, devices are realized via conventional technologies, such as semiconductor-based (PIN diodes or FET), MEMS, digital, etc. Yet, they all face limitations similar to aforementioned filters, i.e., rising complexity, size, insertion loss and power consumption at higher frequencies. \setlength{\parskip}{2pt} 

Phase shifting was one of the earliest application of ferrites, with experiments dating as far back as 1956 \cite{Rodrigue1988}. Passive ferrite phase shifters (including those from YIG) do not use SWs, but usually rely on changes in the permeability of a magnetic medium induced by a continuous or slowly varying magnetic field. This change alters the phase velocity of a propagating EMW's, producing a controllable phase shift. While such devices offer good power handling and broad frequency range, they are still limited by relatively large size and slow response times. Phase shifting can also be achieved via a Faraday effect, which primarily rotates EMW polarization, useful for isolators and circulators, but also leads to a differential phase accumulation between circularly polarized components propagating in a magnetized medium. This approach, however, suffers from the same drawbacks. To improve some of them, namely response time, to sub-microsecond, phase shifters based on pulsed remagnetization can be used, where the magnetic state of ferrite elements within a waveguide is switched by brief pulses through the coil, rather than continuous fields. Based on this approach, Abuelma'atti et al. \cite{abuelma2009variable} demonstrated a variable toroidal ferrite phase shifter with low insertion loss ($<1$~dB) across the 9.5-10.3~GHz range and return loss of 20 dB.

\renewcommand {\thefigure}{16}
\begin{figure*}[!b]
\centering
\includegraphics[width=0.69\paperwidth]{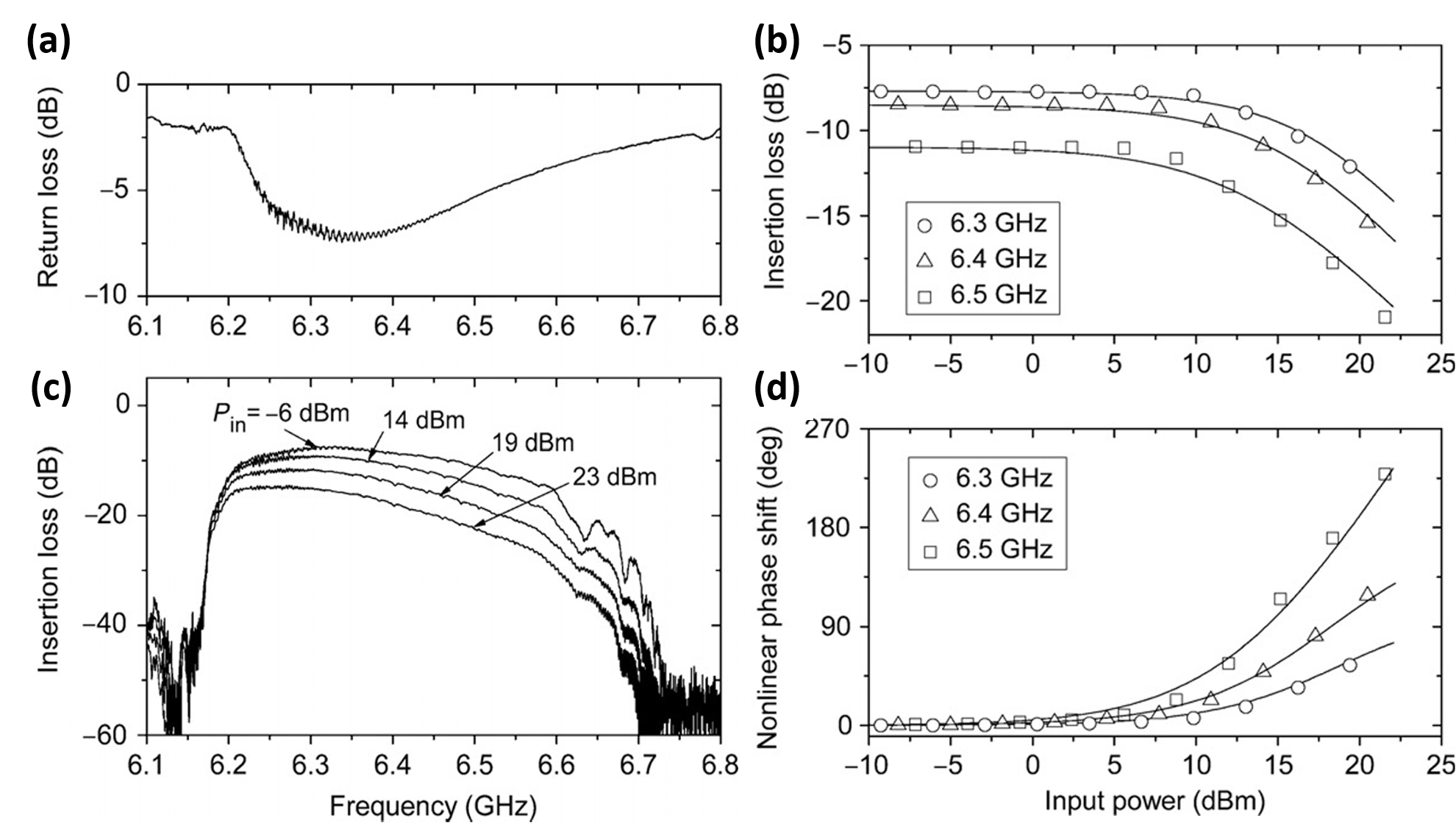}
\caption{Frequency characteristics of return (a) and insertion loss (c) of the nonlinear phase shifter measured for the different input powers \(P_{\textup{in}}\) as indicated. Insertion loss (b) and nonlinear phase shift (d) calculated for the different frequencies of 6.3~GHz, 6.4~GHz, and 6.5~GHz as a function of the input power. The symbols and the solid curves show experimental and theoretical data, respectively. Adapted from \cite{kalinikos2013nonlinear}.}
    \label{f:16}		
\end{figure*}

Magnetostatic spin wave-based delay lines can also act as tunable phase shifters (more on delay lines later in this \textit{Section}). In the device shown by Krug \cite{krug1984microwave}, a YIG delay line was integrated with 5 evenly spaced microstrip taps, each connected to a planar \(\uplambda\)/2 resonator. The antennas together with 20 × 5~mm large and 100 \(\upmu\)m thick YIG film were deposited on alumina substrate. A permanent magnet supplied the magnetic bias field, and beam steering was achieved by adjusting the field with an auxiliary coil. At 3 GHz, the device reached a scan angle of 45° under 109 mT bias field and 100 mA control current. \setlength{\parskip}{2pt} 

An overview of MSW devices was also provided by Kalinikos and Ustinov \cite{kalinikos2013nonlinear}. Operational characteristics of the proposed nonlinear MSSW-based phase shifter are shown in Fig. \ref{f:16}. The prototype device used two short-circuited microstrip antennas, each 50~\(\upmu\)m wide and 2~mm long, evaporated onto a grounded 500~\(\upmu\)m-thick alumina substrate 4.6~mm apart. The active magnetic medium was a 13.6~\(\upmu\)m-thick, 2 mm-wide, 40~mm-long YIG single-crystal film epitaxially grown on a 500 \(\upmu\)m-thick GGG substrate. The film demonstrated a narrow FMR linewidth of 0.05 mT at 5 GHz and a saturation magnetization of 194.7~mT measured with YIG side flipped down to face antennas. Fig.~\ref{f:16} presents a typical return loss and amplitude–frequency characteristics of the nonlinear phase shifter measured under a bias field of 143.1~mT for input powers \(P_{\textup{in}}\) from –6~dBm to +23~dBm. For small input power, SW propagated in a linear regime within the entire operating frequency range. As the input power increased, return loss remained nearly unchanged, but insertion loss increased significantly at 6.34~GHz from –7~dB for powers up to +10~dBm, to –15~dB at +23~dBm, highlighting the device’s nonlinear response.\setlength{\parskip}{2pt}

Further experimental progress in this field was made by Hansen et al. \cite{hansen2009dual} who demonstrated a dual-function SW phase shifter capable of both linear and nonlinear phase control. The device was based on a 6.1 \(\upmu\)m-thick YIG waveguide with lateral dimensions 1.5 x 15 mm$^2$, where excitation and detection were done by 50 \(\upmu\)m-wide microstrip antennas, spaced 4.8 mm apart. A uniform in-plane bias magnetic field of 80 mT was applied along the YIG stripe, while local field modification was achieved via a pulsed current through a 50 \(\upmu\)m-diameter conductor wire centered between the antennas. Linear phase control was realized by creating a local inhomogeneity in the external magnetic field, while nonlinear control was done through the amplitude-dependent shift in the SW dispersion due to magnetization reduction at high power. The study demonstrates that these two mechanisms can operate simultaneously within a single device, offering minimum cross-talk. Such a functionality enables dual-logic operations, enhancing the versatility and performance of SW-based RF and computing systems. \setlength{\parskip}{2pt}

Latest advancements include a nanoscale SW valve and phase shifter, simulated by Au et al. \cite{au2012nanoscale}. The device features a nanomagnet placed atop a Py waveguide ($M_{\mathrm{s}}$ = 64~kA/m, $A_{\mathrm{ex}} = 1.3 \times 10^{-11}$ J/m), with its static magnetization direction controlling the operational mode. A propagating SW can resonantly excite the nanomagnet, which in turn either absorbs energy or shifts the wave’s phase, depending on the magnetization orientation. This enables switchable, non-volatile control over SW transmission, and offers compact, low power and compatible solution for future magnonic circuits. Finally, Louis et~al., \cite{louis2016bias} proposed a bias-free magnonic phase shifter, in which SW propagation is controlled by local modifications of the ground state in a nanodot array with parameters as follows: $M_{\mathrm{s}}$ = 800~kA/m, pillar radius 30~nm, pillar height 60~nm, distance between centers of the nearest neighbors 90~nm. In such a design, domain wall acts as a 'waveguide', while dynamically reconfigurable point defects can locally change its SW dispersion. The remagnetization of a single defect introduces a phase shift up to $\pi$ across a broad frequency range with low insertion loss (up to 1~dB). The concept was numerically validated in a magnonic XOR gate configuration, where the magnetization states of two defects served as logic inputs and the amplitude of the resulting SW was an output, supporting its applied potential. \setlength{\parskip}{2pt}

The field of spin-wave phase-shifters remains less explored than filters, yet with on-going magnonic progress, it can be a competitive R$\&$D direction.

\vspace{4pt}
\subsection{Delay lines} 
\vspace{-2pt}
RF delay lines introduce a controlled time delay (ns- to \(\upmu\)s) into electrical signals. Depending on the choice of propagation medium, they can be realized using coaxial lines, SAW, BAW or integrated circuit components. Such a functionality is crucial for applications requiring precise timing, such as phased-array antennas, Ising machines, radars and multi-band, multi-antenna 5G communication systems. As already mentioned, SAW technology is widely used for frequencies up to $\approx$3 GHz, providing delay lines between 10 and 100 ns. For higher frequencies and delay times (from few hundreds ns to several thousand \(\upmu\)s) BAW-based delay lines are commonly integrated \cite{defranould1984bulk}, yet they suffer from complicated fabrication and increasing damping with increasing frequency. In this context, SW technology offers a promising alternative. Spin systems enable effective operation at any frequency defined by the external magnetic field, including high-frequency range, with delay times in the ns range. The time delay in SW devices can be tuned by selecting magnetic material, adjusting the transducer spacing, and varying the external magnetic field. Furthermore, they are also scalable and compatible with modern fabrication techniques. \setlength{\parskip}{2pt}

In a delay line prototype (Fig. \ref{f:17}), Ishak et al.  \cite{Ishak1988} used micrometer-thick YIG films with parameters $M_\mathrm{s} = 140$~kA/m and $\Delta H =$ -0.3.. -0.6~Oe, resulting in SW delay lines propagation losses of $\approx$ 23 dB/\(\upmu\)s at 9 GHz and 46 dB/\(\upmu\)s at 20~GHz. Thus, for a 200~ns delay, the expected insertion losses would be around 4.6~dB at 9~GHz and 9.2~dB at 20~GHz, assuming propagation loss dominates. These values compare favorably to other delay line types, such as coaxial cables, where a 200~ns delay at 9~GHz would require a 150~ft cable and yield at least 30 dB loss \cite{Ishak1988}.

\renewcommand {\thefigure}{17}
\begin{figure}[hbt] 
\centering
    \includegraphics[width=0.78\columnwidth]{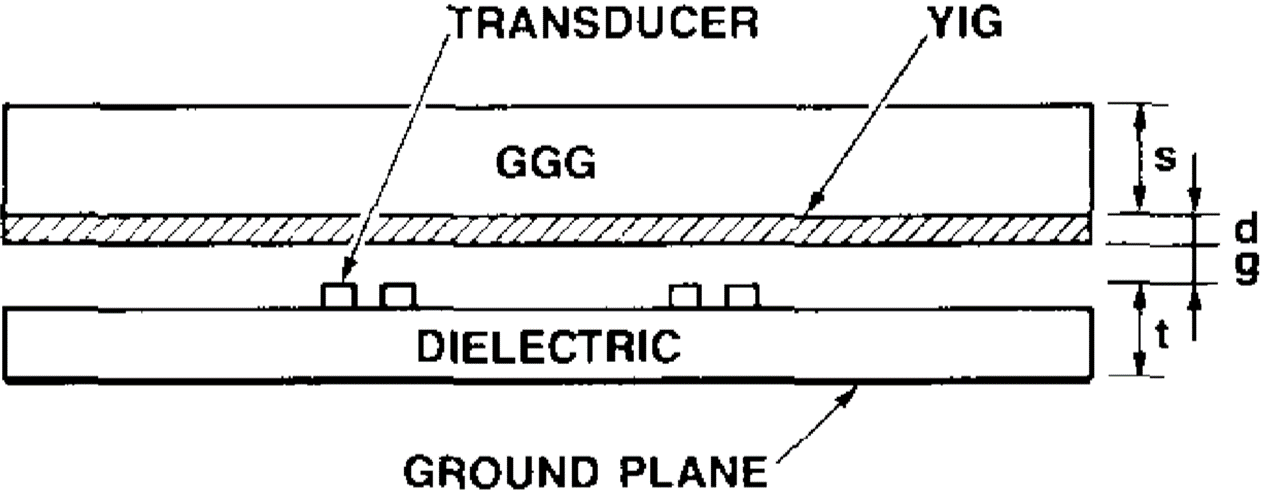}
    \caption{Schematic of a spin-wave delay line. Adapted after \cite{Ishak1988}.}
    \label{f:17}		
\end{figure}

SW delay lines are well-suited for phased-array antennas due to their wide instantaneous bandwidth and tunable dispersion. Significant attention has been given to designs of MSW delay line prototypes with an "up-chirp" MSSW/ FVMSW line and a "down-chirp" BVMSW line in cascade. Both exhibit nearly linear delay-versus-frequency characteristics, but with opposite slopes. When placed in series, the total group delay can be fine-tuned by adjusting the bias magnetic field applied to one or both lines. Following this approach, a prototype MSW device was demonstrated with a tunable delay of 47~ns over a 230-370~MHz bandwidth at 3~GHz, controlled by 55-67~mT bias field applied to the BVMSW line \cite{adkins1984electronically}. The delay profile maintained good linearity, with acceptable rms phase errors in the range of 8.7–12.2°. However, total losses of around 35~dB were quite high, mainly due to SW propagation loss and suboptimal coupling efficiency between transducers and the YIG film. Despite this, the architecture remains attractive for wideband delay-line components.

Variable MSW delay lines were reviewed by Bajpai et al.  \cite{bajpai1985variable}. The experiments were carried on 5~mm-wide and 20~\(\upmu\)m-thick YIG film, separated from a ground plane by a 635~\(\upmu\)m-thick alumina wafer. Golden transducers, each 50~\(\upmu\)m-wide and spaced 1 cm apart, were deposited on the alumina. The applied bias field was rotated in-between two planes: one transitioning from FVMSW to MSSW, and the other from FVMSW to BVMSW. Owing to such a unique angular control, a single-film MSW delay line with a continuously tunable group delay was realized. The time delay of this device was adjustable over $\pm$~20$\%$ range, while maintaining a relatively constant bandwidth of 150~MHz and 20~dB insertion loss within the 8-12~GHz band. 

Among the pioneering demonstrations of nanoscale SW-based delay lines is a work by Li et al. \cite{li2023unidirectional}. The reported devices are realized from a 100-200~nm-thick YIG films with GSG CPWs on top, exciting SWs with a wavelength of 800 or 2000~nm. A robust 30 dB isolation within a field-reconfigurable bandpass range (8-14~GHz) is achieved by utilizing chiral MSSWs. Time-domain analysis yields delays of 23, 47 and 96~ns for 5, 10 and 20~\(\upmu\)m antenna spacing respectively (corresponding to $v_\mathrm{g} \approx 217,~212~\text{and} ~208~\mathrm{m/s}$ at 10 GHz). However, these devices is sensitive to geometry, as increasing ratio between the antenna width and YIG thickness reduces the nonreciprocity, and introduces additional magnon transmission bands. Moreover, demonstrated insertion losses are already more than 50 dB while still below the 5G high-band.

Another key applied research (Fig. \ref{f:KDL}) was recently published by Davídková et al. \cite{davidkova2025spin}. The reported nanoscale delay lines consist of 250~nm-wide CPW transducers with footprints of 2.25~×~100~\(\upmu\)m$^2$, separated by center-to-center distances of 3, 7.5, and 52.5~\(\upmu\)m, and fabricated on a 97~nm-thick YIG film. The delay lines are tested at 4, 9, and 25~GHz for both SSW and BVSW modes -- Table \ref{tab:KD_table}. Time-gating analysis is used to extract delay times from 6.5~ns to 95~ns at 25~GHz, depending on transducer spacing and SW configuration. The measured delays are compared with analytical estimates, and the authors discuss how device parameters, such as film thickness, propagation distance, and SW mode can be used to tune the delay.

\renewcommand {\thefigure}{18}
\begin{figure}[t!]
\centering
\includegraphics[width=1\columnwidth]{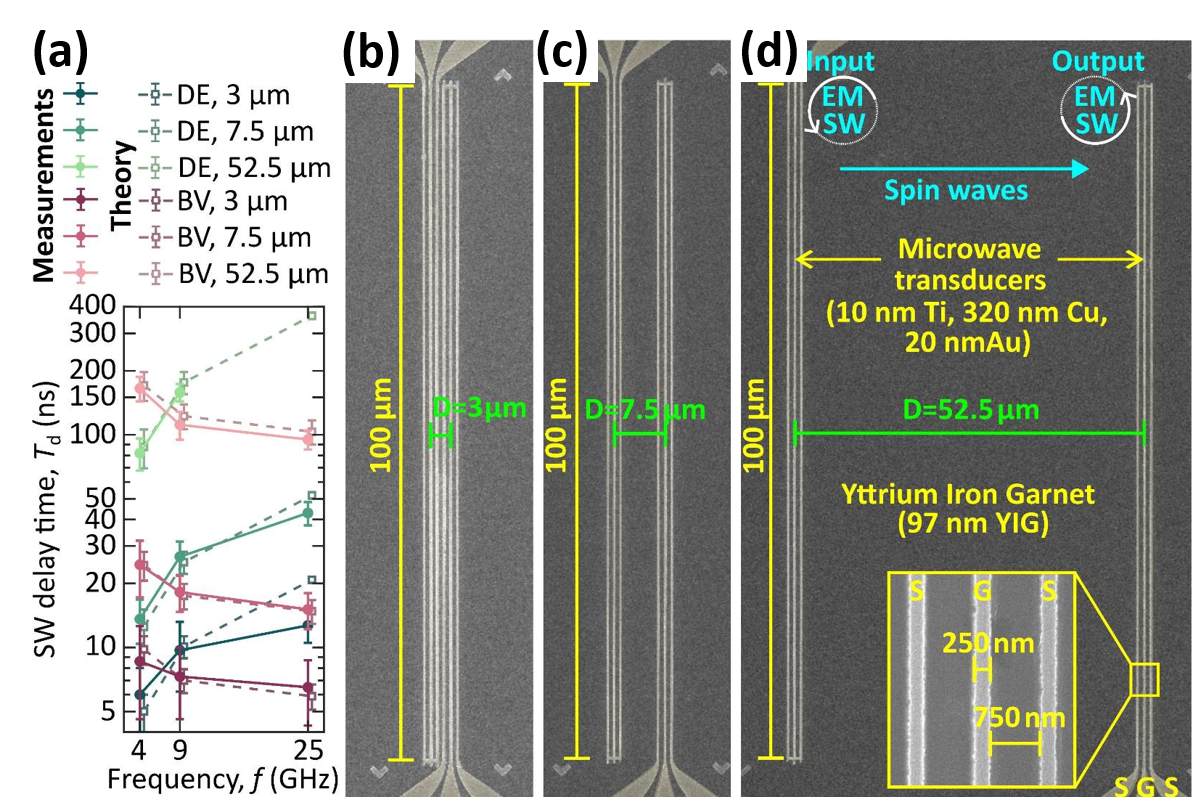}
\caption{(a) The experimental (solid line) and the theoretical delay time (dashed line) calculated for SSW (DE, in green)
and BVSW (BV, in red) and for different distances
between the transducers of 3, 7.5, and 52.5~\(\upmu\)m. (b)-(d) SEM images of SW-based DLs consisting of microwave transducers on the YIG film. The lengths of the transducers are 100~\(\upmu\)m, and both signal(S) and ground(G) conductors are 250~nm wide and 750~nm apart from each other. The center-to-center distance \textit{D} between the transducers is (b) \textit{D} = 3~\(\upmu\)m, (c) \textit{D} = 7.5~\(\upmu\)m, and (d) \textit{D} = 52.5~\(\upmu\)m. Spin-wave propagation between the input and output transducers and the signal conversion between EMW and SW signals are indicated in (d).
Adapted from \cite{davidkova2025spin}.}
\label{f:KDL}		
\end{figure}

\begin{table*}[ht]
\centering
\caption{The applied magnetic field \textit{B}, calculated group velocities $v_{\mathrm{g}}$ at \textit{k} = 3.1 rad/$\upmu$m, theoretical $T^*_\mathrm{d}$ and experimental $T_\mathrm{d}$ delay times for transducers with center-to-center distances of 3 $\upmu$m, 7.5 $\upmu$m and 52.5 $\upmu$m, for Damon-Eshbach (DE) and Backward Volume (BV) modes at 4 GHz, 9 GHz and 25 GHz. Adapted after \cite{davidkova2025spin}.}
\vspace{0.3em}
\renewcommand{\arraystretch}{1.0}
\begin{tabular}{|c|c||cc|cc|cc|}
\hline
\cellcolor{NavyBlue}&\cellcolor{NavyBlue}&
\multicolumn{2}{c|}{\cellcolor{white}\textcolor{black}{\textbf{4 GHz}}} &
\multicolumn{2}{c|}{\cellcolor{white}\textcolor{black}{\textbf{9 GHz}}} &
\multicolumn{2}{c|}{\cellcolor{white}\textcolor{black}{\textbf{25 GHz}}} \\

\cline{3-8} 
\multirow[c]{-2}{*}{\cellcolor{NavyBlue}\textcolor{white}{\textbf{$D$ ($\upmu$m)}}} &
\multirow[c]{-2}{*}{\cellcolor{NavyBlue}\textcolor{white}{\textbf{Parameter (unit)}}}  &
\cellcolor{NavyBlue}\textcolor{white}{\textit{DE}} 
  & \cellcolor{NavyBlue}\textcolor{white}{\textit{BV}} 
  & \cellcolor{NavyBlue}\textcolor{white}{\textit{DE}} 
  & \cellcolor{NavyBlue}\textcolor{white}{\textit{BV}} 
  &\cellcolor{NavyBlue}\textcolor{white}{\textit{DE}} 
  &\cellcolor{NavyBlue}\textcolor{white}{\textit{BV}} \\
\hhline{|==||==|==|==|}

\multirow{2}{*}{} 
& $B$ (mT)      & 71    & 87    & 244   & 255   & 807   & 823 \\
& $v_\mathrm{g}$ (m/s)   & 597.8 & 307.6 & 297.9 & 428.5 & 144.6 & 507.8 \\
\hline

\multirow{2}{*}{3}
 & $T_\mathrm{d}^*$ (ns)   & 5.0   & 9.8   & 10.1  & 7.0   & 20.8  & 5.9 \\
 & $T_\mathrm{d}$ (ns)     & 6.0$\pm$2.0 & 8.6$\pm$4.0 & 9.7$\pm$3.5 & 7.3$\pm$2.7 & 12.7$\pm$8.5 & 6.5$\pm$2.2 \\
\hline

\multirow{2}{*}{7.5}
 & $T_\mathrm{d}^*$ (ns)   & 12.5  & 24.4  & 25.2  & 17.5  & 51.9  & 14.8 \\
 & $T_\mathrm{d}$ (ns)     & 13.6$\pm$3.2 & 24.5$\pm$7.3 & 26.7$\pm$4.8 & 18.2$\pm$3.5 & 42.9$\pm$5.4 & 15.1$\pm$2.9 \\
\hline

\multirow{2}{*}{52.5}
 & $T_\mathrm{d}^*$ (ns)   & 87.8  & 170.7 & 176.3 & 122.5 & 363.2 & 103.4 \\
 & $T_\mathrm{d}$ (ns)     & 82.0$\pm$14.1 & 165.3$\pm$21.8 & 158.4$\pm$14.9 & 111.5$\pm$16.7 & -- & 95.0$\pm$9.8 \\
\hline
\end{tabular}
\label{tab:KD_table}
\end{table*}

Recent improvements in films quality, transducer design, and potential integration with amplifiers could further reduce losses by more than 50\%, making MSW delay lines viable for modern phased-array front-end systems.

\setlength{\parskip}{-2pt}
\subsection{Frequency-selective limiters and signal-to-noise enhancers}
\setlength{\parskip}{-1pt}
Power limiters are crucial devices for protecting RF electronics from large input signals. Their purpose is maintain a constant output level once the input exceeds a defined power threshold. Conventional limiters are usually semiconductors-based (e.g., diodes) due to their planar geometry and CMOS compatibility. However, at high GHz frequencies, these devices suffer from high electrical noise and switching delays. Moreover, when two signals at different frequencies and of different magnitudes are received, the semiconductor-based limiter attenuates these signals equally once a certain power threshold is reached. In contrast, ferrite-based Frequency Selective Limiters (FSLs), which operate through SW absorption or transmission, offer frequency-selective attenuation. High-power signals above threshold are suppressed at specific frequencies, while low-power signals at other frequencies pass unaffected, enhancing signal-to-noise ratio. To date, research has focused mainly on millimeter-scale implementations below 15~GHz according to technological demands. Nonetheless, FSLs started to gain attention from industry for applications in Global Positioning Systems (GPS) \cite{adam2014mitigate}, Mobile User Objective Systems (MUOS) \cite{adam2014mitigate}, and RF front ends for autonomous vehicles \cite{shukla2020adaptive}.

\renewcommand {\thefigure}{19}
\begin{figure*}[h!]
\centering
\includegraphics[width=0.66\paperwidth]{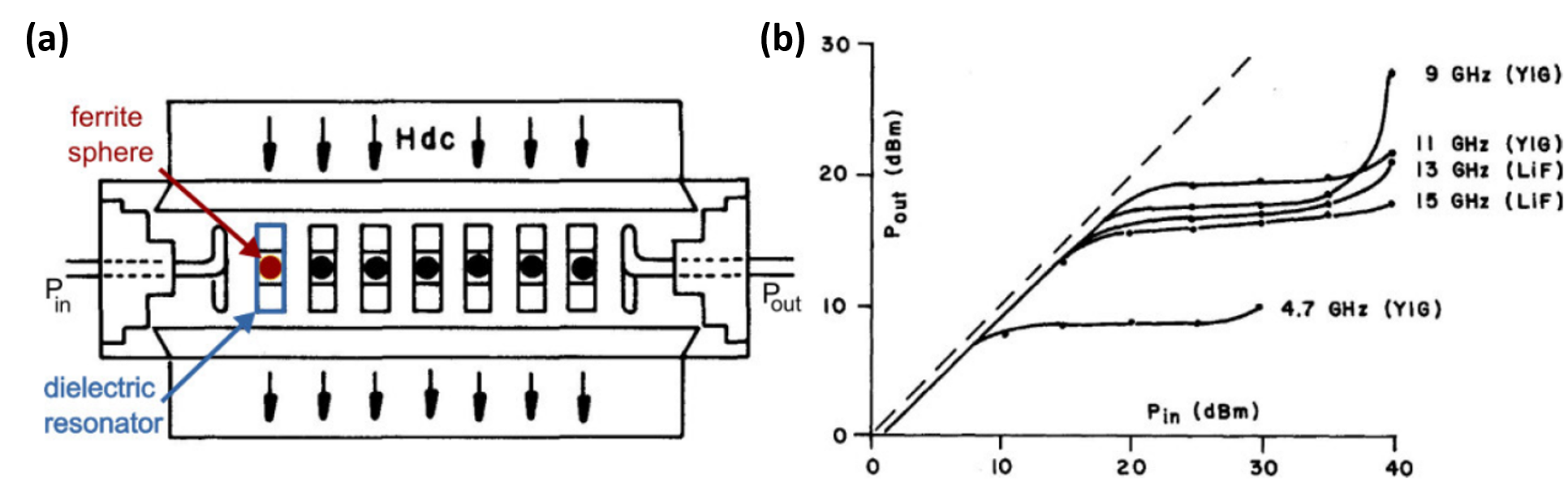}
\caption{(a) Experimental configuration of a power limiter consisting of a ferrite spheres and dielectric resonators. (b) Limiting effect measured at frequencies 13 GHz, 15 GHz for lithium ferrite spheres, and 4.7 GHz, 9 GHz, 11 GHz - for YIG spheres. Adapted from \cite{elliott1974broadband}.}
    \label{f:19}		
\end{figure*}

In 1974, Elliott, Nieh and Craig \cite{elliott1974broadband, ES1974broadband} developed a frequency-selective power limiter (Fig.~\ref{f:19}(a)) based on ferrite spheres placed in a chain of high-Q dielectric resonators. The dielectric resonators, made of strontium titanate, concentrated the dynamic magnetic field at their center, where single-crystal lithium ferrite ($M_\mathrm{s} = 310.35$~kA/m and $\Delta H=$ 0.5~Oe) or YIG spheres ($M_\mathrm{s} = 141.65$~kA/m and $\Delta H=$ 0.3~Oe) were positioned. This configuration lowered the power threshold for nonlinear resonance absorption. The spheres were biased with a permanent magnet, enabling operation in the nonlinear region without external electronics. The limiter achieved less than 1~dB insertion loss, limiting thresholds below +20~dBm, and dynamic ranges above 20~dB across several bands. Power-liming effects are shown in Fig.~\ref{f:19}(b) for lithium ferrite (at 13~GHz and 15~GHz) and for YIG spheres (at 4.7~GHz, 9~GHz and 11~GHz). The device operation is based on the nonlinear response of the ferrites, offering passive, broadband and frequency selective FSL, ideal for front-end receiver protection. 

\renewcommand {\thefigure}{20}
\begin{figure}[!h]
\centering
\includegraphics[width=1\columnwidth]{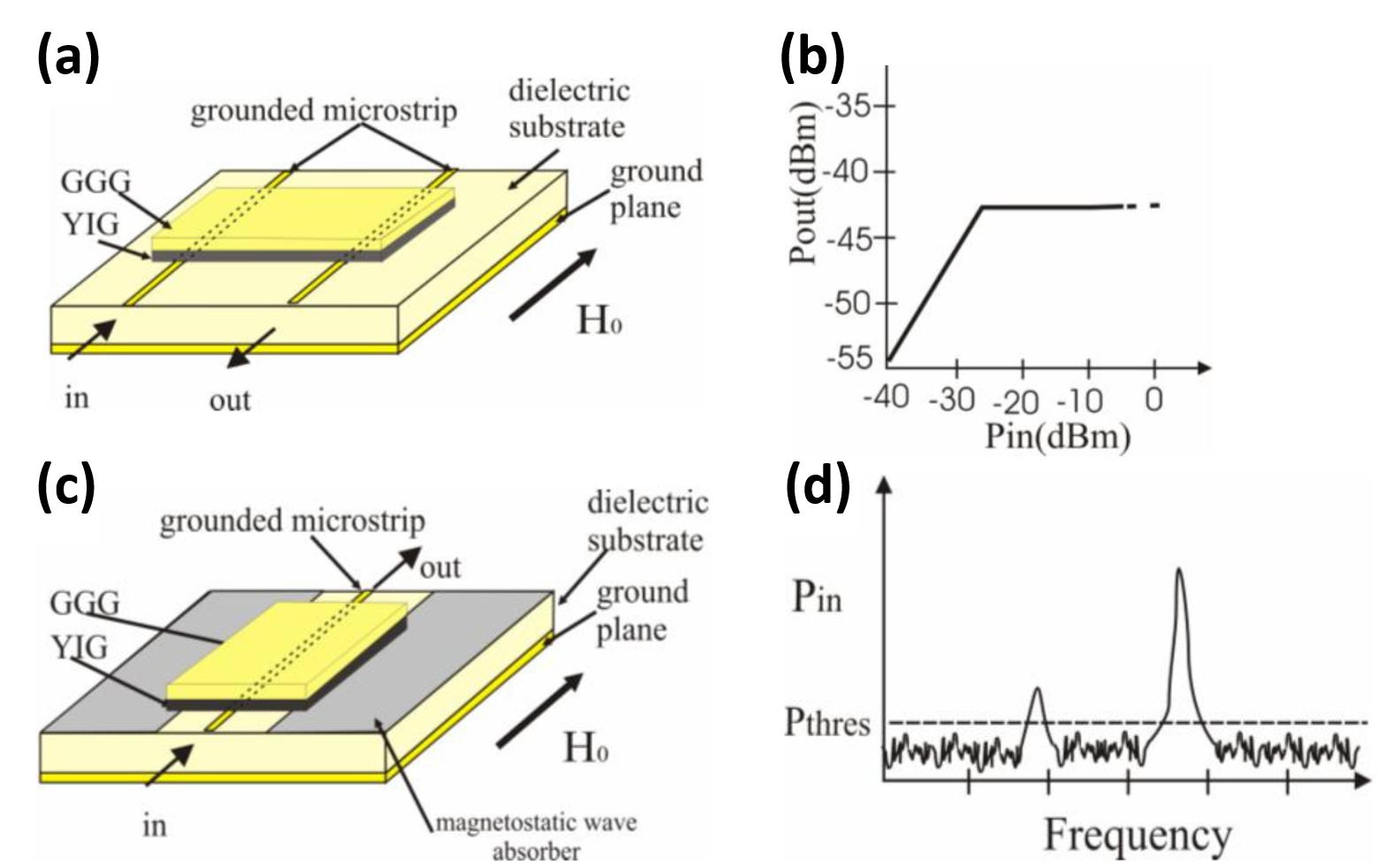}
\caption{(a) Schematics of FSL and (b) measured output power as a function of input power; (c) SNE and (d) input power spectra with the threshold level indicated. Adapted from~\cite{Yttrium}}
\label{f:20}	
\end{figure}

\renewcommand {\thefigure}{21}
\begin{figure*}[!hb]
\centering
\includegraphics[width=0.55\paperwidth]{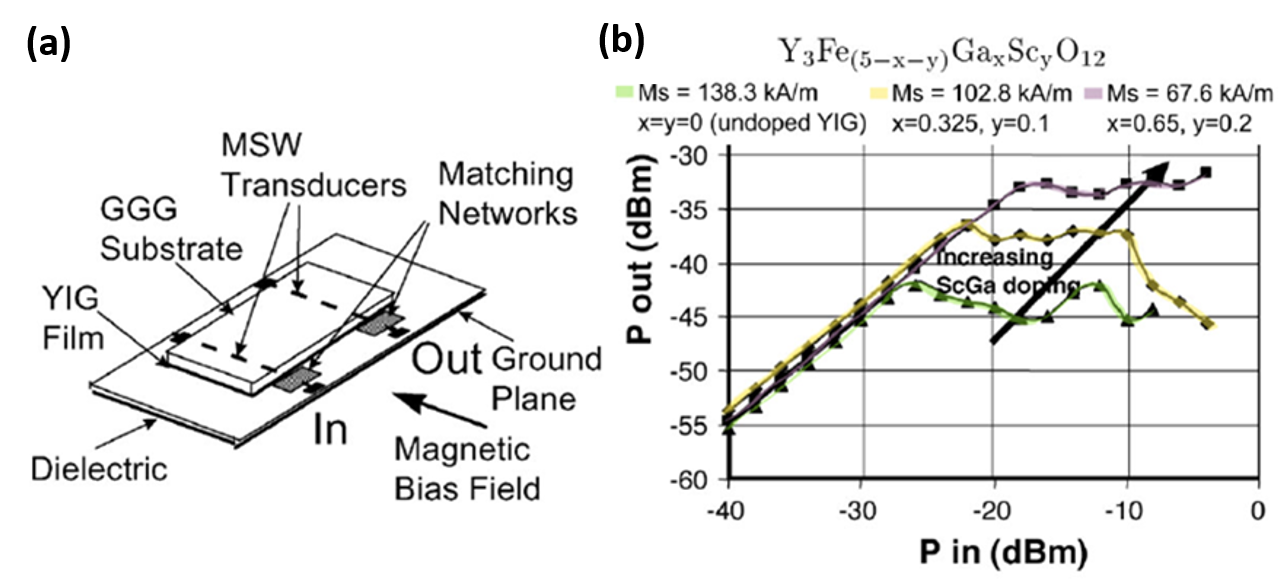}
\caption{(a) Experimental configuration of FSL device (b) Measured limiting effect at unknown frequencies for undoped and doped YIG samples of different saturation magnetization: 138.3 kA/m ({\color{ForestGreen}{green curve}}), 102.8 kA/m ({\color{yellow}{yellow}}) and 67.6 kA/m ({\color{violet}{violet}}). Adapted after \cite{adam2004msw}.}
    \label{f:21}		
\end{figure*}

In 1980s, a YIG-based selective limiter and signal-to-noise enhancer (SNE) prototypes were introduced \cite{Adam1988, adam1980magnetostatic}, demonstrating broadband operation and low-threshold power levels. The FSL suppressed high-power signals while transmitting low-power inputs with minimal loss, while the SNE attenuated weak signals and transmitted stronger ones more efficiently, enhancing signal clarity. These concepts remain relevant for broadband RF receivers and oscillator-based systems. Later, the operating principle of the MSSW-based FSL (Fig. \ref{f:20} (a, b)) and SNE (Fig. \ref{f:20} (c, d)) was detailed by Zavislyak et al. \cite{Yttrium}. In the SNE mode, low-power microwave signals excite MSSWs in the YIG film via a single microstrip, efficiently absorbing part of the input signal energy and dissipating it through magnetic damping, resulting in strong attenuation. As the input power rises above a threshold, nonlinear effects suppress MSSW excitation, reducing energy loss and allowing more signal to pass, hence improving the signal-to-noise ratio. In the FSL mode, MSSWs propagate between two antennas with minimum insertion loss at low-power signals. Exceeding the input threshold power triggers SWs excitation at half the input frequency. These SWs couple energy to the crystal lattice, leading to the dissipation of the excess microwave power as heat and flattening of the output. In both cases, nonlinear effects (\textit{Section II B}) enable power-dependent transmission control.

\renewcommand {\thefigure}{22}
\begin{figure*}[!ht]
\centering
\includegraphics[width=0.57\paperwidth]{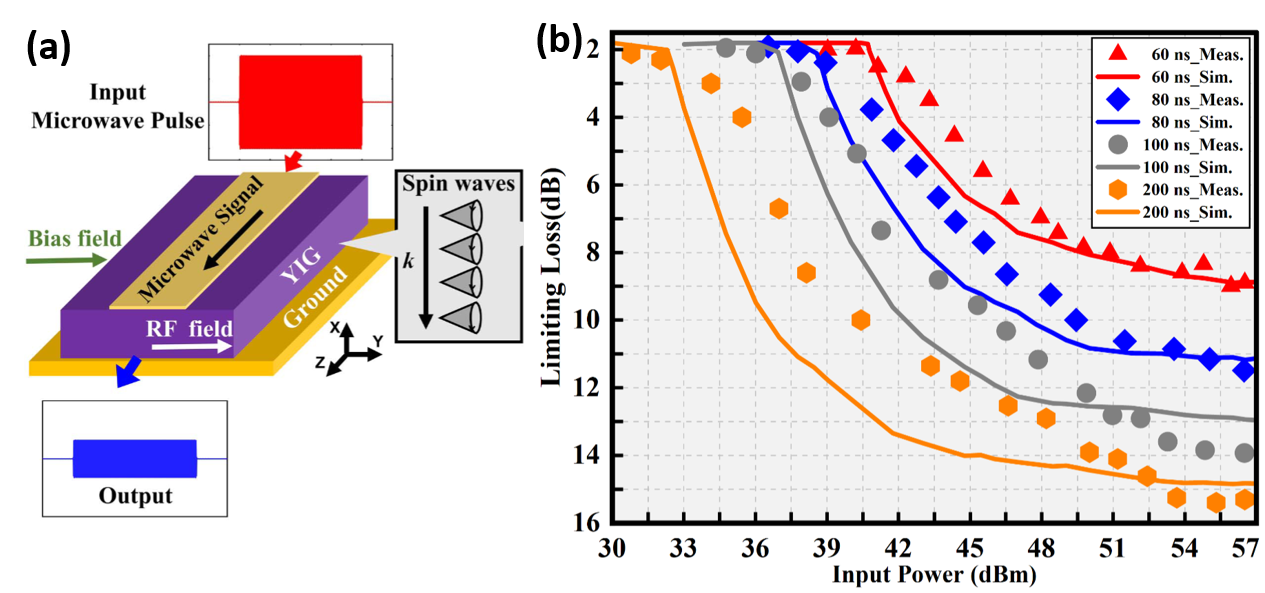}
\caption{(a) Experimental configuration of a YIG-based FSL device. (b) Measured (points) and simulated (curves) data for high-power MW pulses of different durations: 60 ns ({\color{red}{red}}), 80 ns ({\color{blue}{blue}}), 100 ns ({\color{gray}{gray}}) and 200 ns ({\color{orange}{orange}}). Adapted after \cite{yang2022x}.}
    \label{f:22}		
\end{figure*} 

In a 2004 study, Adam and Stitzer \cite{adam2004msw} demonstrated an advanced FSL concept (Fig. \ref{f:21}) with over 100× lower threshold power and improved frequency selectivity compared to stripline-based devices. The MSSW power limiter operated in the 400-800 MHz frequency range under an in-plane bias field of $\approx$1~mT. Three film types were investigated: undoped YIG, and YIG doped with different concentrations of scandium and gallium (ScGaYIG) to tailor saturation magnetization and threshold power as shown in Fig. \ref{f:21}(b).  The FMR linewidth increases with increasing substitution of Sc and Ga from 0.069 to 0.166 mT. A threshold power of +20 dBm, narrow 2 MHz bandwidth, and low intermodulation levels were successfully achieved.\setlength{\parskip}{2pt}

As shown, most FSLs typically use MSSWs, although hybrid MSSW/BVMSW configurations were also developed for lower-frequency operations \cite{Yttrium, kuki1998mssw}. In 2022, Yang et al. \cite{yang2022x} reported on high-power FSL operating at 9.4~GHz via nanosecond microwave pulses (Fig. \ref{f:22}). The polycrystalline YIG ($M_\mathrm{s} = 143.24$~kA/m and $\Delta H =$ 2~Oe) was used to reach a higher power handling and lower cost compared to single-crystal YIG. The device consists from a 300~\(\upmu\)m-thick, 12~mm-wide and 50~mm-long polycrystalline YIG placed on a 17~\(\upmu\)m-thick, 190~\(\upmu\)m-wide gold-plated conductor. An in-plane bias field of $\approx$~82~mT was applied perpendicular to the YIG stripe’s long axis. When high-power pulses were sent, nonlinear effects led to the excitation of SWs at half the input frequency, causing strong power-dependent absorption and power-limiting behavior.  

Just recently, in 2024, the nanoscale SW power limiter was successfully realized \cite{davidkova2025nanoscale}. The device was based on 97~nm-thick YIG film and used CPW antennas, 250~nm-wide and either 10 $\upmu$m or $100~\upmu \mathrm{m}$-long, for SW excitation and detection (Fig.~\ref{f:KD}(f)). Material parameters of the epitaxial ferrite film are as follows: $M_\mathrm{s} = 140$~kA/m, $A_{\mathrm{ex}}=3.6$~pJ/m, $\alpha \approx 2\times 10^{-4}$. Resulting power limiters were tested in a broad frequency range up to 25~GHz for both fundamental in-plane configurations - SSW (here Damon-Eshbach, DE) and BVSW - see Fig.~\ref{f:KD}(a-c). Key parameters, such as insertion loss, power threshold, limiting level and bandwidth, were extracted. The lowest power thresholds, measured via 10 $\upmu$m long transducers, were $-25.5\pm3$ dBm at 4~GHz and 9~GHz for DE, and $0 \pm3.5\,$dBm at 25~GHz for the BV geometry Fig.~\ref{f:KD}(d,e). The lowest loss achieved in this proof-of-concept device was 23 dB. A numerical model was developed to describe the measured power characteristics, providing a pathway to reduce insertion losses. The paper also expertly summarizes advantages and disadvantages of the SW power limiters, theorizing a three-in-one device that functions as FSL, delay line and filter simultaneously.

\renewcommand {\thefigure}{23}
\begin{figure}[hbt]
\centering
\includegraphics[width=1\columnwidth]{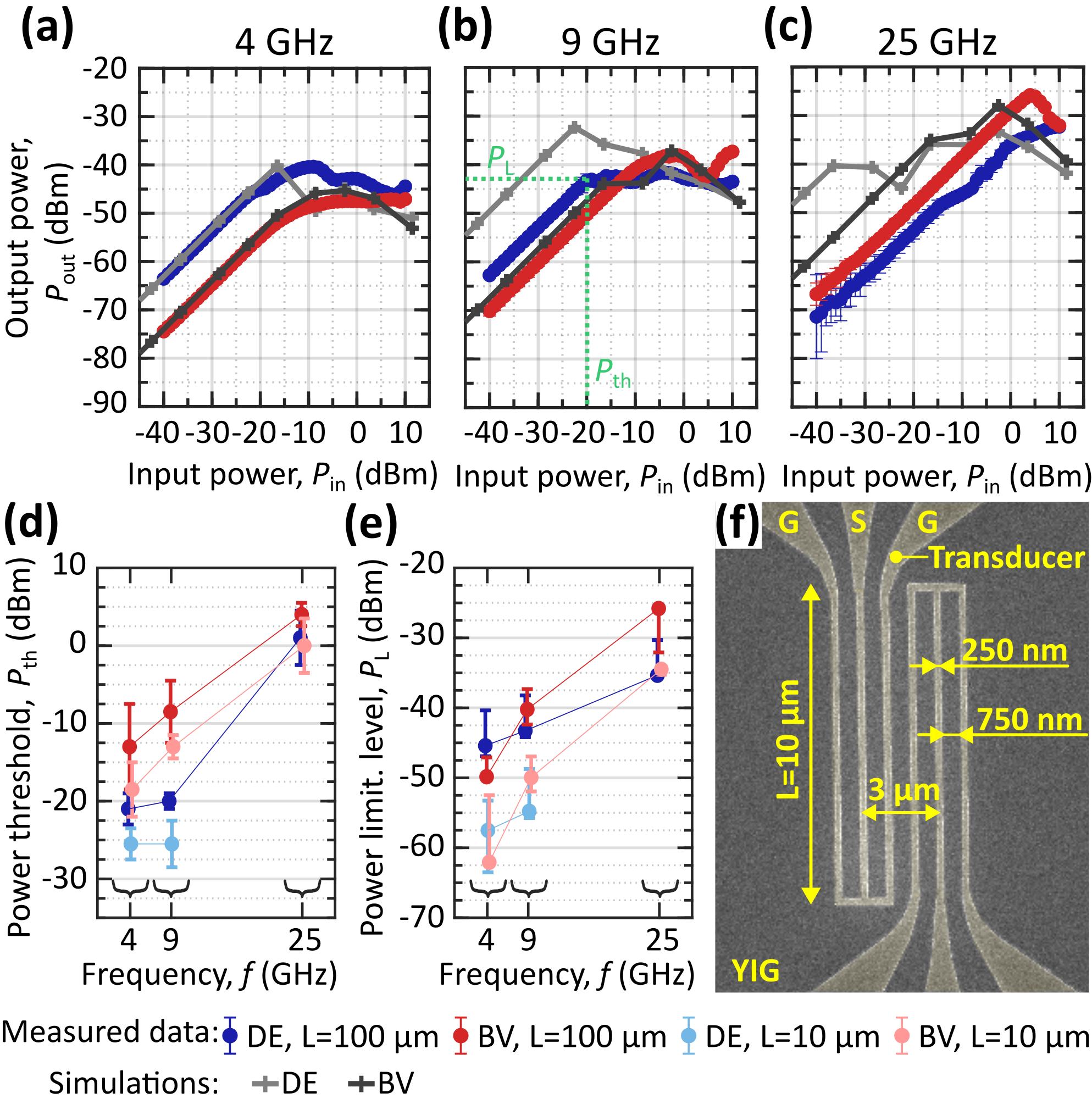}
\caption{(a)-(c) Extracted power characteristic from SW power transmission in 97~nm-thick YIG film measured using CPW antennas of the length L=100~$\upmu$m. The power characteristics were measured and simulated for Damon-Eshbach (DE) configuration in blue and light gray and for Backward volume (BV) in red and dark gray at a)~4~GHz, DE: 71~mT, BV: 87~mT; b) 9~GHz, DE: 244~mT, BV: 255~mT; c) 25~GHz, DE: 807~mT, BV: 823~mT. b) The power threshold and power limiting level for the DE power curve are indicated in green. Extracted d)~power threshold and e)~power limiting level from power characteristics measured using CPW antennas of the length L=100~$\upmu$m (dark colors) and L=10~$\upmu$m (light colors) for both magnetization geometries. f) SEM image of fabricated CPW antennas of the length of L=10~$\upmu$m. Adapted from \cite{davidkova2025nanoscale}.}
  \label{f:KD}		
\end{figure}

Power-limiting effects were also investigated by Wang et al.  \cite{wang2023deeply} in out-of-plane magnetized YIG conduit ($M_\mathrm{s} = 140.7$~kA/m, $A_{\mathrm{ex}}=4.22$ pJ/m, $\alpha \approx 1.75\times 10^{-4}$). The waveguide was 44 nm-thick and 200 nm-wide, with 2~\(\mu\)m-wide strip antenna (10~nm~Ti~/~150~nm~Au) patterned on top of it. At sufficiently high input powers, a nonlinear FMR mode was excited directly beneath the antenna, resulting in the local conversion of energy into short-wavelength, self-normalized SWs (Fig.~\ref{f:24}). In this case, the power-limiting is based on a self-locking nonlinear frequency shift. A record-high nonlinear frequency shift exceeding 2~GHz is achieved, which corresponds to a magnetization precession angle of 55° and leads to excitation of short-wavelength SWs of 200~nm. The results obtained in this paper are of great interest, as they demonstrate the potential for scaling SW-based prototypes down to nanometer size. This opens a pathway toward the integration of ferrite-based RF components into nanoscale circuits, where the optical BLS probe would eventually be replaced by additional RF antennas. 

\renewcommand {\thefigure}{24}
\begin{figure}[hbt] 
\centering
\includegraphics[width=0.99\columnwidth]{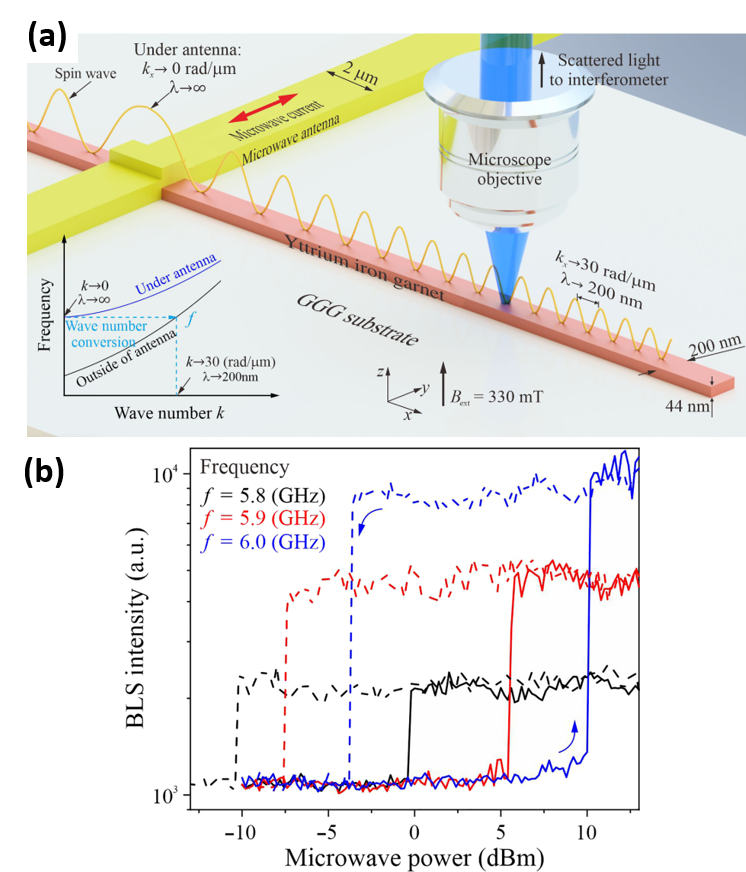}
\caption{(a) Experimental configuration of a device consisting of a 2 \(\upmu\)m-wide antenna placed over 200 nm-wide YIG waveguide under 330 mT out-of-plane external field. The nonlinear shift in the SW dispersion relation is shown at the bottom left. (b) Optically measured SW intensity at different frequencies of 5.8 GHz, 5.9 GHz and 6.0~GHz as a function of the input power upsweep (solid curve) and downsweep (dashed curve). Adapted from \cite{wang2023deeply}.}
\label{f:24}		
\end{figure} 

A subsequent study of a nonlinear shift of FVSWs in nanoscale waveguides was made a year after \cite{wang2024all}, where an all-magnonic repeater was developed, and a large bistable window of 1.1~GHz was observed. The device consists of a 1 \(\upmu\)m-wide YIG waveguide ($M_\mathrm{s} = 140.7$~kA/m, $A_{\mathrm{ex}}=4.22$~pJ/m, $\alpha \approx 1.75\times 10^{-4}$), supplied with a CPW source antenna and 2 \(\upmu\)m-wide pump strip antenna (repeater). The pump antenna enables two stable magnon states with high and low SW amplitudes, and allows one to switch them at a threshold input power of $\approx$~-6 dBm. In a cascade logic system, for example, such a repeater does not only improve the damped input signal, but also regenerates new SWs with amplified amplitude (up to 6x gain) and a normalized phase to connect with the next level logic gate. Proposed design \cite{wang2024all} facilitates simplified and robust magnonic circuitry suitable for a variety of applications. Although the last two articles remain mainly in domain of fundamental academic research, discovered physical phenomena allow the realization of devices and functionalities not previously accessible in SW-based RF applications.

\vspace{1pt}
\subsection{Resonators} 
\vspace{-3pt}

Resonator is a device or structure that naturally oscillates at specific (resonant) frequencies. When excited by an external source at or near its resonant frequency, it can sustain oscillations, often amplifying the signal within a narrow frequency band. Resonators are widely used in RF, optical and acoustic systems, where precise frequency selection, filtering or signal enhancement is needed. MSWs are naturally excellent candidates for the resonators owing to their high group velocity.

\renewcommand {\thefigure}{25}
\begin{figure}[ht]
\centering
\includegraphics[width=0.55\columnwidth]{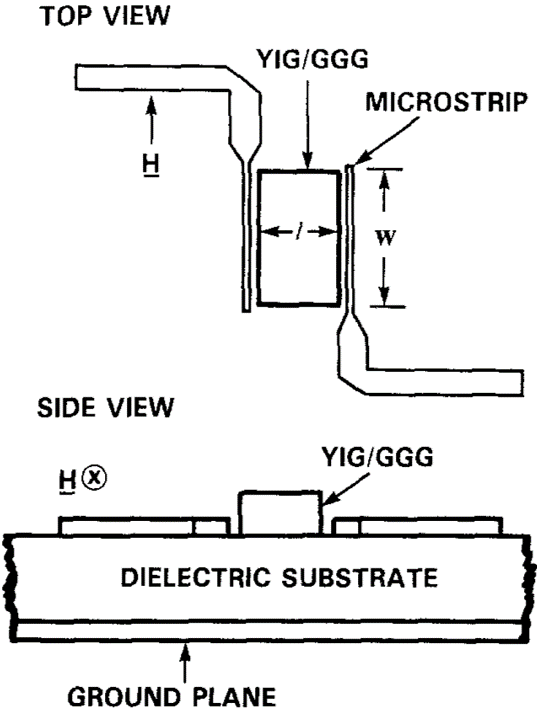}
\caption{Magnetostatic surface wave straight edge resonator. Adapted from \cite{chang1985magnetostatic}.}
\label{f:25}	
\end{figure}

Chang and Ishak \cite{chang1985magnetostatic} presented a straightforward MSSW resonator design using straight-edge reflectors in rectangular YIG film, Fig.~\ref{f:25}. The device consists of a resonant cavity (rectangular piece of YIG/GGG film) placed atop gold microstrip transducers patterned on a dielectric layer, all mounted on an aluminum block for grounding. Positioned between electromagnet poles with the field parallel to the transducers, the device excites MSSWs that reflect onto the other surface at the film’s straight edges. Unlike earlier grating-based designs \cite{castera1983magnetostatic, castera1984state} with complex fabrication, limited tuning < 12~GHz, losses as high as 30~dB, and Q < 1000, this prototype achieved wider 2–16~GHz frequency range, < 10~dB insertion loss, and > 10~dB spurious rejection. Width mode interference, which is a key limitation in MSSW resonator performance, was investigated in a complementary study by the same authors \cite{chang1984effect}. Through modeling and experiment, they showed that spurious resonances arising from higher-order width modes can be mitigated by tailoring the transducer and YIG parameters. Optimized designs achieved insertion losses under 8~dB and spurious rejection > 10~dB in the 3–9~GHz range. The 1 dB compression point was quite low ($\approx$ 20 dBm) for tuning frequency below 4 GHz because of coincident limiting effects in undoped YIG, but it can be enhanced via material engineering (e.g., by using doped YIG).

In 2020, Dai et al. \cite{dai2020octave} have designed, fabricated, and characterized high-performance, on-chip MSW resonators (Fig.~\ref{f:26}) from single-crystal epitaxial YIG film (3 \(\upmu\)m-thick, $M_\mathrm{s} = 139.26$~kA/m) patterned into mesa structures. A thick aluminum coplanar loop-inductor transducers were fabricated around each mesa to individually address and excite RF magnetic fields normal to the film surface, efficiently coupling into BVMSWs. At 4.77 GHz, the 0.68~mm\textsuperscript{2} footprint resonator achieved a quality factor Q > 5000 under an in-plane bias field of 98.7~mT. Frequency tuning from 3.63 to 7.63~GHz was realized by sweeping the in-plane bias field from 70.5 to 186~mT. Importantly, the measured Q remained consistently over 3000 across most of the tuning range, with a peak $f \cdot Q$ product of $2.51 \times 10^{13}$ exceeding values typically reported for acoustic-wave resonators \cite{dai2020octave}.

\renewcommand {\thefigure}{26}
\begin{figure}[ht]
\centering
\includegraphics[width=1\columnwidth]{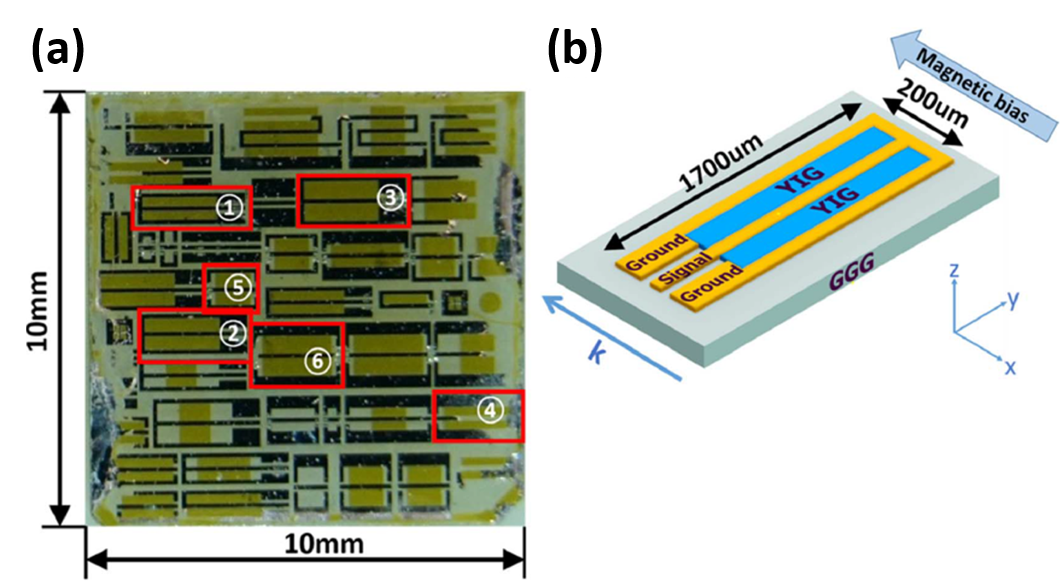}
\caption{(a) Top view of the chip with multiple MSW YIG resonators, that are designed to operate within 4.54 to 4.58 GHz frequency range under constant 90 mT magnetic bias [Devices 1-6]. (b) Scheme of MSW resonator as marked in {\color{red}{red}}. Adapted from \cite{dai2020octave}.}
\label{f:26}	
\end{figure}

Lastly, an innovative design of a passive nanoscale SW ring resonator was proposed by Wang et al. \cite{wang2019nanoscale}, yet its functionality was primary targeting wake-up receivers in power-critical IoT applications. While different in scale and intended function, all works highlight the MSWs potential for compact and tunable resonator architectures with functionality beyond the conventional approaches.

\vspace{1pt}
\subsection{Directional couplers}
\vspace{-3pt}
A directional coupler is a passive RF component designed to split an input signal into two output paths, through line (primary path) carrying the most of the signal power coupled line (secondary path) carrying a smaller, sampled portion.  These devices enable signal monitoring or measurement with minimal disturbance to the main transmission. \setlength{\parskip}{2pt}
 
In SW systems, directional couplers are implemented using multistrip couplers, multiple ferrite-film structures and coupled waveguides. In the former structure, a shorted multi-element grating is combined with microstrip transducers, resembling a SAW multistrip coupler. For instance, a full power transfer between the transducers via MSSWs was achieved at 2.5~GHz in epitaxial YIG film  \cite{castera1980adjustable}. Directional coupling demonstrated frequency-selective behavior due to MSSW dispersion, with coupling strength tunable via the applied magnetic field. Alternatively, in a stacked ferrites structure, as reported in \cite{sasaki1979directional}, 20~\(\upmu\)m-thick YIG films ($M_\mathrm{s} = 139.26$~kA/m) were separated by different dielectric layers of 250-400~\(\upmu\)m thickness. Two modes propagating in the same direction with different propagation constants appear in such a structure, leading to a coupling in an area where two YIG films partially overlap. The proposed directional coupler shows frequency filtering and frequency demultiplexing characteristics. The full power transfer (coupling length) for this prototype also varied depending on the gap between the YIG films (provided by dielectric layer) and the frequency (tuned via applied magnetic field), e.g., for 300~\(\upmu\)m gap at 2.9~GHz the coupling length was around 2.5 mm.

\renewcommand {\thefigure}{27}
\begin{figure}[hbt]
\centering
\includegraphics[width=1\columnwidth]{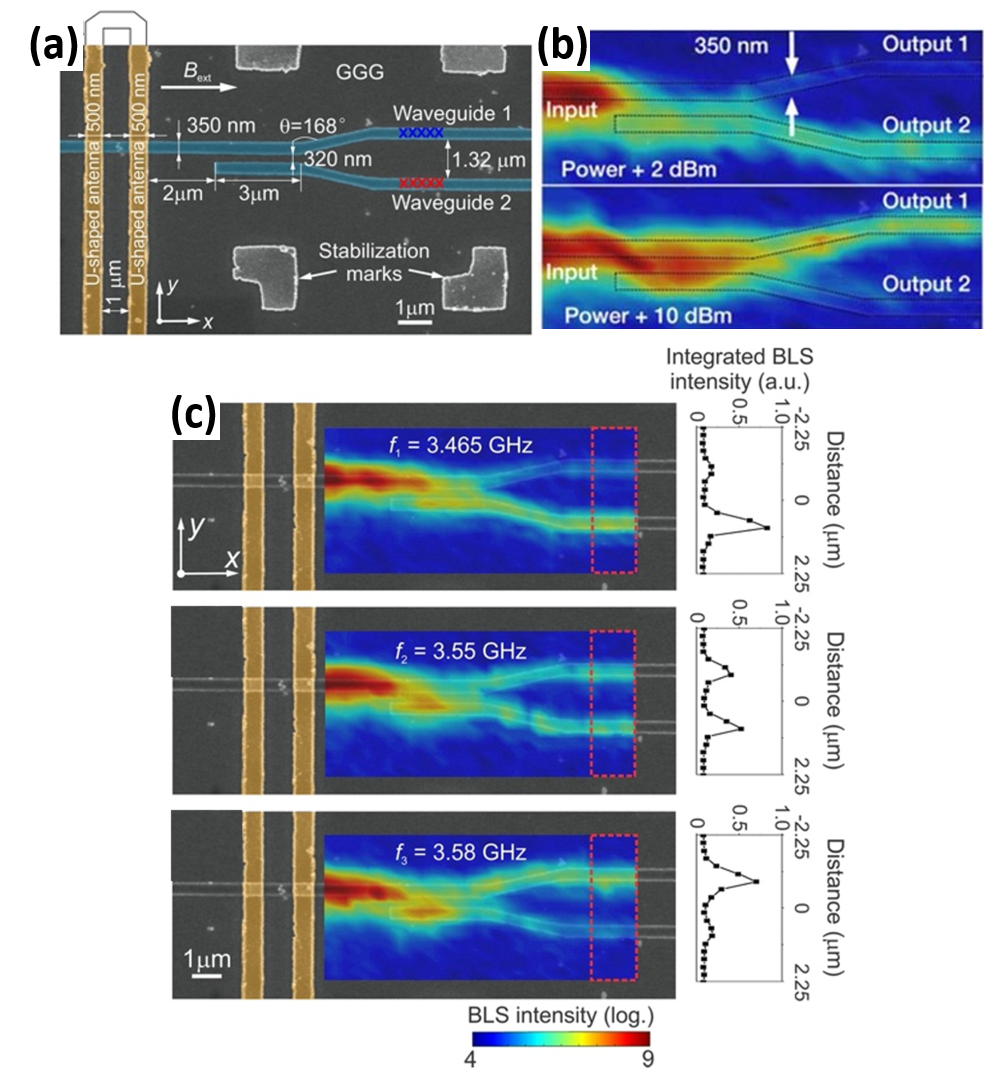}
\caption{Scanning electron microscopy image of the directional coupler (shaded in blue) with the U-shaped antenna. An external magnetic field of 56 mT is applied along the YIG conduits (x-axis) to saturate the directional coupler in BV configuration and RF current with power 0 dBm is applied to the antenna to excite SWs. (b) Nonlinear transfer characteristics of a nanoscale directional coupler visualized via 2D Brillouin Light Scattering (BLS) maps of the SW intensity for a frequency of \textit{f} = 3.52 GHz and different input powers. (c) Two-dimensional BLS maps (the laser spot was scanned over an area of 9.4×4.5 \(\upmu\)m\textsuperscript{2}) of the BLS intensity for different frequencies as indicated on the figure. The right panels show the SW intensity integrated over the red dashed rectangular regions at the end of the directional coupler. Adapted from \cite{Wang2020, Wang2019}}
\label{f:27}	
\end{figure}
 
Due to the advancement in material fabrication and nanoscaling, Wang et al. designed, fabricated and tested a milestone SW directional coupler \cite{Wang2020, Wang2019}  based on single-mode 350~nm-wide, 85~nm-thick YIG waveguides ($M_\mathrm{s} = 142$~kA/m, $\alpha \approx 2.1 \times 10^{-4}$), separated by a narrow gap of 320 nm. The information carried by a SW amplitude, was guided to one of the two coupler's outputs (Fig. \ref{f:27}(a)) according to the signal power (Fig. \ref{f:27}(b)), frequency (Fig. \ref{f:27}(c)) and bias magnetic field. This highlights the coupler’s potential as a universal all-magnonic unit, suitable for functional, low-energy magnonic circuits. Origins of the selective transmission between the two conduits at different powers (Fig. \ref{f:27}(b)) is a nonlinear shift of the dispersion relation. Frequency-selectivity (Fig. \ref{f:27}(c)), on the other hand, depends on a coupling length, which defines the distance where SW energy is completely transferred from one waveguide to another. This proves a potential of such directional coupler for frequency-division demultiplexing, e.g., different frequencies applied to the same input are directed to different outputs. In addition, a magnonic half-adder consisting of two directional couplers was simulated using a micromagnetic solver, and benchmarked against a 7 nm CMOS half-adder \cite{Wang2020}. The results indicate that the proposed concept, miniaturized down to 30~nm in width, 10~nm in thickness, can have a footprint comparable to a CMOS half-adder, while offering an order of magnitude lower energy consumption \cite{Wang2020}. \setlength{\parskip}{2pt}

 A notable prototype of a demultiplexer for magnonic logic networks was experimentally realized by Heussner et al. \cite{Heussner2020} on a 30~nm-thick CoFeB film ($M_\mathrm{s} = 1558$~kA/m, $\alpha \approx 4.3 \times 10^{-3}$. The device relies on the anisotropic dispersion of SWs in a specific magnetic medium, which causes narrow caustic-like beams at different frequencies to propagate at distinct angles. This leads to a frequency-selective spatial separation of the input signals without special patterning, external control or additional power consumption. Two input frequencies (11.2~GHz and 13.8~GHz) were injected through a microstrip antenna and successfully guided to different spatial output ports, as verified via \(\upmu\)-BLS. The total transmission loss was around 18~dB for both output waveguides, dominated by intrinsic SW damping.

Lastly, Ge et al. \cite{ge2025deeply} proposed and validated via micromagnetic simulations a cutting-edge prototype of deeply nonlinear magnonic directional coupler. The device consists of two identical YIG nanowaveguides (100~\(\upmu\)m long, 100~nm wide, 50~nm thick) separated by 40~nm gap and uniformly biased out of plane (300~mT) to support FVMSWs. In the linear regime, dipolar coupling splits the symmetric/antisymmetric modes and enables complete power transfer over the coupling length (set to 8~\(\upmu\)m at 7.05~GHz frequency). When driven to large precession angles $\approx$ 22.5°, the nonlinear frequency shift suppresses the energy transfer between the nanoconduits (decoupling effect), and the SW remains localized in the first waveguide. Such a switchable, power-controlled coupler yields potential for integrated magnonic circuits, such as signal routing, logic operations, and neuromorphic computing.

\vspace{-2pt}
\subsection{RF mixers}
\vspace{-3pt}

Classically, RF mixer is used to convert signals by combining two separated inputs (e.g., a high-power local oscillator and a lower-power RF) to generate sum and difference frequency outputs. This operation enables up-conversion (shifting a signal to a higher frequency) and down-conversion (shifting it to a lower frequency), which are fundamental in modern communication systems for modulation, demodulation, and intermediate-frequency processing. Output power in passive RF mixers is lower than the input due to conversion loss, which is typically above 4.5 dB \cite{RFMixer}. Such devices often rely on nonlinear current-voltage characteristics of semiconductor diodes to produce frequency mixing.  \setlength{\parskip}{2pt}

In magnonics, nonlinear effects that are promising for RF mixer applications mostly originate from the parametric instabilities. In addition to first- and second-order interactions mentioned in \textit{Section II B}, third-order nonlinear effects, such as four-wave mixing, can also facilitate frequency conversion in magnonic systems. For example, in a recent demonstration by Inglis et al.  \cite{inglis2019indirect}, a 7.8~\(\upmu\)m-thick YIG waveguide ($M_\mathrm{s} = 197$~kA/m, $\alpha \approx 5 \times 10^{-5}$) was pumped with two counter-propagating continuous-wave signals at frequency $\textit{f}_1=\textit{f}_2=3.915$ GHz. When a probe frequency $\textit{f}_p = 3.91825$~GHz was introduced via a separate antenna, two signals were seen - one $\textit{f} = 2\textit{f}_p - \textit{f}_1$ driven by a third-order process due to reflections of probe magnons from the waveguide edge, and a new signal at $\textit{f}_c = 2\textit{f}_1 - \textit{f}_p = 3.91175$~GHz, corresponding to a phase-conjugated magnons propagating opposite to the probe. The efficiency of these phase-conjugated signals relies on the applied magnetic field strength and is enhanced when standing SW modes form across the width of the waveguide, increasing the nonlinearity of a region. The experiment was realized in FVMSW geometry to reach an isotropic dispersion.

In general, there are not so many SW analogues of passive RF mixers, as nonlinear magnon interactions are weaker than electronic counterparts and require higher powers or external pumping to manifest. Instead, SW nonlinearity is effectively harnessed in delay lines, FSLs, directional couplers or functionally similar (de)multiplexers and frequency combs. In addition to prototypes described earlier, frequency multiplication was shown by Koerner et al. \cite{koerner2022frequency} in a 20~nm NiFe film, where MHz-range driving field produced nonlinear SW excitations precessing up to 60\textsuperscript{th} harmonic of the pumping frequency and forming a SW frequency comb. This effect arises from a dynamic, periodic and synchronized switching of magnetic texture. Similarly, the work of Xu et al. \cite{xu2023magnonic} reported a magnonic frequency comb from a cascaded nonlinear two-magnon scatterings in a YIG microsphere enabled by a strong microwave pumping and magnetostrictive coupling to mechanical modes. While not mixers in the strict sense, these prototypes highlight the potential of intrinsic SW nonlinearity for realizing all-magnetic frequency conversion in GHz-range.

\vspace{-3pt}
\subsection{Selected spin-wave RF patents} 
\vspace{-3pt}
Innovative spin-wave RF devices are well documented in US and European patent databases, reflecting significance of applied magnonics. According to the Patent Public Search PPUBS Basic, over the last two decades the number of patents that include both "spin-wave" and "RF" has increased approximately ninefold compared to the period from 1960-2000. \setlength{\parskip}{2pt}

Notably, patents related to SW filters often rely on MSWs in magnetic ferrites [e.g., Geho et al. \textit{"Magnetostatic wave device"} 5,985,472, 11/1999 - US5985472A; Geiler \textit{"Frequency Selective Limiter Having an Enhanced Bandwidth"} 11,349,185 B1, 5/2022 - US20220285813A1]. In particular, Murphy et al.  \textit{"Hybrid radio frequency system with distributed anti-jam capabilities for navigation use"} [5,955,987 A, 9/1999 - US5955987A] patented an adaptive RF filter, as an analogue part of a conventional antenna array device based on a classical design of a thin YIG film placed between the stripline transducers. When subjected to an external magnetic field, it operates as a SW delay line with variable impedance. If the input signal power exceeds a tunable threshold, the SWs absorb excess energy, thus limiting high-power RF interference. The power threshold is tunable via ferrite geometry and bias field, enabling robust interference mitigation in critical applications, such as GPS front-ends. Correspondingly, Adam et al. \textit{"Low threshold power frequency selective limiter for GPS"} proposed MSW-based FSL design [6,998,929 B1, 2/2006 - US6998929B1] using a pair of parallel microstrip transducers on a 0.1 to 5.0~\(\upmu\)m-thick and up to 2~mm-wide YIG film. The in-plane biased film supports MSWs transmission between the transducers with above 50~dB input to output isolation of high-power signals and threshold ranging from –75~dBm to –35~dBm. This makes it ideal for jam-resistant RF platforms. \setlength{\parskip}{2pt}

A key step toward on-chip biasing and applied magnonics is the integrated magnonic device patented by Bertacco et al. \textit{Integrated magnonic device"}~[8/2024, EP4420499A1]. The proposed design of a device comprises a permanent micromagnet, a SW medium with source and detector, and a microactuator (MEMS) that repositions the micromagnet relative to the medium, thereby providing the bias field and enabling low-power, reconfigurable operation without bulky electromagnets. Subsequently, Cocconcelli et al. ~\cite{cocconcelli2025self} explored a related self-biased prototype based on CoFeB waveguides.

Just recently, Chumak et al. demonstrated a cutting-edge RF filter design [\textit{"Filtering Device and Process for Filtering Radiofrequency Signals"}, European Patent Application No. EP23200651.0, 2024]. The device uses nanowaveguides rather than planar YIG film, as their reduced lateral dimensions ensure single-mode operation and suppress multi-magnon scattering, providing an additional parameter to tune the power threshold. Carefully spaced gaps between these nanowaveguides are designed for the deposition of permanent micromagnets \cite{autonomous_micromagnet_2011, keller2021batch, kovacs2023physics}. These magnets locally magnetize the YIG, enabling reconfigurable frequency according to applications.

Alternatively, SW FSL can be realized from magnonic crystals adapted to selectively filter specific spectral components of SWs (e.g., bandwidth, central frequency, phase shift, etc) as they propagate through a periodically modulated magnetic medium. A case in point here is a patent by Ciubotaru et al. [\textit{"Tunable magnonic crystal device and filtering method"} 10,033,078 B2, 07/2018 – US20170346149A1]. 
Another important invention was made by Aquino et al.  [\textit{"Short-wavelength spin wave transducer" } US 2023/0299451A1], who designed transducers to generate ultra-short-wavelength SW for microwave and mm-wave magnonic devices without structural nanopatterning. 

These developments highlight the growing potential of spin-wave technology in next-generation RF systems, offering improved efficiency, miniaturization and performance compared to traditional electronic counterparts.
 
 \vspace{2mm}
\noindent\rule{\columnwidth}{0.4pt}
 \vspace{1mm}
 
\section{Advantages and challenges of spin-waves for RF applications}
\subsection{Advantages}
\vspace{-3pt}
The field of magnonics attracts significant interest from both scientific community and industrial R\&D due to the unique combination of the intrinsic spin-wave properties: \setlength{\parskip}{3pt}

\textcolor{NavyBlue}{{\scalebox{0.75}{$\bullet$}}} {\bf {Scalability of the devices from centimeters down to nanometers.}} Since the minimum size of wave-based elements is defined by the operating wavelength \(\lambda\), the lateral dimensions of SW devices can range from cm, as in classical MSW-based components \cite{Serga2010, Wang2020, adam2004msw}, to sub-100 nm, as recently demonstrated \cite{Wang2020, Heinz2020}. The smallest possible SW wavelength is limited only by the lattice constant of a material, and is thus theoretically in the ångström range. As highlighted by Davidkova et al. \cite{davidkova2025nanoscale}, SW technology, naturally confined to magnetic medium, allows for an easy planar design, facilitating on-chip integration. For example, a chip containing multi-channel frequency-selective limiters can have a footprint below 1~×~1~mm$^2$, with individual device areas as small as 10~×~100~\(\upmu\)m$^2$. \textbf{Therefore, one of the major advantages of SW-based technologies} over conventional RF devices \textbf{is the ability to miniaturize a structure by at least three orders of magnitude, almost independently of operating frequency}. For better understanding, please refer to Table \ref{TableII}, where the wavelengths \(\lambda\) of both electromagnetic and spin waves (dipolar-dominated magnetostatic SW and exchange-dominated SW) are calculated for a 5 \(\upmu\)m-thick YIG film in BV geometry, along with the corresponding bias magnetic fields. Reciprocally, it can also be stated that \textbf{SW devices are universal units, capable of adapting to a wide frequency range without requiring physical scaling} - see Table \ref{TableIII} and subsequent discussion. In contrast to SAW and BAW approaches, which rely on miniaturization to reach higher frequencies, SW devices have an edge in compactness, frequency and scalability, making them particularly attractive for mobile RF applications. 

\begin{table*}[!t]
\caption{Comparison of the frequency-dependent electromagnetic wave (EMW) and spin wave (SW) wavelengths \( \lambda\) calculated for 5 \(\upmu\)m-thick YIG film in backward volume configuration.}
\vspace{-0.3\baselineskip}
\centering
\begin{tabular}{ |m{2.8cm}|m{3cm}|m{4cm}|m{3cm}|m{3cm}| } 
 \hline
\rowcolor{NavyBlue} \multicolumn{1}{|c|}{\textcolor{white}{Frequency $f$}}&\multicolumn{4}{|c|}{\textcolor{white}{Wavelength \(\lambda\) }} \\
 \hline
   & EMW & Magnetostatic (dipolar) SW & Exchange SW & Magnetic field\\
 \hline
\textbf{1 GHz} & 30 cm & $\sim$ 10 \(\upmu\)m & $\sim$ 200 nm & 10 mT\\
 \hline
 \textbf{10 GHz} & 3 cm &  & $\sim$ 70 nm & 10 mT\\
 \hline
  &  & $\sim$ 10 \(\upmu\)m & $\sim$ 200 nm & 300 mT\\
 \hline
 \textbf{50 GHz} & 0.3 cm &  & $\sim$ 30 nm & 10 mT\\
 \hline
  &   &   & $\sim$ 40 nm & 300 mT\\
 \hline
 &   & $\sim$ 10 \(\upmu\)m  & $\sim$ 200 nm & 1.75 T\\
 \hline
\end{tabular}
\label{TableII}	
\end{table*}

\vspace{2pt}
\textcolor{NavyBlue}{{\scalebox{0.75}{$\bullet$}}}{\bf { Ability to tune spin wave wavelength, frequency and velocity independently.}} SW characteristics can be controlled through a variety of parameters, including the choice of magnetic material, geometry of the structure/sample, orientation of the bias magnetic field \cite{Gurevich1996, Stancil} and antennas design. These factors define key operational parameters of SW devices. For instance, the center frequency of a magnonic filter is governed by the SW dispersion; the size of transducers and the overall device footprint is determined by the SW wavelength and the required bandwidth; and the delay time is given by SW group velocity, which can be adjusted by altering transducer spacing and/or the magnetic film thickness. The threshold power in nonlinear devices can also be tuned via the choice of magnetization configuration, transducer design, and materials properties \cite{davidkova2025nanoscale}. Moreover, dynamic control over the passband and bandwidth can be achieved using variable magnetic or electric fields, particularly in hybrid architectures. Hence, the ability to tune multiple parameters independently provides strong flexibility in adapting SW devices to various RF applications.  \setlength{\parskip}{3pt}

\textcolor{NavyBlue}{{\scalebox{0.75}{$\bullet$}}}{\bf { Low noise and tunable power threshold.}}  Following the previous point, we want to highlight the lower noise levels in SW devices at high frequencies vs semiconductor counterparts. A high signal-to-noise ratio, combined with low power thresholds, makes SW technology particularly attractive for sensitive RF front-end applications such as GPS \cite{davidkova2025nanoscale}.  Furthermore, the implementation of nanoscale waveguides would add additional control over the power threshold by suppressing multimagnon scattering, as proposed by Chumak et al. in a recent patent [EP23200651.0]. 

\setlength{\intextsep}{9pt}
\begin{wraptable}{L}{4.73cm}
\caption{Dependency of SW wavelength on frequency (bias field).}
\vspace{-0.3\baselineskip}
\centering
\begin{tabular}{ |c|c|c| } 
 \hline
\rowcolor{NavyBlue} \multicolumn{1}{|c|}{\textcolor{white}{Frequency $f$}}&\multicolumn{2}{|c|}{\textcolor{white}{Wavelength \(\lambda\) }} \\
 \hline
\textbf{1 GHz}  & 2 \(\mu\)m & 30 mT\\  
 \hline
\textbf{10 GHz} & 2 \(\mu\)m & 350 mT\\ 
 \hline
\textbf{50 GHz} & 2 \(\mu\)m & 1.78 T\\  
 \hline

\end{tabular}
 \label{TableIII}
\end{wraptable}
\vspace*{-\baselineskip} 

An example of SW flexibility is shown in Table~\ref{TableIII} for an exemplary 5~\(\upmu\)m-thick YIG film in BV geometry. SW wavelength of 2~\(\upmu\)m can be excited at different frequencies depending on  bias fields.   \setlength{\parskip}{4pt}

\textcolor{NavyBlue}{{\scalebox{0.75}{$\bullet$}}}{\bf { Compatibility with existing industrial technology and standards.}} 
Following the train of thoughts, it can concluded that \textbf{inexpensive and industry-compatible conventional photolithography can be used for fabricating SW RF devices targeting any operational 5G frequency range}. Magnonic RF technology, based on propagating SWs, has consistently relied on \textbf{commercially viable, CMOS-compatible planar fabrication methods} (e.g., via the chiplet concept \cite{Chiplet}). The transition SAW → SW is technologically more straightforward than SAW → BAW, making SW solutions highly attractive for integration into current RF system manufacture. \setlength{\parskip}{4pt}

 \textcolor{NavyBlue}{{\scalebox{0.75}{$\bullet$}}}{\bf { No need in isolation.}} Spin waves are intrinsically confined within the magnetic material and, unlike SAWs or BAWs, do not propagate into the substrate or leak into adjacent circuits. This inherent confinement makes \textbf{SW technology more compatible with industrial integration and less susceptible to parasitic losses}. Additionally, Bragg reflectors commonly used in SAW can be replaced by the flat edge of the magnetic waveguide, resulting in perfect reflection of the spin waves.\setlength{\parskip}{4pt}

\textcolor{NavyBlue}{{\scalebox{0.75}{$\bullet$}}}{\bf { Wide frequency range from sub-GHz to THz.}} Modern information and communication technology systems, such as cellular networks, Bluetooth and Wi-Fi operate in the 4G/5G frequency bands. Boolean computation is also typically performed in the GHz range, with a benchmark overclocked frequency reaching up to 8.5 GHz in cryobath \cite{guinness}.  While most laboratory SW experiments mostly stay within 1-20 GHz, the magnon spectrum spans several orders of magnitude - from hundreds of MHz up to the promising THz regime. The frequency is limited only by the Brillouin zone boundaries of the chosen magnetic material and the strength of the applied bias field. For example, in YIG, the material of choice for many RF applications, the magnon Brillouin zone is limited at about 7~THz \cite{plant1977spinwave}. Thus, the \textbf{same magnonic RF application can fully support current 4G/5G systems and are promising for future 6G technologies}.  \setlength{\parskip}{4pt}

\textcolor{NavyBlue}{{\scalebox{0.75}{$\bullet$}}}{\bf { Low energy transport loss and heat dissipation.}} Typical materials used for SW-based applications are insulating ferrites (e.g., YIG, BaM). Since a magnon current does not involve the motion of electrons, it is free of Joule heat dissipation, enabling efficient signal transmission. Therefore, SWs can propagate over distances larger than centimeter  \cite{Serga2010}, while an electron-spin current has a much shorter span, typically a few micrometers, due to a finite spin diffusion length \cite{brataas2020spin}. The energy losses per unit time (e.g., reported 23~and 46~dB/\(\upmu\)s propagation losses at 9 and 20~GHz, respectively \cite{Ishak1988}) are mostly driven by the Gilbert damping \(\alpha\) and the operating frequency. However, it is often important to translate this loss into the loss per distance, which is also defined by the SW group velocity. The velocity is consequently tuned by the saturation magnetization of a magnetic material and the film thickness for dipolar (magnetostatic) SWs, and by the exchange stiffness for exchange SWs.  \setlength{\parskip}{4pt}

\textcolor{NavyBlue}{{\scalebox{0.75}{$\bullet$}}}{\bf { Pronounced (controllable) nonlinear phenomena.}} In order to process digital information, nonlinear elements are essential, as they allow one signal to manipulate another, similar to how semiconductor transistors operate in electronics. SWs have a variety of pronounced nonlinear effects that can be applied to control one magnon current by another, enabling signal suppression or amplification \cite{Gurevich1996,Stancil,Serga2010,wigen1994}. Such magnon-magnon interactions have already been used to demonstrate magnon directional couplers and transistors \cite{Wang2020, Chumak2014}, paving the way for fully integrated all-magnonic circuits. For RF applications, this inherent nonlinearity can be advantageous, since it allows the realization of power limiters and signal-to-noise enhancers. On the other hand, it can also impose limitations by reducing the maximum operational RF power. Despite this trade-off, the versatility of nonlinear magnon physics facilitates control and suppression of parasitic nonlinear processes, making it a powerful tool in the design of robust SW-based components.  \setlength{\parskip}{4pt}

\textcolor{NavyBlue}{{\scalebox{0.75}{$\bullet$}}}{\bf { Fast reconfigurability of parameters.}} The SW dispersions can be controlled, for example, by the magnetization orientation, the strength of applied magnetic field or via electric current-induced effects (e.g., Oersted fields or spin-orbit torques), enabling reconfigurable signal-processing device with nanosecond switching times. In principle, the same SW device can support multiple functionalities, and both the operational mode and key parameters (e.g., the center frequency of a passband filter) can also be dynamically reconfigured on nanosecond timescales through the same control mechanisms. \setlength{\parskip}{4pt}

\textcolor{NavyBlue}{{\scalebox{0.75}{$\bullet$}}}{\bf { Broad scope of RF functionalities and their integration.}} Spin waves have been successfully implemented \cite{Harris2012, Yttrium, Adam1988, Glass1988, Ishak1988, Rodrigue1988, Schloemann1988}, or proposed \cite{Chumak2022, Barman2021} for a variety of passive and active RF components: \textbf{bandpass and bandstop RF filters, resonators, phase shifters, delay lines, (de-)multiplexers, directional couplers, isolators and Y-circulators, power limiters, signal-to-noise enhancers}. For more complex RF systems, these functionalities can be integrated or dynamically reconfigured within a universal device. 

\begin{table*}[!b]
\caption{Comparison of the phase shifter technologies. Reprinted after \cite{Harris2012}.}
\vspace{-0.3\baselineskip}
\begin{tabular}
{ |m{2.3cm}|m{2.5cm}|m{2.5cm}|m{2.5cm}|m{2.5cm}|m{2.5cm}| } 
 \hline
\rowcolor{NavyBlue} \multicolumn{1}{|l|}{\textcolor{white}{ }}&\multicolumn{1}{|l|}{\textcolor{white}{Ferroelectric}}&\multicolumn{1}{|l|}{\textcolor{white}{MEMS}}&\multicolumn{1}{|l|}{\textcolor{white}{Semiconductor}}&\multicolumn{1}{|l|}{\textcolor{white}{Ferrite (waveguide)}}&\multicolumn{1}{|l|}{\textcolor{white}{Ferrite (microstrip)}} \\
 \hline
Cost & Low & Low & High & Very high & Low\\
 \hline
Reliability & Good & Good & Very good & Excellent & Excellent \\
 \hline
Power handling & > 1 W & < 50 mW & > 1 W & $\sim$ 1 kW & > 10 W \\
 \hline
Switching speed & $\sim$ ns (limited if high voltage) & 10 - 100 \(\upmu\)s & < ns (low power) & 10 - 100 \(\upmu\)s & < 10 \(\upmu\)s \\
 \hline
Radiation tolerance & Excellent & Excellent & Poor (good if hardened) & Excellent & Excellent\\
 \hline
DC power consumption & $\sim$ 1 \(\upmu\)W  & Negligible  & < 10 mW & $\sim$ 10 W ($\sim$ 1 W latching) & < 10 \(\upmu\)W\\
 \hline
Microwave loss & $\sim$ 5 dB/360$^{\circ}$ K band & $\sim$ 2.3 dB/337.5$^{\circ}$ Ka band & 2 dB/bit band Ka band = 8 dB
& < 1 dB / 360$^{\circ}$ X band  & < 2 dB / 360$^{\circ}$ C-Ku bands \\
 \hline
Size & Very small & Small & Small & Large & Small\\
 \hline
\end{tabular}
\label{TableIV}
\end{table*}

\vspace{-7pt}
\subsection{Areas of concern in RF applications}
\vspace{-6pt}

\textcolor{NavyBlue}{{\scalebox{0.75}{$\bullet$}}} {\bf { Insertion loss.}} As discussed earlier, there are two concepts of SW RF applications: the one in which the signal remains in the form of an electromagnetic wave, while magnetic media are used to absorb or re-emit the microwave signal. In this case, the insertion losses are usually in the range of 1~to~3~dB (see Figs.~\ref{f:8}-\ref{f:9}) and are mainly determined by the quality of the microstrip-based microwave engineering and by the reflection of the RF signal from the region where the magnetic structures are placed. Proper RF design of the microstrip lines and the dielectric spacing between the metals and the magnetic material provides enough freedom to optimize the transmission characteristics and minimize the parasitic losses. The second approach is based on the excitation of propagating SWs, their transmission over a certain distance and a subsequent detection via a back-conversion into EMWs. In this case, the insertion loss is higher, typically 6-20 dB. Among the lowest reported insertion losses are those, shown by Zavislyak, Bobkov et al. \cite{Yttrium, bobkov2002microwave} and by Devitt et al. \cite{devitt2025spinwave} – around 2.5~dB (Fig.~\ref{f:11}), and by Wu et al. \cite{wu2012nonreciprocal} – around 1.7-2.4~dB (Fig.~\ref{f:12}). As discussed in \textit{Section III}, total transduction efficiency for the former exceeds 80\%. Other examples of relatively small insertion losses ($\approx$~6-7~dB) are given in Fig.~\ref{f:16} \cite{kalinikos2013nonlinear}, in the context of RF power-dependent performance, and in \cite{freire2003new}, where a novel method for the computation of the transducer insertion loss is discussed. 

\textcolor{NavyBlue}{{\scalebox{0.75}{$\bullet$}}} {\bf { High power capacity.}} To increase operational power capacity in SW RF devices, it is important to distinguish between the two main approaches to SW-based applications, as outlined above. In the first approach, where energy remains within the EMW, the operational RF power can range from tens of watts to kilowatts – see Table \ref{TableIV} \cite{Harris2012}. In contrast, for SW devices based on the propagating waves, increasing the input power may lead to the excitation of nonlinear SW effects, as observed in power limiter devices. Although this ability enables useful functionalities, such as aforementioned power limiting and signal-to-noise enhancement, it is generally undesirable for most passive RF components, as it reduces the dynamic range, degrades device performance and may lead to design trade-offs required for maintaining linearity. One of the most widely used magnonic materials (YIG) and its doped forms, has pronounced nonlinear properties, as they are stronger for films with smaller magnetic losses, defined by FMR linewidth. For example, as shown in Fig.~\ref{f:21}(b), deviations from a linear input-to-output power relation can already start at input powers of around –20~dBm. Based on academic studies \cite{Serga2010}, it is possible to ensure operation in the linear regime up to 10~dBm input power, see Fig.~\ref{f:28} \cite{chumak2012direct}. 

\renewcommand {\thefigure}{28}
\begin{figure}[!h]
\centering
\includegraphics[width=0.94\columnwidth]{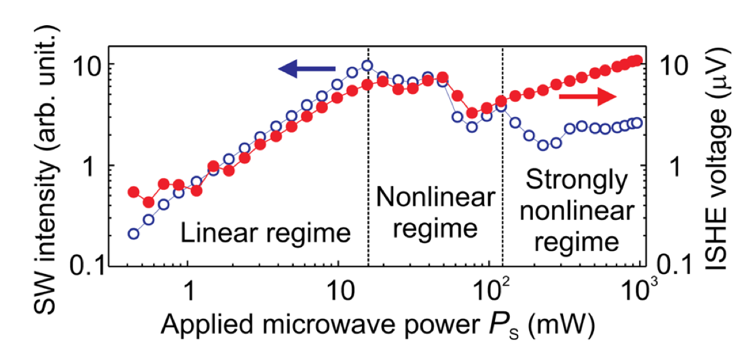}
\vspace{-0.35\baselineskip}
\caption{ Transmitted spin-wave intensity (\textcolor{NavyBlue}{open circles}) and ISHE voltage (filled circles) as functions of applied microwave power \(P_{\textup{s}}\). Adapted from \cite{chumak2012direct}.}
\label{f:28}	
\end{figure}

\textcolor{NavyBlue}{{\scalebox{0.75}{$\bullet$}}} \textbf{ Linearity}. As mentioned above, linearity is a key requirement for some RF components, in particular passive SW RF devices, which are often called nondispersive delay lines \cite{Ishak1988}. These lines maintain a nearly constant group delay across a specified bandwidth, and are considered as potential replacements for phase shifters and coaxial cables at RF frequencies. Applications also include delay-line frequency discriminators, which require low-loss delay lines with time delays between 50 and 500~ns.\setlength{\parskip}{5pt}

\textcolor{NavyBlue}{{\scalebox{0.75}{$\bullet$}}} \textbf{ Magnetic field integration and localization}. One of the main engineering issues of magnonic technology is the need for magnetic bias field to enable SW operation. While generating such a field is manageable in isolated devices or in the laboratory set, localizing it to individual components in the industry-viable integrated system becomes increasingly difficult, as multiple on-chip magnonic devices may require different bias levels. Magnetic fields from closely spaced magnets can overlap, causing reduced signal fidelity, frequency shifts and unwanted SW mode coupling. Additionally, magnets add bulk and require spacing to minimize interactions, which undermines scalability.\vspace{-2pt} 

\textcolor{NavyBlue}{{\scalebox{0.75}{$\bullet$}}} \textbf{ High frequency - high magnetic field trade-off}.
Beyond the sheer need of bias field, another issue lies in its magnitude required to reach high-frequency regime, as emphasized by Davidkova et al. \cite{davidkova2025nanoscale} in a recent cutting-edge study. To operate in a frequency region above 20~GHz, a magnetic biasing field above 700-800~mT is required. While this may not pose significant challenges in laboratory settings, it becomes impractical for industrial implementation, particularly in compact and energy-efficient RF systems. \setlength{\parskip}{6pt}

\subsection{Selected mitigation strategies}
\vspace{-7pt}
There is no recipe of a universal solution to all existing issues in applied RF magnonics - a step towards addressing one challenge might be a step backwards for the other, as discussed further. Yet, recent advancements (\textit{Section II B.}) and growing interest from both academia and tech sector, fueled by market demands, suggest that a compromise solution is within reach. The most immediate pathway for SW technology to enter commercial RF systems is in developing direct, one-to-one replacements for existing components. For example, SW-based phase shifter module that meets the same port configuration and performance metrics of traditional RF device could be adopted almost instantly with minimum changes. \setlength{\parskip}{1pt}

Below, we will address separate potential solutions that address one specific area of concern: \setlength{\parskip}{11pt}

\ul{Issue: Insertion loss.}  \setlength{\parskip}{3pt}

Here we discuss only the approach in which the propagating spin waves are excited and detected by the two spatially separated transducers, since it suffers most from the insertion losses. Even in the state-of-art devices, reported values typically range between 1.7~dB to 2.5~dB. Addressing this issue involves several strategies, including: \setlength{\parskip}{4pt}

\textcolor{NavyBlue}{{\scalebox{0.75}{$\bullet$}}} {\bf {Increasing the thickness of YIG films.}} The excitation and detection efficiencies are primarily defined by the saturation magnetization of the magnetic material and by the volume of magnetic material interacting with antenna's electromagnetic field. Choosing an optimal thickness can significantly increase the transducer coupling efficiency. Moreover, the group velocity of MSWs scales nearly linearly with YIG film thickness. As a result, thicker films reduce propagation time between transducers, thus decreasing SW propagation loss.\setlength{\parskip}{4pt}

\textcolor{NavyBlue}{{\scalebox{0.75}{$\bullet$}}} {\bf {Designing more efficient transducer antennas.}} As discussed in \textit{Sections II B} and \textit{Section III}, transducers, especially in microscale devices, should be carefully optimized to minimize ohmic losses and, consequently, reduce overall insertion losses. This can be accomplished either by reverting to a large external matching networks at a cost of bandwidth or by utilizing a proper circuit-level model that accounts for micromagnetic dynamics. A notable example is the aforementioned work by Erdélyi et al. \cite{erdelyi2025design}, who developed an optimization model for transducers, predicting insertion losses below 5~dB across a 100~MHz frequency range. At the microscale, ohmic losses are largely dissipated as Joule heating, even with matched characteristic impedance of the antennas. High transduction efficiency is feasible only if these losses remain negligible compared to radiation resistance. For short SW wavelengths that require narrow waveguides, increased resistance inevitably leads to higher insertion losses, creating a trade-off between scalability and efficiency. However, for SW wavelengths above several micrometers, low-loss operation is feasible  and can be achieved if, for example, to follow the models' \cite{erdelyi2025design, connelly2021efficient, bruckner2025micromagnetic} design rules. Therefore, even from the standpoint of transducer, magnonic RF devices hold strong potential, particularly in frequency ranges where traditional RF components are limited. \setlength{\parskip}{4pt}

\textcolor{NavyBlue}{{\scalebox{0.75}{$\bullet$}}} {\bf {Selection of magnetic material with higher value of saturation magnetization.}} An alternative strategy involves magnetic materials with higher saturation magnetization than YIG, e.g., metals (Py, CoFeB,..) or hexaferrites (BaM). In principle, this can enhance both the efficiency of SW excitation/detection and MSWs group velocity. However, this approach comes with a significant trade-off, namely increased magnetic damping. These materials typically exhibit much higher magnetic losses than YIG, which still holds the benchmark for ultra-low damping and long SW propagation. As a result, while antennas performance may improve, overall signal transmission may suffer. However, this may change with the development of epitaxial growth methods capable of producing high-quality, high-$M_\mathrm{s}$, low-loss BaM films. \setlength{\parskip}{4pt}

\textcolor{NavyBlue}{{\scalebox{0.75}{$\bullet$}}} {\bf {Engineering the distance between the spin-wave transducers.}} The simplest way to reduce insertion losses due to SW decay is to decrease the distance between the input and output transducers. However, its obvious drawback is associated with the direct electromagnetic leakage \cite{greil2023secondary} when the distance between the antennas becomes sufficiently short (i.e., the RF signal is not transmitted as a SW, but as a direct EMW). The negative effects include reduced out-of-band attenuation for RF filters, degradation of signal quality, etc. Nevertheless, careful antenna design can mitigate parasitic crosstalk. For example, the meander antennas discussed above have highly localized microwave emission and minimal radiation beyond their active area, effectively suppressing leakage. In delay-line structures, reducing the transducer separation shortens the delay time. However, this can be compensated by tuning other parameters, e.g., decreasing the SW group velocity via reduced film thickness.  \setlength{\parskip}{4pt}

\textcolor{NavyBlue}{{\scalebox{0.75}{$\bullet$}}} {\bf {Advanced numerical simulation tools.}} To further accelerate the development and optimization of SW transducers, a conceptually novel approach to numerical simulations is required. As discussed earlier in \textit{Section II B}, a promising direction involves combining the micromagnetic tools with Maxwell solvers to simulate eddy currents. For further enhancement, the simulation model might be fused with machine learning-based inverse-design frameworks. This approach allows a vast exploration of data, potentially uncovering high-performance geometries that would be difficult to identify through manual design alone. The development of such a hybrid approach is currently on-going in the magnonics community. \setlength{\parskip}{4pt}

\textcolor{NavyBlue}{{\scalebox{0.75}{$\bullet$}}} {\bf {Optimizing magnetic film geometry.}} The layout of the magnetic thin film can also be optimized to improve devices efficiency. For example, in studies \cite{kalinikos2013nonlinear}, where device's losses were not a primary concern, authors used a simple YIG film that demonstrated acceptable 6-7~dB insertion loss at 6.34~GHz (Fig.~\ref{f:16}). However, the geometry and positioning of the sample is very different in the insertion loss-optimized device of Zavislyak et al. (Fig.~\ref{f:11}), or in a concept shown by Wu et al. (Fig.~\ref{f:12}) with 2-4x smaller losses. \setlength{\parskip}{4pt}

\textcolor{NavyBlue}{{\scalebox{0.75}{$\bullet$}}} {\bf {Minimizing eddy currents.}} A magnetic layer/normal metal system, despite being a standard (magnon) spintronic model, comes with enhanced SW damping as a result of spin pumping \cite{tserkovnyak2002spin}. The contribution of Ohmic losses from SW-induced eddy currents in the heavy-metal layer to the SW damping in magnetic layer has been experimentally shown by Serha et al. \cite{serha2022low}, with mitigation strategies proposed by Bunaev et al. \cite{Bunyaev2020}. One such strategy involves placing a highly conductive metal plate parallel to the structure surface to redirect eddy currents away from the magnetic layer, thereby partially suppressing the enhancement of Gilbert damping. To be effective, the capping layer should exceed the skin depth of the plate material. \setlength{\parskip}{16pt} 

\ul{Issue: High power capacity.}  \setlength{\parskip}{4pt}

Again, we will only discuss the approach concerning SWs propagating between the antennas. These devices can usually operate with limited applied microwave powers due to multi-magnon and parametric scattering processes \cite{Gurevich1996}. Potential ways to address the challenge include:\setlength{\parskip}{3pt}

\textcolor{NavyBlue}{{\scalebox{0.75}{$\bullet$}}} \textbf{Suppressing nonlinear multi-magnon scattering processes.} The effective strategy to minimize nonlinear scatterings at high MW powers is to decrease the thickness of magnetic film. All multi-magnon interactions are governed by energy and momentum conservation laws. The probability that these conditions are met is determined by the density of available magnon states. Therefore, when the dispersion is diluted due to the mode quantization in ultra-thin films or ultra-narrow nanowaveguides, there are fewer possibilities for fulfilling the energy and momentum conservation laws, resulting in a suppression of multi-magnon scattering. This effect as experimentally demonstrated by Jungfleisch et al. \cite{jungfleisch2015thickness} in nm-thick YIG/Pt heterostructures in the inverse spin Hall effect measurements. The broadening of the FMR linewidth with increasing microwave power, an indicator of nonlinear magnon interactions, was significantly reduced in thinner films of 20 nm. Importantly, only slightly thicker films of around 240 nm displayed strong nonlinear enhancement of damping due to a greater number of magnon modes available for scattering. Therefore, careful engineering of the YIG film thickness can effectively suppress parasitic scattering processes while preserving high transducer efficiency.\setlength{\parskip}{4pt}

\textcolor{NavyBlue}{{\scalebox{0.75}{$\bullet$}}} \textbf{Engineering of magnetic waveguides and use of multiple waveguides}. Assuming that the critical factor is not the absolute value of the energy transferred by the SW, but rather its density (i.e., the energy per unit volume) a straightforward solution would be to use SW films/waveguides of greater width and thickness to distribute the energy over a larger volume of magnetic media. Alternatively, an array of non-interacting MSW waveguides can be used to distribute the input energy among multiple identical channels. This strategy allows higher total input power while keeping the local energy density in each individual waveguide below the nonlinear threshold. \setlength{\parskip}{4pt}

\textcolor{NavyBlue}{{\scalebox{0.75}{$\bullet$}}} \textbf{Engineering of spin-wave transducers with small magnetic field strength.} Another strategy to delay the onset of nonlinear effects involves designing transducers that for the RF signal of the same applied microwave power will induce weaker magnetic field. This can be achieved by adding a spacer between antenna and magnetic medium, increasing the antenna size or modifying its geometry to spread the RF field over a larger area. A weaker local magnetic field induces a smaller precession angle, which in turn reduces the SW amplitude. As a result, the system reaches the threshold for nonlinear processes at higher absolute input power levels \cite{Gurevich1996}. This may support high-power transducer operation without compromising linearity, yet due to inefficient coupling can introduce higher signal losses. \setlength{\parskip}{4pt}

\textcolor{NavyBlue}{{\scalebox{0.75}{$\bullet$}}} \textbf{Selection of magnetic material with higher value of Gilbert damping.} The majority of nonlinear scattering processes have a threshold, where the energy supplied into a SW mode exceeds the rate at which it is dissipated \cite{Gurevich1996}. In practice, this means that materials with higher magnetic damping dissipate energy faster, so it takes more input power to reach the nonlinearity. For device architectures where information is conserved within the EMW, the large damping should not have many drawbacks and commercially attractive materials, such as Py, can be used. However, in systems based on SW propagation, increased damping leads to reduced SW amplitude and shorter propagation lengths, which is generally undesirable. This can be partially mitigated by increasing the SW velocity, e.g., by using thicker magnetic films. In the case of low-damping insulators like YIG, one potential approach to increase the nonlinear power threshold without abandoning the material system entirely is to use polycrystalline or doped YIG. These materials are typically more affordable and have higher Gilbert damping compared to high-quality single-crystal YIG, hence raising the power level required to trigger nonlinearity (but at the cost of increased propagation loss). \setlength{\parskip}{16pt}

\ul{Issue: High linearity.} \setlength{\parskip}{3pt} 

Engineering spin-wave dispersion is essential for achieving linear behavior in RF devices, where signal fidelity and phase coherence are critical. As discussed above, the development of nondispersive, wide-band MSW delay lines with linear characteristics has been the focus of previous studies \cite{Ishak1988}. Many successful approaches were demonstrated, which can now be adapted to new materials and modern device architectures. \setlength{\parskip}{4pt}

\textcolor{NavyBlue}{{\scalebox{0.75}{$\bullet$}}} \textbf{Engineering dispersion parameters.} A detailed list of parameters affecting SW dispersion for various magnetization configurations is provided in \textit{Section I C}. Beyond these, dispersion can also be modified by introducing controlled nonuniformities in the magnetic medium via ion implantation, thermal gradients, lithographic patterning, etc. One effective approach is the spatial modulation of the saturation magnetization on submicrometer scale, as shown by Bunyaev et al. \cite{bunyaev2021engineered} in Co-Fe nanoelements. The disks fabricated with longer e-beam waiting time were irradiated more with Ga ions, leading to a reduction of magnetization and exchange stiffness, hence engineering the dispersion. A simpler, experimentally established method involves tilting the bias magnetic field relative to the film normal in FVMSW-based delay lines. Bajpai et al. \cite{bajpai1983delay} showed that this technique enables precise tuning of delay time across a broad frequency range. \setlength{\parskip}{4pt}

\textcolor{NavyBlue}{{\scalebox{0.75}{$\bullet$}}} \textbf{Adjusting the YIG film thickness and the YIG/ground plane separation.} Using such an approach, it is possible to broaden the band over which the group delay remains flat, and then use the bias field to center this band around the target frequency \cite{ishak1983magnetostatic, Adam1988}. Additionally, by tuning the bias field, the group delay of composite devices can be dynamically controlled. For example, Sethares et al. \cite{sethares1980msw} demonstrated $\pm$8.4~ns delay variation over a 200~MHz bandwidth centered at 3~GHz using a composite MSSW/BVMSW structure. To minimize total loss, both delay lines must have low insertion loss. Other approaches to controllable nondispersive delay lines \cite{Ishak1988} include multiple YIG films separated by dielectric layers, graded periodicity metal strip reflective arrays or application of magnetic field gradient transverse to the film plane. Following the latter, Bajpai et al.  \cite{bajpai1983delay} obtained a variable delay line prototype. By applying magnetic field at an angle to 20 \(\upmu\)m-thick YIG film in a FV geometry, the time delay was adjusted over a $\pm$20\% delay range and had a bandwidth of around 50~MHz at X-band. \setlength{\parskip}{4pt}

\textcolor{NavyBlue}{{\scalebox{0.75}{$\bullet$}}} \textbf{Using MSSW/BVMSW composite device.} Dispersive distortions in the operational characteristics of SW-based prototypes originate from the nonzero second derivative in the dispersion curve, as SWs at different frequencies propagate with slightly different group velocities. This leads to signal distortion, particularly in broadband RF applications. Considering the second derivatives have different signs in the BVMSW and MSSW configurations, it is possible to engineer partial dispersion compensation. For example, cascading BVMSW and MSSW lines in a single transmission path leads to equal dispersions, but opposite in slope over the common operating bandwidth, and a flattened group delay, as demonstrated by Sethares et al.~\cite{sethares1980msw}. This produces a nondispersive tunable delay unit with 165-180~ns delay over a 6\% bandwidth at 3~GHz, which can be adjusted by a bias field variation of one of the lines.\setlength{\parskip}{4pt} 

\textcolor{NavyBlue}{{\scalebox{0.75}{$\bullet$}}} \textbf{The use of exchange spin waves of high wavenumbers.} Classical magnonic RF devices operate with (magnetostatic) SWs of relatively small wavenumbers (still orders of magnitude higher than the wavenumbers of electromagnetic waves). However, modern techniques allow for the operation with exchange-dominated SWs with much higher wavenumbers \cite{Chumak2022, Barman2021}. In the exchange regime, the quadratic dispersion becomes increasingly quasi-linear at large wavenumbers, offering advantages for broadband, nondispersive SW-based RF devices, as the group velocity remains nearly constant across a wide frequency span. Nevertheless, a possible practical limitation arises from the high frequencies associated with exchange waves. In materials like YIG, linear region only appears at very short wavelengths, which often corresponds to frequencies in the 100~GHz range or higher. Therefore, careful theoretical analysis is required.\setlength{\parskip}{4pt}

\textcolor{NavyBlue}{{\scalebox{0.75}{$\bullet$}}} \textbf{The use of active delay line devices and solitons.} Another strategy for realizing a nondispersive MSW delay line is via magnetic solitons, as they overcome dispersion spreading \cite{kalinikos1991envelope, serga2005parametric}. Magnetic solitons are highly stable and once generated by sufficiently strong microwave pulses, they can propagate over long distances at constant velocity maintaining their shape (they do not spread their energy). This stability is caused by the compensation of the nonlinear and dispersive effects. \setlength{\parskip}{16pt}

\ul{Issue: Magnetic field integration and localization}  \setlength{\parskip}{3pt}

There is intensive ongoing research within the magnonics community aimed at resolving or minimizing the negative effects of this issue.

\textcolor{NavyBlue}{{\scalebox{0.75}{$\bullet$}}} \textbf{Development and integration of micromagnets.} Most key SW materials and developed prototypes still require a magnetic field for operations. To adapt acquired knowledge for industrial application, recent efforts have been focused on the development of on-chip micromagnets. For example, Keller et al. \cite{keller2021batch} successfully fabricated high-performance permanent NdFeB micromagnets (30-50~\(\upmu\)m) on a large Si wafer (100~mm in diameter). These magnets achieved high out-of-plane coercivity (up to 2.2~T), with stray magnetic fields 80-100~mT at 30~\(\upmu\)m above the surface. Microscale dimensions and compatibility with SW-based device nanofabrication (lithography, etching, and sputtering) make them promising candidates for integrated on-chip biasing. The issue of field localization was addressed by Zanini et al. \cite{autonomous_micromagnet_2011} with a design and fabrication of chessboard-patterned NdFeB micromagnets (feature size 50 - 200~\(\upmu\)m). Magnets were integrated in a device with microfluidic channels for selective trapping of magnetic particles (\(\upmu_{\mathrm{0}} M_{\textup{s}}\) $=$ 0.6 T). Particles as small as 1 \(\upmu\)m in diameter were captured just after 20 ns with 95\% of efficiency. Importantly, presented micromagnets offered highly-localized fields and minimal stray interference, which is essential for compact lab-on-chip systems and for potential magnonic biasing. To reduce reliance on critical rare-earth materials, Kovács et al. \cite{kovacs2023physics} applied machine learning combined with micromagnetic simulations to design Nd-lean permanent magnets. Their model accurately predicts key properties such as coercivity, magnetization, and thermal stability, offering a path toward sustainable high-performance micromagnets for magnonics. A leading example on the integrated micromagnets is a self-biased magnonic device by Cocconcelli et al. \cite{cocconcelli2025self}. A standalone compact (100~$\times$ 150~$\upmu$m$^2$) proof-of-concept SW phase shifter consists of CoFeB waveguide with two antennas, flanked by permanent SmCo micromagnets and flux concentrators. A tunable MSSW propagation in 3-8~GHz range was achieved, with phase shifts up to 120°at 6~GHz. Reconfiguration was done by modifying distance between the flux concentrators and the micromagnets, leading to a change of localized magnetic fields  11 - 20.5~mT. Even higher frequencies can be reached through optimized micromagnet engineering.  \setlength{\parskip}{4pt}

\textcolor{NavyBlue}{{\scalebox{0.75}{$\bullet$}}} \textbf{Self-biased systems.} To completely avoid magnets, self-biased magnetic structures can be considered based on intrinsic magnetic interactions and anisotropies. We have already discussed a notable work by Louis et~al., \cite{louis2016bias} on bias-free magnonic phase shifter in a \textit{Section III C}. Fundamental results on the design of bias-free waveguides were also reported by Haldar and Adeyeye \cite{haldar2021functional}. Two types were realized: one using exchange-coupled Py/Co-Pd multilayers for isotropic FVSWs, and another using dipolar-coupled rhomboid nanomagnet chains for MSSWs. The SW propagation over 1–2~$\upmu$m with operational frequencies 4.2 to 8.2~GHz was achieved, and SW transmission through 32\textdegree~bends and microwave-assisted MSSW gating were demonstrated. Finally, we would like to mention external-field-free spin Hall nano-oscillators, realized by Gupta et al. \cite{gupta2024magnetization} using uniaxial anisotropy and geometry-driven demagnetization. Here, a change of nanoconstriction width (20–100~nm) modifies the internal effective field, enabling control over frequency (2-10~GHz), amplitude, and threshold current. \setlength{\parskip}{2pt}

Altogether, these studies demonstrate that magnet-free, bias-free magnonic devices are not only possible, but increasingly viable for future low-energy, scalable signal processing systems.\setlength{\parskip}{15pt}

\ul{Issue: High magnetic field}  \setlength{\parskip}{3pt}

As stated earlier in \textit{Section II} and in \cite{davidkova2025nanoscale}, this issue can be solved by using alternative materials with strong crystallographic anisotropy, like Ga-substituted YIG with perpendicular magnetic anisotropy or M-type hexaferrites, depending on the target frequencies. Pure and doped barium hexaferrites (BaM) are particularly promising for applications approaching the 6G range. They offer a range of high operational frequencies up to 100 GHz and exhibit low SW losses in this regime, even under moderate magnetic fields <~1~T.\vspace{-2mm}

\noindent\rule{\columnwidth}{0.4pt}
\vspace{-2mm}
\section{\label{Summary}Summary}
\vspace{-3mm}

This review highlights the current landscape and future outlook of spin-wave technology as a pathway to compact, energy-efficient, and high-performance components for modern RF and signal-processing systems. Core prototypes, including filters, delay lines, power limiters, mixers and phase shifters, have been experimentally demonstrated, yet challenges persist in insertion loss, antenna efficiency, and nonlinear thresholds. YIG remains the benchmark for low-loss magnonics, while alternative materials (e.g., BaM hexaferrite, Py, doped YIG) are being explored to extend functionality beyond its limitations. Looking ahead, innovations in antenna integration, on-chip magnetic biasing, and micromagnetic modeling will be essential for translating spin-wave prototypes into scalable RF technologies. Crucially, their ability to operate across GHz to sub-THz frequencies, combined with ultra-low power consumption and scalability makes spin-wave devices well-suited for meeting the strict latency, size and broadband reconfigurability demands of next-generation 5G and 6G communication systems.
\vspace{25pt}

\section*{Acknowledgments}
The project is funded by the Austrian Science Fund (FWF) project PAT 3864023 IMECS [10.55776/PAT3864023] and the FWF ESPRIT Fellowship Grant ESP 526-N TopMag [10.55776/ESP526]. The authors acknowledge Rostyslav O. Serha for his productive discussions and valuable contributions to the topic of quantum magnonics.

\section*{Author contributions}
KOL developed the review concept and authored the initial draft of the manuscript. KD provided crucial support through scientific discussions and insight into cutting-edge research. JM contributed a unique perspective from the tech industry, assessing the practical application potential. AVC conceptualized the review framework, proposed innovative concepts, and spearheaded the SW RF application research direction. All authors collaboratively contributed to the refinement and finalization of the manuscript.

\section*{Competing interests}
The authors declare no competing interests.


\textbf{Data availability:}
All cited data is an intellectual property of the original authors, and we claim no rights on them. The data supporting the original section of the report dedicated to discussions
are available from the corresponding author upon reasonable request.
\emergencystretch=5em
\bibliographystyle{naturemag}
\bibliography{Review_main}


\end{document}